\documentclass[12pt]{article}
\usepackage{enumerate}
\usepackage{natbib}
\usepackage{latexsym}
\usepackage{amsfonts}
\usepackage{graphicx,amscd,amsmath,amssymb,amsfonts,verbatim,bm,psfrag,epsf}

\usepackage{mathrsfs}
\usepackage{bbm}
\usepackage{multirow}
\usepackage{enumerate}
\usepackage{tikz}
\usetikzlibrary{patterns}
\usetikzlibrary{trees}
\usepackage{pgfplots}
\usepgfplotslibrary{fillbetween}
\usetikzlibrary{shapes.misc}
\usetikzlibrary{shapes.geometric}
\usepackage{setspace}
\pgfplotsset{compat=1.16}
\pgfplotsset{soldot/.style={color=blue,only marks,mark=*}}
\pgfplotsset{holdot/.style={color=blue,fill=white,only marks,mark=*}}
\usetikzlibrary{math}
\usetikzlibrary{patterns}
\usetikzlibrary{svg.path}
\def\bO{\mathbf{O}}
\def\bX{\mathbf{X}}
\def\bo{\mathbf{o}}

\def\q{\mathsf{q}}
\def\qone{\mathsf{q}_1}
\def\qtwo{\mathsf{q}_2}

\newcommand{\blind}{0}

\begin{document}

\def\spacingset#1{\renewcommand{\baselinestretch}%
{#1}\small\normalsize} \spacingset{1}

%%%%%%%%%%%%%%%%%%%%%%%%%%%%%%%%%%%%%%%%%%%%%%%%%%%%%%%%%%%%%%%%%%%%%%%%%%%%%%

\if0\blind
{
	\title{\bf Bayesian Clustering for Continuous-Time Hidden Markov Models}
\author{Yu Luo\thanks{
		Department of Mathematics, Imperial College London, United Kingdom} ,\hspace{.2cm}
	David A. Stephens\thanks{
		Department of Mathematics and Statistics, McGill University, Canada}, \hspace{.2cm}
	David L. Buckeridge \thanks{
		Department of Epidemiology, Biostatistics and Occupational Health, McGill University, Canada},\hspace{.2cm}
}
	\date{ }
  \maketitle
} \fi

\if1\blind
{
  \bigskip
  \bigskip
  \bigskip
  \begin{center}
    {\LARGE\bf Bayesian Clustering for Continuous-Time Hidden Markov Models}
\end{center}
  \medskip
} \fi

\bigskip
\spacingset{1.5}
\begin{abstract}
	We develop clustering procedures for longitudinal trajectories based on a continuous-time hidden Markov model (CTHMM) and a generalized linear observation model.  Specifically in this paper,  we carry out finite and infinite mixture model-based clustering for a CTHMM and achieve inference using Markov chain Monte Carlo (MCMC).  For a finite mixture model with prior on the number of components, we implement reversible-jump MCMC to facilitate the trans-dimensional move between different number of clusters. For a Dirichlet process mixture model, we utilize restricted Gibbs sampling split-merge proposals to expedite the MCMC algorithm. We employ proposed algorithms to the simulated data as well as a real data example, and the results demonstrate the desired performance of the new sampler.
\end{abstract}

\noindent%
\vfill

%\newpage
 % DON'T change the spacing!
	\section{Introduction}
Continuous-time Markov processes on a finite state space have been widely used to represent longitudinal data, especially if the times between successive observations are irregular. Likelihood inference for the infinitesimal  generator of a continuous-time Markov jump process has been studied by \cite{billingsley1961statistical}.  However, in reality, the process is only observed at discrete time points, and inference for the generator of the process becomes difficult. This problem arises in a range of practical settings ranging from public health surveillance \citep{luocthmm2018a} to molecular dynamics \citep{hobolth2009simulation}. Inference for discretely-observed continuous-time Markov processes has been explored by \cite{bladt2005statistical} in both likelihood and Bayesian frameworks.

Continuous-time Markov processes on a finite state space have been widely used to represent longitudinal data, especially if the times between successive observations are irregular. Likelihood inference for the infinitesimal  generator of a continuous-time Markov jump process has been studied by \cite{billingsley1961statistical}.  However, in reality, the process is only observed at discrete time points, and inference for the generator of the process becomes difficult. This problem arises in a range of practical settings ranging from public health surveillance \citep{luocthmm2018a} to molecular dynamics \citep{hobolth2009simulation}. Inference for discretely-observed continuous-time Markov processes has been explored by \cite{bladt2005statistical} in both likelihood and Bayesian frameworks.

In most settings, observed data are either not direct observations of the Markov process, or the process is observed with measurement errors.  In such cases, we may consider a latent process as an unobserved `trajectory' recording the true but unobserved state of nature.  In this setting, a hidden Markov model (HMM) is suitable, and assumes that the Markov property is imposed on the unobserved process. There is a broad interest in the application of continuous-time HMMs (CTHMMs) \citep[see for example][]{jackson2003multistate,lange2015joint}, with the majority of research devoted to frequentist approaches; however,  recently \cite{luocthmm2018a} and \cite{williams2019bayesian} constructed  Bayesian CTHMMs using different missing data likelihood formulations for the underlying Markov chain.

If the cohort is assumed homogeneous with respect to the stochastic properties of the latent model, a common continuous-time model may be fitted to the entire data set.  However, it is plausible that the study base comprises different sub-cohorts that have distinct stochastic properties.  In this paper, we develop model-based clustering procedures to cluster individuals according to their sub-cohort, and to study the pattern in each cluster.  In Section \ref{sec:cthmm}, we review the development of a continuous time HMM/generalized linear model (CTHMM-GLM) for non-equidistant longitudinal data. Section \ref{sec:mbc} discusses model-based clustering approaches; Section \ref{sec:prior} compares the prior distribution of finite and infinite mixture models that are commonly used for clustering models. A reversible-jump MCMC approach to clustering under finite mixture models is presented in Section \ref{sec:fincluster}. Section \ref{sec:infin} presents Dirichlet process mixture model-based clustering implemented using restricted Gibbs sampling split-merge proposals. Simulated examples and a real data example to examine the performance of model-based clustering are presented in Sections  \ref{sec:sim} and \ref{sec:app} respectively, and we discuss possible in Section \ref{sec:dis}.

\section{The CTHMM-GLM model}
\label{sec:cthmm}
We presume that, for a single individual, a sequence $\{O_1,\ldots,O_T\}$ of health status-related variables is observed at time points $\{\tau_1,\dots,\tau_T\}$.  A latent process $\{X_s\}$ ($s \in \mathbb{R}^+$), representing the health status for a condition of interest, is assumed to be a continuous-time Markov chain (CTMC) with parameters $\left(\pi ,Q \right)$, where $\pi$ is the initial distribution and $Q$ is the infinitesimal generator, taking values on the finite state space $\{1,2,\ldots,K\}$.  The assumption of a finite state space for the latent process is justified in many real cases, such as the real example in \cite{luocthmm2018a}, and corresponds to conceptual stages in the progression of the underlying disease process. However such discrete models are undoubtedly an approximation in general.   The observation process $O\left| {X_s,B} \right.$ is presumed to follow an exponential family, with
\begin{equation}
\label{exp}
f\left( {O_t} | X_{\tau_t} = k \right) = \exp \left[ \left\{O_t \theta_{t,k} - b (\theta_{t,k})\right\}/\phi^2 + c ( {{O_t},\phi }  ) \right].
\end{equation}
A GLM with link function specified as $g( u( \theta_{t,k} )) = {\mathbf{Z}_{{t}}}^\top{\beta_{k}}$ can be incorporated if there are  explanatory variables $\mathbf{Z}\in\mathbb{R}^D$, where $u(\theta_{t,k}) = \mathbb{E}\left( {{O_t}\left| X_{\tau_t} = k \right.} \right)$ and $\beta_{k}$ is a coefficient vector for state $k$. Let $S_t=\left(S_{t,1},\ldots,S_{t,K}\right)^\top$ be an indicator random vector with $S_{t,k}=1$ if $X_{\tau_t}=k$ and 0 otherwise. We can rewrite the linear predictor into a matrix form as $g ( u ( \theta_{t,k} ) )   = \mathbf{Z}_{{t}}^\top B S_t$, with  $B = (\beta_{d,k})$ for $d=1,\ldots,D$ and $k =1,\ldots,K$, a coefficient matrix containing all GLM coefficients for each latent state. \cite{luocthmm2018a} developed likelihood and Bayesian inference procedures for this model using the expectation-maximization (EM) algorithm and MCMC approaches. In the formulation, observations are indexed using an integer index (that is, $O_t$), and the latent process using a continuous-valued index (that is, $X_{\tau_t}$). In this paper, we assume that the measurement process itself (that is, the collection of times $\tau_t, t=1,\ldots,T$) is not informative about the system either in its hidden or observed components.

This model, developed in \cite{luocthmm2018a} (the Supplement gives details of the likelihood construction and the Bayesian hierarchical model), is a parametric model for which the likelihood is reasonably complex, albeit one that is simplified in its representation using a latent process.  The model is presumed to apply to all individuals in the study, who are presumed a random sample from the target population.  In this paper we consider an extension to allow for systematic heterogeneity to be exhibited by sub-populations of subjects. 

\section{Model-based clustering}
\label{sec:mbc}

The contribution of this paper is to develop model-based clustering for data presumed to be generated by the complex model in Section \ref{sec:cthmm}; specifically we develop clustering approaches for the latent trajectories based on the observed data and the presumed hidden Markov structure.

Model-based clustering is typically achieved via parametric likelihood- or density-based calculations, with the number of clusters selected using information criteria, such as AIC or BIC \citep{fraley1998many}. This approach was explored extensively by \cite{Luothesis} in the context of CTHMMs for modeling health trajectories. However, in such calculations,  the number of clusters has had to be pre-specified. Bayesian model determination approaches have been a longstanding focus of interest in Bayesian inference \citep[see, for example,][]{carlin1995bayesian,green1995reversible,godsill2001relationship}. Two approaches are typically adopted to address this issue; first, reversible-jump Markov chain Monte Carlo (MCMC) \citep{green1995reversible} exploits trans-dimensional Metropolis-Hastings (MH) moves, allowing movement across parameter spaces of different dimensions; secondly Bayesian nonparametric procedures based on the Dirichlet process are also widely used -- these models are often termed as \textit{infinite} mixture models, where a prior is placed on the space of discrete random measures, and where the models are limiting versions of exchangeable finite mixture models.  Dirichlet process models are now widely used with implementation facilitated via MCMC; key relevant references in a well-established literature include \cite{escobar1995bayesian,maceachern1998estimating,neal2000markov,ishwaran2001gibbs,jain2004split}.

The principal challenge in our setting is that the core model on which the clustering will be based is driven by the unobserved continuous time trajectory represented by $\{X_s\}$.  We meet this challenge by implementing MCMC algorithms based on the complete data likelihood described in Section \ref{sec:cthmm}.

% for Dirichlet processes; methods based on Gibbs sampling can be implemented with conjugate prior distributions %\citep{escobar1995bayesian,neal2000markov,ishwaran2001gibbs,jain2004split} although procedures using more %general priors have also been extensively studied, following the work of \cite{neal2000markov}.
\subsection{Clustering via mixture models}
\label{sec:prior}
The general principle behind model-based clustering is to cluster individuals based on the component model parameters that determine the mixture form. The basic formulation of the model envisages that the population is composed of distinct sub-populations each with distinct stochastic properties.  In the case of the CTHMM formulation, this corresponds to each subpopulation having a potentially different parameter $\Theta = (\pi, Q, B)$.  The estimated parameter for each cluster will provide a subpopulation-level summary, and will allow comparisons within and between clusters.

Let $C_n$ be the cluster membership indicator for individual $n$ with cluster-specific model parameters $\Theta_{C_n}=\left\{\pi_{C_n}, Q_{C_n}, B_{C_n} \right\}$.  The likelihood contribution for this individual is $\mathcal{L}\left(O_n,X_n \left|\Theta_{C_n}\right. \right)$. With $M$ denoting the number of mixture components, a prior probability, $p_0\left(\mathbf{C}\left|M\right.\right)$, would be assigned to each cluster membership partition $\mathbf{C}=\left\{C_1,\ldots,C_N\right\}$, resulting in the posterior form
\[
p\left(\mathbf{C},\Theta\left|\bO,\bX, M\right.\right)\propto p_0\left(\mathbf{C}\left|M\right.\right)\prod_{n=1}^{N}\mathcal{L}\left(O_n,X_n \left|\Theta_{C_n} \right. \right)
\]
where $\mathbf{C}$ is the a specific partition of the $N$ individuals into $M^*$ non-empty clusters. For practical inference, the key is whether the form of $\mathcal{L}\left(O_n,X_n \left|\Theta_{C_n} \right. \right)$ can be easily computed, and the cluster-specific parameter $\Theta_{C_n}$ can be integrated out. This would greatly simplify the MCMC algorithm. In our model specification, conditional on a proposed clustering model, we will adopt the formulation of \cite{luocthmm2018a} for the CTHMM, so therefore differences in inference will be driven by the specific clustering model adopted. In next sections, we compare aspects of the finite mixture and Dirichlet process mixture models, noting their similarities and where they differ.

\subsection{Mixture of finite mixtures}
\label{sec:fmm}
Conventionally, in a finite mixture model, the number of clusters is determined by fitting an $M$-component mixture model with various $M$ and then using information criteria to choose among them.  A natural extension is to regard the number of components, $M$, as a random variable, specifying the prior as $M\sim p_0\left(M\right)$, a mass function on $\left\{1,2,3,\ldots\right\}$,  resulting in the joint prior distribution of the form $p_0\left(\mathbf{C}, M\right) = p_0\left(\mathbf{C} |M \right) p_0\left(M\right)$. However,  there is a crucial distinction between the number of components $M$ in the mixture model and the number of clusters $M^*$ in the data which is defined as the number of components used to generate the observed data, or the number of “filled” mixture components. By specifying a prior on the number of components $M$, we implicitly place a prior on $M^*$.  We typically specify mixture weights
\[
\varpi_1,\ldots,\varpi_M |M \sim {Dirichlet}\left(\delta,\ldots,\delta\right)
\]
with $\delta=1$, making the weight distribution uniform, and then draw  cluster labels $\mathbf{C}$ independently from the multinomial distribution with $ \mathbb{P}\left( {C_n = m} \left|M,\varpi_1,\ldots,\varpi_M\right.\right)=\varpi_m $.   The resulting conditional distribution of $\mathbf{C}$ given that there are $M$ components has the form
\begin{equation}
\label{finp0}
p_0\left(\mathbf{C}\left|M,\delta\right.\right)=\frac{\Gamma\left(M\delta\right)}{\Gamma\left(N+M\delta\right)}\prod_{m=1}^{M}\frac{\Gamma\left(N_m+\delta\right)}{\Gamma\left(\delta\right)}
\end{equation}
where  $N_{m}$ is the number of subjects placed in component $m$ by this procedure.

\cite{miller2018mixture} discussed similarities and differences between the Dirichlet mixture model and a mixture of finite mixtures (MFMs). They showed that when $N \to \infty$, the distribution of  the number of clusters $M^*$ behaves similarly to the number of components $M$, and under the posterior they also behave similarly. In addition, they also illustrated that this prior is  a simple exchangeable partition distribution closely paralleling that of the Dirichlet process, with the distribution of the random partition
\[
p_0\left(\mathcal{Z}\left|\delta\right.\right)= V_N\left(M^*\right) \prod_{m=1}^{M}\delta^{N_m}
\]
where $$V_N\left(M^*\right)=\sum\limits_{M=1}^{\infty}\frac{M\left(M-1\right)\cdots\left(M-M^*+1\right)}{\delta M (\delta M +1) \cdots\left(\delta M +N-1\right)} p_0\left(M\right),\;\; \mathcal{Z} =\{E_m:\left|E_m\right|>0\}$$ 
and $E_m=\{n:C_n=m\}$ for $m\in \{1,2,\ldots\}$. Therefore, posterior samples can be generated via a direct application of MCMC for the Dirichlet process mixture, and the only difference is that the new element is placed in an existing cluster $m$ with probability $\propto \left(N_m+\delta\right)$ or a new cluster with probability $\propto \delta\frac{V_N\left(M^*+1\right)}{V_N\left(M^*\right)}$. In Section \ref{sec:fincluster}, we will explicitly introduce this model in the context of CTHMM and inference based on reversible-jump MCMC  \citep{green1995reversible} with an efficient proposal.  In addition, we can also directly apply the algorithm developed in Section \ref{sec:infin} for Dirichlet mixture models to this model by  replacing $\alpha$ with $\delta\frac{V_N\left(M^*+1\right)}{V_N\left(M^*\right)}$ and $N_m$ with $N_m+\delta$.

\subsection{Dirichlet mixture models}
In the finite mixture formulation, allowing $M \longrightarrow \infty$, given a finite sample size $N$, will still yield a valid prior distribution albeit with some $N_m$'s equal zero. This leads to another choice of a prior distribution, the Dirichlet process  \citep{ferguson1973bayesian,antoniak1974mixtures,lo1984class}.  Specifically, for a Dirichlet process prior, the prior distribution of the partition has the form
\begin{equation}
\label{infinp0}
p_0\left(\mathcal{Z}\left|\alpha\right.\right)= \frac{\alpha^M\Gamma\left(\alpha\right)}{\Gamma\left(N+\alpha\right)} \prod\limits_{m=1}^{M}\Gamma\left(N_m\right)
\end{equation}
and the probability mass function of $M$ and the conditional distribution of $\mathbf{C}$ given $M $ are \citep{green2001modelling}
\begin{equation}
\label{eq:DPpriors}
p_0\left(M\right)=\left| S_{N,M}\right| \frac{\alpha^M\Gamma\left(\alpha\right)}{\Gamma\left(N+\alpha\right)} \qquad \qquad p_0\left(\mathcal{Z}\left|M\right.\right)=\frac{\prod\limits_{m=1}^{M}\Gamma\left(N_m\right)}{\left| S_{N,M}\right|}
\end{equation}
where $\left| S_{N,M}\right|=\sum_{\text{all }\mathcal{Z}} \prod\limits_{m=1}^{M}\Gamma\left(N_m\right)$,
is the absolute value of a Stirling number of the first kind, the number of ways of partitioning $N$ items into $M$ non-empty subsets.

In Section \ref{sec:infin}, we will use the Dirichlet process prior in the context of CTHMMs. This representation marginalizes out parameters of the model, which is particularly useful in terms of the computation. Specifically, a split-merge MCMC for this model will be introduced to facilitate the computation of updating cluster memberships.

\subsection{Calibrating the prior on the number of clusters}

In order to make our analysis under the finite and Dirichlet mixture models as comparable as possible, we aim to match their cluster specifications.  For the parameters in the CTHMM itself, it is possible to match the specification exactly, so we focus on implied the cluster structures.  The Dirichlet process prior on $M$, given by the first expression in \eqref{eq:DPpriors}, can be computed directly for any $N$; this is numerically challenging when $N$ is large, so rather than compute it exactly we adopt a Monte Carlo strategy, as sampling the number of clusters is straightforward: under this prior $M^* \stackrel{d}{=} \sum_{i=1}^N \mathbbm{1} \left\{ U_i > \frac{\alpha}{\alpha + i - 1} \right\}$,
where $U_1,\ldots,U_N$ are independent $Uniform(0,1)$ random variables, and $\mathbbm{1}\{ . \}$ is the indicator function. Table \ref{prid} shows $p_0(M^*)$ computed for $\alpha=0.5$ and $N=1000, 3000, 5000, 10000, 25000$, with Monte Carlo simulation being used to compute for the two largest sample sizes.
\begin{table}[ht]
	\caption{\label{prid} Prior distribution for number of clusters, $p_0(M^*)$ for Dirichlet mixture models (DMM)  with $\alpha=0.5$ and mixture of finite mixtures (MFMs) with $\delta=1$ and $M \sim Poisson(\alpha log(N))+1$.}
	\begin{tabular*}{41pc}{@{}c@{\hskip3pt}c|@{\hskip3pt}c@{\hskip3pt}c@{\hskip3pt}c@{\hskip3pt}
			c@{\hskip3pt}c@{\hskip3pt}c@{\hskip3pt}c@{\hskip3pt}c@{\hskip3pt}c@{\hskip3pt}c@{\hskip3pt}c@{\hskip3pt}c@{\hskip3pt}
			c@{\hskip3pt}c@{\hskip3pt}c@{\hskip3pt}}	
		\hline
		& \multicolumn{15}{c}{$M^*$}\\\cline{3-16}
		$N$ & Model	& 1& 2 & 3 & 4 &5 & 6 & 7 &8 & 9& 10 & 11 & 12 & 13 & $\ge$14 \\
		\hline
		\multirow{2}{*}{1000}& DMM&0.028&0.105&0.191&0.225&0.194&0.131&0.073&0.034&0.014&0.005&0.001&0.000&0.000&0.000\\
		&MFM&0.037 &0.119& 0.188& 0.193& 0.183& 0.126& 0.090& 0.038& 0.021& 0.004 &0.001& 0.000 & 0.000& 0.000\\
		\multirow{2}{*}{3000} & DMM&0.016&0.069&0.146&0.200&0.202&0.160&0.104&0.058&0.027&0.011&0.004&0.001&0.000&0.000\\
		& MFM&0.024 &0.075& 0.160& 0.174& 0.198& 0.153& 0.104& 0.058& 0.030& 0.015& 0.008 & 0.001 &  0.000& 0.000\\
		
		\multirow{2}{*}{5000} &DMM&0.013&0.057&0.127&0.185&0.200&0.169&0.118&0.070&0.036&0.016&0.006&0.002&0.001&0.000\\
		
		&MFM&0.012 &0.068& 0.126 &0.181& 0.177& 0.169& 0.122& 0.076 &0.045& 0.012 &0.008 & 0.002 &0.002  &0.000\\
		
		\multirow{2}{*}{10000}&DMM&0.009&0.044&0.104&0.164&0.192&0.178&0.134&0.086&0.048&0.024&0.010&0.004&0.001&0.000\\
		&MFM&0.005& 0.044& 0.098& 0.168& 0.195& 0.168& 0.129& 0.085& 0.065& 0.023& 0.013& 0.006& 0.001& 0.000\\
		
		\multirow{2}{*}{25000}&DMM&0.006&0.030&0.079&0.137&0.176&0.180&0.151&0.107&0.067&0.037&0.018&0.008&0.003&0.001\\
		
		&MFM& 0.009& 0.034& 0.082 &0.139& 0.169& 0.166 &0.154& 0.106& 0.074& 0.031 &0.022& 0.012& 0.002 & 0.000 \\
		\hline
	\end{tabular*}
\end{table}
The expected number of clusters under this prior is
\[
\sum_{i=1}^N \frac{\alpha}{\alpha + i - 1} = \alpha \left(\psi(\alpha+N) - \psi(\alpha) \right) \bumpeq \alpha \log \left(1 + \frac{N}{\alpha} \right)
\]
where $\psi(.)$ is the Digamma function. Thus the bulk of the mass in this distribution is concentrated on low values of $M^*$ even when $N$ is very large, and the expectation increases very slowly with $N$. To make approaches comparable, in the simulation and real data analysis, we will use this $p_0(M^*)$ as the prior in the finite mixture model discussed in Section \ref{sec:fmm}.

While under MFMs, we can use the similar procedure to sample the number of clusters by replacing $\alpha$ with $\delta\frac{V_N\left(M^*+1\right)}{V_N\left(M^*\right)}$ and $N_m$ with $N_m+\delta$. We use the  Monte Carlo simulation to generate the empirical distribution of $p_0\left(M^*\right)$ (shown in Table \ref{prid}). Under the prior of $M \sim Poisson(\alpha log(N))+1$, the prior of $M^*$ is similar to $Poisson(\alpha log(N))+1$ as well as with  Dirichlet mixture models. In terms of cluster sizes, as is evident from the form in \eqref{eq:DPpriors}, the Dirichlet process prior favours few, large clusters \citep[see, for example,][]{green2001modelling,miller2018mixture}, whereas under the symmetric prior with $\delta=1$, the finite mixture model specification puts higher mass on clusters of the same order of magnitude.

\section{Computation for the Mixture of Finite Mixtures CTHMM}

We deploy reversible-jump MCMC procedures to allow the number of clusters to be inferred via the posterior distribution under a finite mixture model formulation. Specifically, we use a split-combine move to update the number of clusters and then implement fixed dimension MCMC.
\label{sec:fincluster}
We start with finite mixture model-based clustering for CTHMM in this section. We have the following hierarchy:
\begin{align*}
M &\sim p_0\left(M\right),  \ M \in \left\{1,2,3,\ldots\right\}\\
\varpi_1,\ldots,\varpi_M\left|M\right. &\sim \text{Dirichlet}\left(\delta,\ldots,\delta\right)\\
\mathbb{P}\left(C_n=m\left|\varpi_1,\ldots,\varpi_M, M\right.\right)& =\varpi_m, m=1,\ldots, M; n=1.\ldots N \\
X_{n} \left| \Theta, C_n \right. &\sim \text{CTMC}\left(\pi_{C_n} ,Q _{C_n}\right)\\
O_n \left| {X_{n},\Theta,C_n} \right. & \sim{\text{Exponential Family}}\left( {{B_{C_n}}} \right)
\end{align*}
The complete-data likelihood for subject $n$ is
\[
\mathcal{L}\left(O_n,X_n|C_n,\Theta\right)=\prod \limits_{m=1}^M {\left[ \varpi_m \mathcal{L}\left(O_n,X_n \left|C_n=m,\Theta_{m} \right. \right)\right]^{\mathbbm{1}\left(C_n=m\right)}}.
\]
A subject is assigned to cluster $m$ according to the probability
\begin{equation}
\label{postfin}
\mathbb{P}\left(C_n=m\left|{O_n,X_n},\Theta\right.\right)=\frac{{{\varpi_m}\mathcal{L}\left( {O_n,X_n\left| {C_n = m},\Theta_{m}\right.} \right)}}{{\sum\limits_{l = 1}^M {{\varpi_l}\mathcal{L}\left( {O_n,X_n\left| {C_n = l},\Theta_{l} \right.} \right)} }}.
\end{equation}
In reality, the model parameters, $\Theta$, and the values of the latent states, $X_n$, are not known, and must be inferred from the observed data.

\subsection{Reversible-jump MCMC}

The reversible-jump algorithm \citep{green1995reversible} can be deployed to implement trans-dimensional MCMC for the finite mixture model with ${M^*}$ unknown. We study split/combine moves for pairs of states similar in spirit to the split/merge moves of \cite{richardson1997bayesian} and \cite{dellaportas2006multivariate}. If the model consists of $M$ components with $K$ latent states ($K$ is assumed fixed) in each component, then the model can be viewed as a CTHMM-GLM with $K\times {M}$ states.  Infinitesimal generator $Q$ can be expressed as a block diagonal matrix with diagonal blocks $Q_1,\dots,Q_m,\dots,Q_{M}$, where $Q_m$ is the $K \times K$ infinitesimal generator matrix for cluster $m$ $\left(m=1,\dots,{M}\right)$. This constraint prevents the transition of subjects between clusters across time. One iteration of the algorithm includes
\begin{enumerate}
	\item a move that considers \textit{splitting} a cluster into two, or \textit{combining} two clusters into one;
	\item an update of the cluster label for each individual according to the posterior probability \eqref{postfin}, given parameters in each cluster, and the latent states;
	\item an update of the model parameters using standard MCMC moves for each cluster with the number of clusters $M$ fixed, specifically proposals that
	\begin{itemize}
		\item update latent state indicators $S_{n,t}$;
		\item update the parameters associated with the observation process $B$;
		\item update the initial distribution $\pi$;
		\item update the infinitesimal generator $Q$.
	\end{itemize}
	For any empty cluster from Step 2, we generate model parameters from prior distributions.
\end{enumerate}
For the split and combine moves, we will implement a reversible jump algorithm \citep{green1995reversible}.  We carry out this move on the \textit{marginalized} model, where the cluster labels and latent processes are marginalized out from the calculation, and use the likelihood
\[
\mathcal{L} (\bo|\Theta ,M ) = \prod_{n=1}^N \left\{ \sum_{m=1}^M \varpi_m \mathcal{L}(O_n|\Theta_m) \right\}
\]
where $\Theta$ denotes the collection of $\varpi$ and CTMC parameters.

Consider a proposal from the current state $(M,\Theta)$ to a new state $(M',\Theta')$ using the proposal density
$\q\left(M',\Theta';M,\Theta\right) = \qone\left(M';M\right)\qtwo\left(\Theta';\Theta\right)$, that is, using independent proposals for the two components. The acceptance probability for this proposal is given by
\begin{equation*}
\begin{aligned}
\alpha\left(M',\Theta';M,\Theta\right)
&=\min \left(1, \frac{\qone\left(M;M'\right)\qtwo\left(\Theta ;\Theta'\right) p\left(M',\Theta'\left|\bo\right.\right)}{\qone\left(M';M\right)\qtwo\left(\Theta';\Theta \right)p\left(M,\Theta \left|\bo\right.\right)}\right)\\
\end{aligned}
\end{equation*}
where $p\left(M,\Theta\left|\bo\right.\right)$ is the posterior distribution of $(M,\Theta)$ given the observed data $\bo$, which can, up to proportionality, be decomposed into the marginal (or `incomplete data') likelihood of the data $\mathcal{L} (\bo| \Theta,M )$ times the prior distribution for $(M,\Theta)$, with prior distribution as $p_0$;
\[
p\left(M,\Theta\left|\bo\right.\right) \propto \mathcal{L} (\bo| \Theta,M ) p_0(\Theta|M) p_0(M).
\]
Our algorithm relies upon the ability to compute the marginal likelihood efficiently for any $\Theta$; however, this is a standard `forward' calculation for continuous-time hidden Markov models. The reversible-jump algorithm is for the most part standard, and the only noteworthy elements are the trans-dimensional moves.  We discuss these in more detail below.

\subsection{Split and combine move}
To construct efficient split and combine moves, we adopt the idea of centered proposals \citep{brooks2003efficient} designed to produce similar likelihood contributions for the current and proposed parameters. The combine move is designed to choose a cluster, $m$ say, at random and select another cluster $i$ such that $\left\|B_i-B_m\right\|_2$ is smallest for $i\ne m$. The reverse split move is to randomly select a cluster, $m$  to split into two clusters, say $m$ and $m^*$, and check if the condition, $\left\|B_{m*}-B_m\right\|_2 <\left\|B_j-B_m\right\|_2$ for $j\ne m$. If this condition is not met, then the split move is rejected directly.

\subsubsection{Split move}
We consider an update that changes the number of clusters from $M \to M+1$. Without loss of generality, we aim to split the $M^\text{th}$ cluster with parameters $\Theta_{M}=\{\pi_{M},Q_{M},B_{M}\}$ into two clusters, with corresponding parameters $\Theta{'}=\{\pi{'},Q{'},B{'}\}$ and $\Theta{''}=\{\pi{''},Q{''},B{''}\}$.  To implement the idea of centering proposals, we use a deterministic proposal for $Q$ and $\pi$, and let $Q{'}=Q{''}=Q_{M}$ and $\pi^{'}=\pi^{''}=\pi_{M}$.  For observation model parameter $B$, we can use a similar proposal: for $k=1,\ldots, K$, let $\beta_{1,k}^{'}=\beta_{M,1,k}$ and $\beta_{1,k}^{''}\sim \mathcal{N}\left(\beta_{M,1,k},c^2\right)$, with $\beta_{j,k}^{'}= \beta_{j,k}^{''}=\beta_{M,j,k}$ for $j=2,\ldots, D$.  For $\varpi$, let $w\sim Beta(2,2)$ and set $\varpi{'}=w  \varpi_{M}$ and $\varpi{''}=(1-w) \varpi_{M}$. The proposed move is from $M$ to $M+1$ clusters, with the new parameters $\Theta^{p}=\left\{\pi^{p},Q^{p},B^{p}\right\}$ with
\[
\pi^{p}=\left(\pi{'},\pi{''}\right)\qquad Q^{p}=\left(\begin{array}{cc}
Q{'} & 0\\
0&Q{''}\\
\end{array}\right) \qquad B^{p}=\left(B{'},B{''}\right).
\]
If we denote the posterior ratio as
\[
r\left(M+1,(\Theta^{p},\varpi',\varpi'');M,(\Theta_{M},\varpi_M)|\bo\right) = \dfrac{p\left(M+1,(\Theta^{p},\varpi',\varpi'')\left|\bo\right.\right)}{p\left(M,(\Theta_{M},\varpi_M)\left|\bo\right.\right)}.
\]
Then, the acceptance probability for this proposal is
\small{
	\begin{equation}
	\label{as}
	\begin{aligned}
	%\alpha\left(M+1,\Theta^{p} \right. & \left. ;M,\Theta_{M}\right) \\
	%=&
	&\min \left(1, \frac{\q\left(Q_{M};Q^{p}\right)}{\q\left(Q^{p};Q_{M}\right)} \frac{\q\left(B_{M};B^{p}\right)}{\q\left(B^{p};B_{M}\right)}  \frac{\q\left(\pi_{M};\pi^{p}\right)}{\q\left(\pi^{p};\pi_{M}\right)}
	\frac{\q\left(\varpi_{M};\varpi{'},\varpi{''}\right)}{\q\left(\varpi{'},\varpi{''};\varpi_{M}\right)} r\left(M+1,(\Theta^{p},\varpi',\varpi'');M,(\Theta_{M},\varpi_M)|\bo\right) \right)\\[6pt]
	&\qquad = \min \left(1, \dfrac{d_{M+1} \varpi_{M}}{b_M   p_{\varpi}(w )p_{\beta}(\beta_{1,k}^{''}) } r\left(M+1,(\Theta^{p},\varpi',\varpi'');M,(\Theta_{M},\varpi_M)|\bo\right) \right)
	\end{aligned}
	\end{equation}}\normalsize
where $b_M$ is the probability of choosing the split move and $d_{M+1}=1- b_M$ is the probability of choosing the combine move, and $p_{\beta} (\cdot)$ is the Normal density with mean $\beta_{1,K}$ and variance $c^2$ and $p_{\varpi} (\cdot)$ is the $Beta(2,2)$ density.

\subsubsection{Combine move}
For the combine move, we consider an update from $M+1 \to M$ clusters. Again, without  loss of generality, we consider combine the $(M+1)^\text{th}$ and $M^\text{th}$ clusters into one cluster. We first find the stationary probabilities, $s_{M}$ and $s_{M+1}$ associated with $Q_{M}$ and $Q_{M+1}$ respectively. To combine $Q_{M}$ and  $Q_{M+1}$ into $Q^{'}$, the operation is as follows:
\[
q_{i,k}^{'}=\frac{s_{M,i}}{s_{M,i}+s_{M+1,i}}\times q_{M,i,k}+\frac{s_{M+1,i}}{s_{M,i}+s_{M+1,i}}\times q_{M+1,i,k}, i\ne k=1,\ldots,K
\]
and $q_{M,k,k}=-\sum_{i \ne k}q_{M,i,k}$ for $k=1,\ldots,K$. For the observation process parameter $B$,
\[
\beta_{i,k}^{'}=\frac{s_{M,i}}{s_{M,i}+s_{M+1,i}}\times \beta_{M,i,k}+\frac{s_{M+1,i}}{s_{M,i}+s_{M+1,i}}\times \beta_{M+1,i,k}, i, k=1,\ldots,K
\]
For the initial distribution $\pi$,
\[
\pi_{k}^{'}=\frac{s_{M,i}}{s_{M,i}+s_{M+1,i}}\times \pi_{M,k}+\frac{s_{M+1,i}}{s_{M,i}+s_{M+1,i}}\times \pi_{M+1,k},  k=1,\ldots,K
\]
and rescale the sum to 1. For mixture weight update, we set $\varpi^{'}=\varpi_{M}+\varpi_{M+1}$. The acceptance probability for the move from $M+1$ to $M$ clusters is
\begin{equation}
\label{am}
\min\left( 1,\frac{b_M p_{\varpi}(w ) }{d_{M+1}\varpi^{'}}  r\left(M,\Theta{'};M+1,\Theta_{M},\Theta_{M+1}|\bo \right)\right).
\end{equation}

\section{Dirichlet Process Mixture CTHMM}
\label{sec:infin}

In the Dirichlet process mixture model, for some $\alpha > 0$ and distribution $G_0$, we assume that $\Theta_m =\left\{\pi_m, Q_m,B_m \right\} \sim G_0$ for $m=1,2,\ldots$ again represent the component-specific model parameters.  We then assume $\widetilde\Theta_n \sim G \left(.\right)$, and if $v_m \sim Beta(1,\alpha)$, $m=1,2,\ldots$
\[
G (\widetilde\Theta_n )=\sum\limits_{m=1}^{\infty}\varpi_m \mathbbm{1}{(\widetilde\Theta_n=\Theta_m)} \qquad \text{with} \qquad \varpi_m=v_m\prod_{k<m}(1-v_k)
\]
We then suppose that cluster label $C_n$ is defined so that $C_n = m$ implies that $\widetilde\Theta_n=\Theta_m$.  Within cluster $m$, for the latent process, $X_{n} | \widetilde\Theta_n, C_n=m \sim \text{CTMC}  (\pi_m ,Q_m   )$, and for the observations $O_n | {X_{n}, \Theta_n,C_n=m} \sim{\text{Exponential Family}} ( {{ B_m}} )$. We assume \textit{a priori} for the equilibrium probabilities, $ \pi \sim  Dirichlet \left(\alpha_1,\ldots,\alpha_K\right)$, and for the off-diagonal elements in $Q$, $ q_{lm} \sim Gamma\left(a_{lm},b_{l}\right)$ for $1\le l\ne m \le K$; this prior is conjugate with the complete data likelihood -- the representation using the latent trajectory as auxiliary data.  Conjugacy can be relaxed using the approaches outlined in \cite{neal2000markov}, but here we restrict attention to the conjugate case.

\subsection{Split-merge MCMC algorithm for DMM}
Inference for the Dirichlet process mixture model is often carried out using MCMC \citep{escobar1995bayesian,maceachern1998estimating,neal2000markov,jain2004split,jain2007splitting}.  For this model, Gibbs sampling via a P\'{o}lya urn scheme is often used, but this can be slow mixing when clustering a large volume of data, and it is non-trivial to parallelize. This limitation has motivated the development of MCMC algorithms which partially address the inherently sequential nature of the updates. \cite{celeux2000computational} noted that the standard MCMC sampler tends to stay within the neighborhood of the local mode. As a consequence, it is less likely to move to a new mixture component even with well-separated components because of the low probability of moving to an intermediate state: updating a group of subjects simultaneously can help resolve this problem.  \cite{green2001modelling} introduced a split-merge update under the reversible-jump MCMC framework, and also showed how to construct the split proposal. Subsequently, \cite{jain2004split,jain2007splitting} extended this to a Metropolis-Hastings (MH) sampling scheme with split-merge updates: the extended approach suggested splitting the component in a deterministic manner by employing restricted Gibbs sampling, which would increase the probability of forming a new component.

We will use the split-merge approach to obtain the posterior samples.  We first focus on the case where the observation model is covariate-free and then the time-varying covariates are categorical only, and use the original parameterization $\theta$ in the exponential family as in \eqref{exp} instead of $B$.  The algorithm proceeds as follows, with superscript $i=0,1,\ldots$ denoting iteration number.

\begin{itemize}
	
	\item \textbf{Initialization:} Randomly sample cluster labels $\mathbf{C}^{0}$ from $\{1,\dots,M^{0}\}$ for each subject, where $M^{0}$ is an arbitrary positive integer with $\left|\mathbf{C}^{0}\right|=M^{0}$. Starting with initial values (for example, those values can be obtained from the EM algorithm) $\pi^{0}_{C^0}$, $Q^{0}_{C^0}$, $\theta^{0}_{C^0}$ and $\phi^{0}_{C^0}$, compute the `forward' and `backward' quantities $a_{n,t,k}$ and $b_{n,t,k,j}$ using the forward-backward algorithm (see the Supplement for definitions of these quantities).
	
	\item \textbf{Update latent state indicators:} For each $n$ and $t$, generate the random vector $S_{n,t}^{i}$ from the multinomial distribution with the parameter set $a_{n,t}^{i}=(a_{n,t,1}^{i},\ldots,a_{n,t,K}^{i})$ where $S_{n,t}=\left(S_{n,t,1},\ldots,S_{n,t,K}\right)$ is an indicator random vector with $S_{n,t,k}=1$ if $X_{\tau_t}=k$ and 0 otherwise.
	\item \textbf{Simulate the path of the latent process:} For each individual, the path simulation follows in two steps.  First, draw the current and next state ($X_{n,\tau_{n,t}}$ and $X_{n,\tau_{n,t+1}}$) from a multinomial distribution with the parameter matrix $b_{n,t,k,j}$, then simulate ${N_{n,l,m}}\left( {{\Delta _{n,t}}}\right)$ and ${R_{n,l}}\left( {{\Delta _{n,t}}}\right)$ from the Markov processes with infinitesimal generator $Q_{C_n}^{i-1}$ through the intervals $\left[\tau_{n,t},\tau_{n,t+1}\right)$ initiated at $X_{n,\tau_{n,t}}$ and end point $X_{n,\tau_{n,t+1}}$ sampled previously.
	
	\item \textbf{Update $\mathbf{C}$ by split-merge \citep{jain2004split}}: Details of this procedure can be found in the Supplement. 

	\item \textbf{Update the component parameters $\theta$, $Q$ and $\pi$ }:	Assume that $\mathbf{C}^i=\left\{C_1^{i},\ldots,C_N^{i}\right\}$.
	\begin{enumerate}
		\item \textbf{Update $\mathbf{Q}$}: Update the ${N_{l,m}}\left( {{\Delta _{n,t}}}\right)$ and ${R_{l}}\left( {{\Delta _{n,t}}}\right)$ from the updated label component generator $Q_C$, and then $q_{l,m\left|C\right.}^{i}$  associated with component $C$ from  a Gamma distribution with shape parameter $\Lambda_{l,m}^{C}$ and rate parameter $\Upsilon_{l}^{C}$ where $\Lambda_{l,m}^{C}=\sum\limits_{n:C_n^i=C} {\sum\limits_{t = 1}^{T_n} { {N_{l,m}}\left( {{\Delta _{n,t}}} \right)} }+a_{lm}$ and  $\Upsilon_{l}^{C}=\sum\limits_{n:C_n^i=C} {\sum\limits_{t = 1}^{T_n} { {R_{l}}\left( {{\Delta _{n,t}}} \right)} }+b_l$.
		\item \textbf{Update $\mathbf{\theta}$}: For prior $\pi_0\left(\theta\right)$, generate the $\theta_C^{i}$ from the conditional posterior distribution using only the individuals with cluster label $C$ via a Gibbs/MH step.
		
		\item \textbf{Update $\mathbf{\pi}$}: For conjugate $Dirichlet\left(\alpha_1,\ldots,\alpha_K\right)$ prior,  $$\pi^{i}_C \sim Dirichlet  \left(\sum\limits_{n:C_n^i=C}{S_{n,1,1}^{i}}+\alpha_1,\ldots,\sum\limits_{n:C_n^i=C}{S_{n,1,K}^{i}}+\alpha_K\right)
		$$
	\end{enumerate}
\end{itemize}	

\subsection{Acceptance probability for the split-merge proposal}

The acceptance probability for the proposal to update $\mathbf{C}$ for split and merge moves takes the form
\begin{equation}\label{eq:Cupdate}
a\left(\mathbf{C}^{*},\mathbf{C}\right)=\min \left\{1,\frac{q\left(\mathbf{C}\left|\mathbf{C}^{*}\right.\right)}{q\left(\mathbf{C}^{*}\left|\mathbf{C}\right.\right)}\frac{\pi_0\left(\mathbf{C}^{*}\right)}{\pi_0\left(\mathbf{C}\right)}\frac{\mathcal{L}\left(\mathbf{C}^{*}\right)}{\mathcal{L}\left(\mathbf{C}\right)}\right\}
\end{equation}
The prior distribution of $\mathbf{C}$ is the product over the partition of $ \left\{C_1,\ldots,C_N\right\}$.  Therefore
\begin{equation*}
\frac{\pi_0\left(\mathbf{C}^{\text{split}}\right)}{\pi_0\left(\mathbf{C}\right)}=\alpha \frac{\left(N_{C_d^{\text{split}}}-1\right)!\left(N_{C_e^{\text{split}}}-1\right)!}{\left(N_{C_d}-1\right)!}
\quad  \frac{\pi_0\left(\mathbf{C}^{\text{merge}}\right)}{\pi_0\left(\mathbf{C}\right)}=\frac{1}{\alpha} \frac{\left(N_{C_d^{\text{merge}}}-1\right)!}{\left(N_{C_d}-1\right)!\left(N_{C_e}-1\right)!}
\end{equation*}
where $N_{j}$ denotes the count of the number of subjects with label $j$ in the configuration. For MFMs, $\alpha$ and $N_{j}$ will be replaced by $\delta \frac{V_N\left(M^*+1\right)}{V_N\left(M^*\right)}$ and $N_{j}+\delta$, respectively.

As there is only one way to assign all $k\in S$ into one component, then the proposal density $q (\mathbf{C}^{\text{merge}}\left|\mathbf{C}\right. )=  q (\mathbf{C}\left|\mathbf{C}^{\text{split}}\right.)=1$. For $q\left(\mathbf{C}^{\text{split}}\left|\mathbf{C}\right.\right)$, the probability is the product of transition probabilities from the last launch state $\mathbf{C}^{l}$ to the final proposed state $\mathbf{C}^{\text{split}}$, that is, $q (\mathbf{C}^{\text{split}}\left|\mathbf{C}\right. )=\prod_{k \in S}\mathbb{P} (C_k^{\text{split}}\left|C_{-k}\right. )$
which is calculated by the final P\'{o}lya urn scan in the split update %in \eqref{dmsp22}
. Each time, the $C_k$ is incrementally modified during the P\'{o}lya urn scan and the updated $C_k$ is used in the subsequent restricted Gibbs sampling computation.

Given the current labels for $f\in \mathcal{M}\cup \left\{d,e\right\}$, the likelihood contribution is given by
\begin{equation}
\label{likc}
\begin{aligned}
{\mathcal{L}\left(C_f\right)}&={\prod_{f:C_f=j_d}\int \mathcal{L}_f\left(\Theta\right)dH_{f,j_d}\left(\Theta\right)}\times {\prod_{f:C_f=j_e}\int \mathcal{L}_f\left(\Theta\right)dH_{f,j_e}\left(\Theta\right)}
\end{aligned}
\end{equation}
where $H_{f,j}$ is the posterior distribution of $\Theta$ based on $G_0$ and all subjects such that their label $C_g=j$ but $g<f$.  As parameters $\theta$, $\pi$ and $Q$ are separable in $\mathcal{L}_f\left(\Theta\right)$, $\mathcal{L}_f\left(\Theta\right)$  can be written as
\begin{equation}
\label{liklihood}
\mathcal{L}_f\left(\Theta\right)=\prod\limits_{t=1}^{T_f}\prod_{k=1}^{K}f\left(O_{f,t}\left|S_{f,t,k}\right.\right)^{S_{f,t,k}}\prod_{k=1}^{K}\pi_{k}^{S_{f,1,k}}\prod\limits_{t=1}^{T_f}\prod\limits_{l\ne m }\mathcal{L}\left(q_{lm}\left|\Delta_{f,t}\right.\right).
\end{equation}
In the split step, the likelihood of the cluster label configuration is
\begin{equation*}
\begin{aligned}
{\mathcal{L} (C_f^{\text{split}} )}&=\prod_{f:C_f^{\text{split}}=C_d^{\text{split}}}\int{\mathcal{L}_f\left(\Theta\right)dH_{f,C_d^{\text{split}}}\left(\Theta\right)} \times \prod_{f:C_f^{\text{split}}=j_e}\int \mathcal{L}_f\left(\Theta\right)dH_{f,j_e}\left(\Theta\right)
\end{aligned}
\end{equation*}
For the merge step, the likelihood of the cluster label configuration is
\begin{equation*}
\mathcal{L}\left(C^{\text{merge}}\right)=\prod_{f:C_f^{\text{merge}}=j_e}\int{\mathcal{L}_f\left(\Theta\right)dH_{f,j_e}\left(\Theta\right)}
\end{equation*}
The term $q\left(\mathbf{C}\left|\mathbf{C}^{\text{merge}}\right.\right)$ is the product of the transition probabilities from the last launch state to the original `split' state, but there is no actual sampling step since the `split' states are already known. Full details of the calculation of a Poisson example are given in the Supplement.

\section{Examples}
\label{sec:sim}
We examine the performance of the proposed algorithm in simulation. In all examples, data are generated from the finite mixture model with a fixed number of clusters.  We place $Gamma(2.5,5)$ priors on the off-diagonal elements of $Q$,  $\mathcal{N}\left(-2,1\right)$, $\mathcal{N}\left(0,1\right)$ and $\mathcal{N}\left(2,1\right)$ priors for states 1,2, and 3 respectively in the Gaussian case and $Gamma\left(5,10\right)$, $Gamma\left(10,10\right)$ and $Gamma\left(20,10\right)$ respectively in the Poisson case for the components of $B$, and a $Dirichlet\left(10,\ldots,10\right)$ prior for $\pi$. For MFMs, we take $\delta=1$ and $M \sim Poisson\left(0.5log(N)\right)+1$.

\subsection{Example 1}
\label{infmmsim1}
In the first example, a three-cluster CTHMM-GLM is considered, with each cluster having three latent states and transition matrices
{\small \begin{equation}
	\label{simQ}
	Q_1=\left(\begin{array}{rrr}
	-2.5 & 2.0&  0.5\\
	0.5 & -1.5  &1.0 \\
	0.1&  0.9 & -1  \\
	\end{array}
	\right)
	\;
	Q_2=\left(\begin{array}{rrr}
	-1.20 & 1.00&  0.20\\
	1.40 & -1.50  &0.10 \\
	0.05&  0.20 & -0.25  \\
	\end{array}
	\right)
	\;
	Q_3=\left(\begin{array}{rrr}
	-0.50 & 0.49&  0.01\\
	0.25 & -0.30  &0.05 \\
	0.01&  0.10 & -0.11  \\
	\end{array}
	\right)
	\end{equation}}
with associated coefficient matrices
\begin{itemize}
	\item Gaussian case: $B_1=\left(-4,0,5 \right)$, $B_2=\left(-5.5,0.5,5.5\right)$, $B_3=\left(-5,1,4.8\right)$.
	
	\item Poisson case: $B_1=\left(-2 , 1.2 , 3	\right)$, $B_2=\left(-1 ,1 , 2.5\right)$, $B_3=\left(-1.5 , 1.1 , 2.8	\right)$.
\end{itemize}
The initial distributions for three clusters are $\pi_1=\left(0.5,0.4,0.1\right)^{\top }$, $\pi_2=\left(0.3,0.5,0.2\right)^{\top }$  and $\pi_3=\left(0.45,0.45,0.1\right)^{\top }$. In the Gaussian outcome model, the residual error standard deviation $\sigma$ is set to equal 1.  In the split-merge Gibbs sampler, the intermediate restricted sampling scanned three times before performing the actual split or merge update. We initiate the model with one cluster. Data are generated by constructing the continuous Markov chain from the generator $Q_i$ for cluster $i=1,2,3$, a continuous-time realization $\left\{X_s,0 \le s \le 15\right\}$, and uniformly sampled $T-1$ time points between 0 and 15.  Data are generated with 1000 subjects for each cluster.

Results are shown in Table \ref{sim10} (an extended version is given in the Supplement), and are based on the three-cluster iterations taken from a total of 2000 iterations. MFM-RJ represents MFMs using reversible jump MCMC, and MFM-SM represents MFMs using split-merge algorithm. For Dirichlet mixture models (DMMs), in all cases, the posterior modal number of clusters was three, and this is especially clear when $T=50, 100$ over 90\% of iterations resided in the three-cluster model.  For a summary output, cluster membership is assigned to the subject according to its posterior mode conditional on the three-cluster iterations.   When $T=30$ for Normal case, there were 23.70\% and 14.85\% of two and four clusters respectively, while for the Poisson case, 43.40\% of iterations contained four clusters. In general, restricted Gibbs sampling using split-merge proposals performed well; The norm differences in parameters between true values and the posterior means were small, and misclassification rates were low. It was more difficult to cluster trajectories, and the misclassification rate increased, and as sample size decreased, the Gaussian cases had higher misclassification than the Poisson cases. For MFMs, the posterior mode of the number of clusters is three for all cases. MFM-SM has a lower percentage of 3-cluster iterations as the split-merge algorithm encourages more trans-dimensional moves to explore the parameter space, while our reversible jump MCMC yields more accurate in terms of the cluster membership because it is updated  in each sweep. In \cite{miller2018mixture}, they included both the split-merge algorithm and the P\'{o}lya urn scheme in each MCMC step to encourage trans-dimensional moves as well as increase the accuracy of clustering. For a small sample size, this approach is feasible, but in our case when the sample size is large and the model structure is complex, it will be extremely computationally intensive to include both steps.

%In general, we notice there are fewer three-cluster iterations compared to DMMs, while a very few cases are misclassified in FMMs. This is because reversible-jump MCMC updates the number of clusters and the cluster labels %separately, while split-merge MCMC employs the MH step to update them simultaneously.

\begin{table}[ht]
	\caption{\label{sim10}Example 1: Inference for simulated data with three clusters and three latent states.}
	
	{\centering
		\begin{tabular*}{40pc}{@{\hskip5pt}@{\extracolsep{\fill}}c@{}c@{}c@{}c@{}c@{}c@{}c@{}@{}c@{\hskip5pt}}
			
			\hline
			%\multicolumn{6}{c}{EM}\\\cline{1-6}
			&& \multicolumn{3}{c}{Gaussian} &\multicolumn{3}{c}{Poisson}\\ \cline{2-8}
			%\midrule[2pt]
			%\multicolumn{8}{c}{EM algorithm with tolerance 0.05 }\\\cline{1-8}
			&&$T=30$ &$T=50$  & $T=100$ &$T=30$& $T=50$ & $T=100$\\
			\hline
			\multirow{3}{*}{\% 3-cluster iterations} &
			MFM-RJ &62.35\%& 54.40\% & 81.05\%  & 66.85\%  &80.65\%  & 89.95\%\\
			&MFM-SM  & 45.25\% &46.85\%  & 50.25\%   & 48.65\%  &56.60\%   & 58.15\% \\
			&DMM &59.65\%&  99.25\% & 99.75\% & 56.40\% &93.05\% &98.25\%\\
			\hline
			\multirow{3}{*}{Misclassification rate}& MFM-RJ& 1.20\% & 0.60\% &0.00\% & 4.00\%&2.30\%& 0.70\%\\
			&MFM-SM  &15.60\% & 12.70\%  &9.10\%  &12.10\%  & 7.30\% & 2.40\%\\
			&DMM& 20.97\% &19.16\% &12.9\%& 14.10\% &6.53\%&2.40\%\\
			\hline
		\end{tabular*}
	}
\end{table}

\subsection{Example 2}
\label{sec:ex2}
In the second example, data are generated from a finite mixture model with different infinitesimal generators $Q$ within each cluster, but with other parameters $B$ and $\pi$ identical to Cluster 1 from Example 1. The observation process is taken to be the Gaussian distribution only. Data are generated with 1000 subjects in each cluster with 3000 subjects in total.  The restricted Gibbs sampling procedure with split-merge proposals (with five intermediate Gibbs steps) for component parameters $Q$ only was implemented (see Supplement for the algorithm).

Results for this simulation are shown in Table \ref{sim9} with an extended version in the Supplement: again, results presented are conditional on three-cluster iterations with total 2000 iterations. As this is a more difficult clustering problem, the percentage of three-cluster iterations decreased compared to the previous example. However, for DMMs, posterior modes of the number of clusters were still 3 apart from when $\sigma=2$, in which case the posterior was more diffuse; for the DMM also, the posterior distribution of the number of clusters was also more dispersed in this case. For $\sigma=0.5$, 17.05\% of iterations had four clusters, while for $\sigma=1$, there were 4.10\%. When $\sigma=2$, 18.15\%, 38.25\% and 6.05\% of iterations had two, four and five clusters respectively. The misclassification rate for $\sigma=2$ is around 40\%, but the posterior mean estimates were still close to the true values.  While for MFMs, the  posterior modes of the number of clusters were 3 for all cases. Specifically for MFM-RJ with $\sigma=0.5$, the model spent more iterations in two clusters before stabilizing while the model stabilized in three clusters faster, as for $\sigma=0.5$ cluster separation is greater.  When $\sigma=2$, the algorithm tends to move to higher dimensions, with 20.95\%, 19.80\% 18.10\% and 10.15\% of iteration in four to seven clusters respectively.  Similar to the previous example, MFM-RJ had the smallest misclassification rates conditional on three-cluster iterations, but MFM-SM yielded more trans-dimensional moves. This example illustrates that separating clusters via $Q$ is more challenging than for the other parameters. 

\begin{table}
	\caption{\label{sim9}Example 2: Simulation study with three clusters.  Each cluster has three latent states, the same $B$ and $\pi$ parameters, but different $Q$ matrices.}
	\centering
\small{	\begin{tabular*}{41pc}{@{\hskip1pt}@{\extracolsep{\fill}}c@{}c@{}c@{}c@{}c@{}c@{}c@{}c@{}c@{}c@{\hskip1pt}}
			\hline
			\multirow{2}{*}{\# of cluster}& \multicolumn{3}{c}{$\sigma=0.5$} & \multicolumn{3}{c}{$\sigma=1$}& \multicolumn{3}{c}{$\sigma=2$}\\\cline{2-10}
			& MFM-RJ & DMM& MFM-SM & MFM-RJ & DMM & MFM-SM& MFM-RJ & DMM & MFM-SM \\
			\hline
			1& 1.20\% & 0.05\%& 0.10\%&0.65\%&0.10\% & 0.10\%& 0.05\%& 0.15\%&0.25\%\\
			2 & 38.50\%&0.20\%& 26.05\% &0.55\%& 0.05\% & 5.70\% &0.75\%&18.15\%&13.95\%\\
			\textbf{3} & \textbf{58.05\%}&\textbf{80.90\%}& \textbf{54.10 \%}& \textbf{61.05\%}&  \textbf{95.75\%} & \textbf{46.30\%} & \textbf{26.15\%} &\textbf{37.40\%}&\textbf{ 37.20 \%} \\
			4 &1.45\% &17.05\%& 17.25\% &31.95\%  & 4.10\%& 32.60\%& 20.95\%&38.25\%& 28.55\%  \\
			5 & 0.80\%&1.80\%&  2.55\% & 5.80\%& 0.00\%&11.95\% &19.80\%&6.05\%& 12.80\% \\
			6 & 0.00\%& 0.00\%&2.30 \% &0.00\%&0.00\%& 7.80\% &1.80\%& 0.00\%& 6.10\% \\
			$\ge 7$ & 0.00\%& 0.00\%&  0.00\%& 0.00\% &  0.00\% &1.60\%& 14.20\%&   0.00\%&1.15\% \\
			\hline
			\% Misclassification & 4.90\% & 19.20\% &15.40\% &9.50\% & 22.13\% & 19.30\%&24.20\% & 24.70\%&  25.30\% \\
			\hline
	\end{tabular*}}
\end{table}

\subsection{Example 3}
\label{poiscatex}
In the third example, again using a three cluster generating model, we add a three level, time-varying factor covariate $Z_1\sim {Multinomial}\left(1;1/4,1/4,1/2\right)$ at each observation time point to modify a Poisson outcome model, with coefficient matrices
{\small
	\[
	B_1=\left(\begin{array}{rrr}
	-2 & 1.2 & 3 \\
	-0.3 & 0 & 0\\
	0.5& -0.1 & -0.1\\
	\end{array}
	\right)
	\qquad
	B_2=\left(\begin{array}{rrr}
	-1 &1 & 2.5 \\
	0.4 & -0.2 & -0.5\\
	-0.1 &0&-0.4\\
	\end{array}
	\right)
	\qquad
	B_3=\left(\begin{array}{rrr}
	-1.5 & 1.1 & 2.8 \\
	1 & 0.1 & -0.1\\
	-0.5 & 0.1 &-0.5\\
	\end{array}
	\right)
	\]}Parameters $Q$ and $\pi$ are as in Example 1.  Data are generated with 300, 500, 200 subjects in Cluster 1, 2, and 3 respectively with 1000 subjects in total. The intermediate restricted Gibbs sampling scanned twice before performing the actual split or merge update. In this case, we place $Gamma\left(11,10\right)$, $Gamma\left(55,10\right)$ and $Gamma\left(165,10\right)$ for state 1,2,3 as prior distributions for $B$.

Results summarizing the three-cluster iterations for this example are shown in Table \ref{simex3} (full version in the Supplement). We observed similar results in terms of misclassification rates for the Poisson cases in Example 1. For DMMs, as we have more parameters in this case, the posterior distribution of the number of clusters was again more dispersed (see plots in the Supplement).  For $T=30$, the number of clusters fluctuated between three and four clusters, with 37.2\% and 56.8\% of total iterations respectively. When $T$ increases to 50 and 100, the mode becomes three with less posterior variation observed -- for $T=50,100$, only 13.55\% and 2.95\% of 2000 iterations had four clusters respectively. For $T=100$, there were a few iterations with more than five clusters initially, but the number of clusters soon stabilized at three. For MFMs, all cases have posterior modal three. MFM-RJ cases tended to stay in one model for a longer time, which leads to more 3-cluster iterations.  While for MFM-SM cases, they had more dimension changes but suffered from higher misclassification rates, which is consistent with previous examples.

\begin{table}[ht]
	\caption{\label{simex3}Example 3: Simulation study with three clusters with one time-varying factor covariate. Each cluster has three latent states and Poisson observation process.}
	
	{\centering
		\begin{tabular*}{40pc}{@{\hskip5pt}@{\extracolsep{\fill}}c@{}c@{}c@{}c@{}@{}c@{\hskip5pt}}
			\hline
			%\multicolumn{6}{c}{EM}\\\cline{1-6}
			%\midrule[2pt]
			%\multicolumn{8}{c}{EM algorithm with tolerance 0.05 }\\\cline{1-8}
			&	&$T=30$ &$T=50$  & $T=100$ \\
			\hline
			%$\left\| {\pi_1-\hat \pi_1} \right\|$  & 0.08& 0.06 & 0.07\\
			%\left\| {B_1-\hat B_1} \right\|$ & 0.99 & 0.70 &0.08\\
			%$\left\| {Q_1-\hat Q_1} \right\|$ &0.34 & 0.94&0.37\\
			%$\left\| {\pi_2-\hat \pi_2} \right\|$  & 0.10 & 0.05& 0.04\\
			%$\left\| {B_2-\hat B_2} \right\|$ & 0.21& 0.11 &0.04\\
			%$\left\| {Q_2-\hat Q_2} \right\|$ &0.47 & 0.47&0.39\\
			%$\left\| {\pi_3-\hat \pi_3} \right\|$  & 0.12 & 0.10 & 0.12\\
			%$\left\| {B_3-\hat B_3} \right\|$ & 0.42 & 0.36 &0.19\\
			%$\left\| {Q_3-\hat Q_3} \right\|$ &0.39 & 0.39&0.31\\
			\multirow{2}{*}{\% 3-cluster iterations}	&MFM-RJ& 97.55\%  & 98.65\% & 98.55\%\\
			&MFM-SM & 65.90\% & 65.90\% &87.50\%\\
			&DMM& 37.20\%&  85.05\% & 95.60\%\\
			\hline
			\multirow{2}{*}{\% Misclassification}	&MFM-RJ&  3.50\% & 1.90\%  & 0.60\%\\
			&MFM-SM & 12.90\% &7.00\%  &3.60\%\\
			&DMM&11.50\% &6.20\% &2.90\%\\
			\hline
		\end{tabular*}
	}
\end{table}

\section{Real example: Health surveillance of COPD patients}
\label{sec:app}
Our real health trajectories example relates to healthcare surveillance for the chronic condition COPD in greater Montreal, Qu\'{e}bec, Canada.  In 1998, a 25\% random sample was drawn from the registry of the R\'{e}gie de l'assurance maladie du Qu\'{e}bec (RAMQ, the Qu\'{e}bec provincial health authority) with a residential postal code in the census metropolitan area of Montreal. Subsequently at each year, a 25\% random sample of residents new to Montreal within the previous year were sampled to maintain the representative cohort. If people died or changed their residential address outside of Montreal, the follow-up was ended. The data include outpatient diagnoses and procedures submitted through RAMQ billing claims, and procedures and diagnoses from inpatient claims. Using established case-definitions based on medical diagnostic codes \citep{blais2014quebec,lix2018canadian}, 76,888 COPD patients were enrolled with an incident event occurring after a minimum of two years at risk with no events.  Patients were followed from January 1998, starting from the time of their first diagnosis, until December 2014.  Physicians only observed these patients during medical visits, which occurred when patients chose to interact with the healthcare system, and at which information, including the number of prescribed medications, is collected.  However, as this information was only available for patients with drug insurance, we restrict the cohort to patients over 65 years old with COPD, as prescription data are available for all of these patients.

In our analysis, we  fit and compare MFMs and DMMs. Specifically, we use reversible-jump MCMC for MFMs, as from simulation studies it yielded more accurate results with a reasonable amount of trans-dimensional moves. The number of states in each cluster was fixed to four to match the convention in the COPD literature \citep{GOLD}, with the states termed mild, moderate, severe and very severe.  We fit the model with parameters $Q$, $\theta_\text{HOSP}$, $\theta_\text{SPEC}$, $\theta_\text{GP}$, $\theta_\text{ER}$ and $\pi$ in each cluster, where $\theta_{U}$ ($U=$ HOSP, SPEC, GP, ER) represents the log mean parameter for the number of drugs prescribed in a Poisson model for each healthcare utilization. Each algorithm was initiated in the one-cluster model.  Using elicitation via simulating trajectories, subjective prior distributions for off-diagonal elements of $Q$ is $Gamma\left(20,500\right)$, chosen with consideration for plausible holding and transition times within each state, and $Gamma\left(1.5,10\right)$, $Gamma\left(30,10\right)$, $Gamma\left(60,10\right)$ and $Gamma\left(100,10\right)$ for state 1, 2, 3, and 4 respectively for $B$ and $Dirichlet\left(10,\ldots,10\right)$ for $\pi$. Finally, we place $p_0(M)$ in Table \ref{prid} with $N=25000$ as the prior distribution on the number of clusters for the finite mixture model.

In the analysis to compare the performance of the two models and demonstrate the feasibility of clustering trajectories,  we implemented this method on a data set comprising of 24,712 COPD patients.  We present results for both models based on 2000 MCMC samples with 100 burn-in iterations after initialization using an EM algorithm fit of the one-cluster model as described in \cite{luocthmm2018a}. For both models, the mode of the posterior distribution of the number of clusters was four, and hence the results presented are from the four-cluster iterations. The Supplement gives details of the effective sample sizes of the posterior samples for $Q$ and $B$ in each cluster: these were satisfactory. Parameters $\theta_\text{HOSP}$, $\theta_\text{SPEC}$, $\theta_\text{GP}$ and $\theta_\text{ER}$ are converted to a contrast (relative-risk) parameterization in each cluster for both models. The posterior mean of the exponential of the coefficients for the four-cluster model with the number of patients in each cluster are shown in Tables in the Supplement. The numbers of drugs across different healthcare utilizations are different within each state. All four clusters have distinct numbers of prescribed drugs in each state. In general, specialist visits on average has smaller number of drugs prescribed compared to general practitioner visits; Cluster 1 has on average a greater number of drugs prescribed. Cluster 2 is the largest cluster, and has on average a greater number of drugs prescribed than the other two clusters.  The posterior distribution for the Dirichlet mixture model varied slightly more than for finite mixture models as a result of the nature of the assumptions on cluster structure.

Recall that one of the principal computational challenges in this problem is that at each iteration we essentially need to sample and store complete trajectories for each of the almost 25,000 patients in the study; however, these posterior sampled trajectories can themselves be useful in gathering inference concerning time spent in each state, time to transition into more severe states and so on.  In the Supplement, we show figures of posterior mean transition patterns over 5 years for the four clusters. In both plots, we observed similar transition patten; we summarize as follows: for State 1, patients in Cluster 2 are the most likely to hold in State 1, while Cluster 4 has the greatest probability for patients to exit this state; for State 2, all the clusters have a similar transition pattern; in State 3, patients in Cluster 2 have a relatively higher probability transition to State 4 compared to other clusters; Cluster 2 has notably different patterns from other clusters as patients tend to stay in State 4 but are more likely to transition to State 4 from State 3 -- it is believed that those patients are likely in a progression from middle to end stage COPD.

\begin{figure}[ht]
	\centering
	\caption{Real data analysis: finite mixture model (left) and  DP mixture model (right), pair plots of real component of eigenvalues of posterior samples for $Q$. Black, red, green and blue represent Clusters 1,2,3,4 respectively.}
	\begin{minipage}[b]{0.45\textwidth}
		\includegraphics[width=\textwidth]{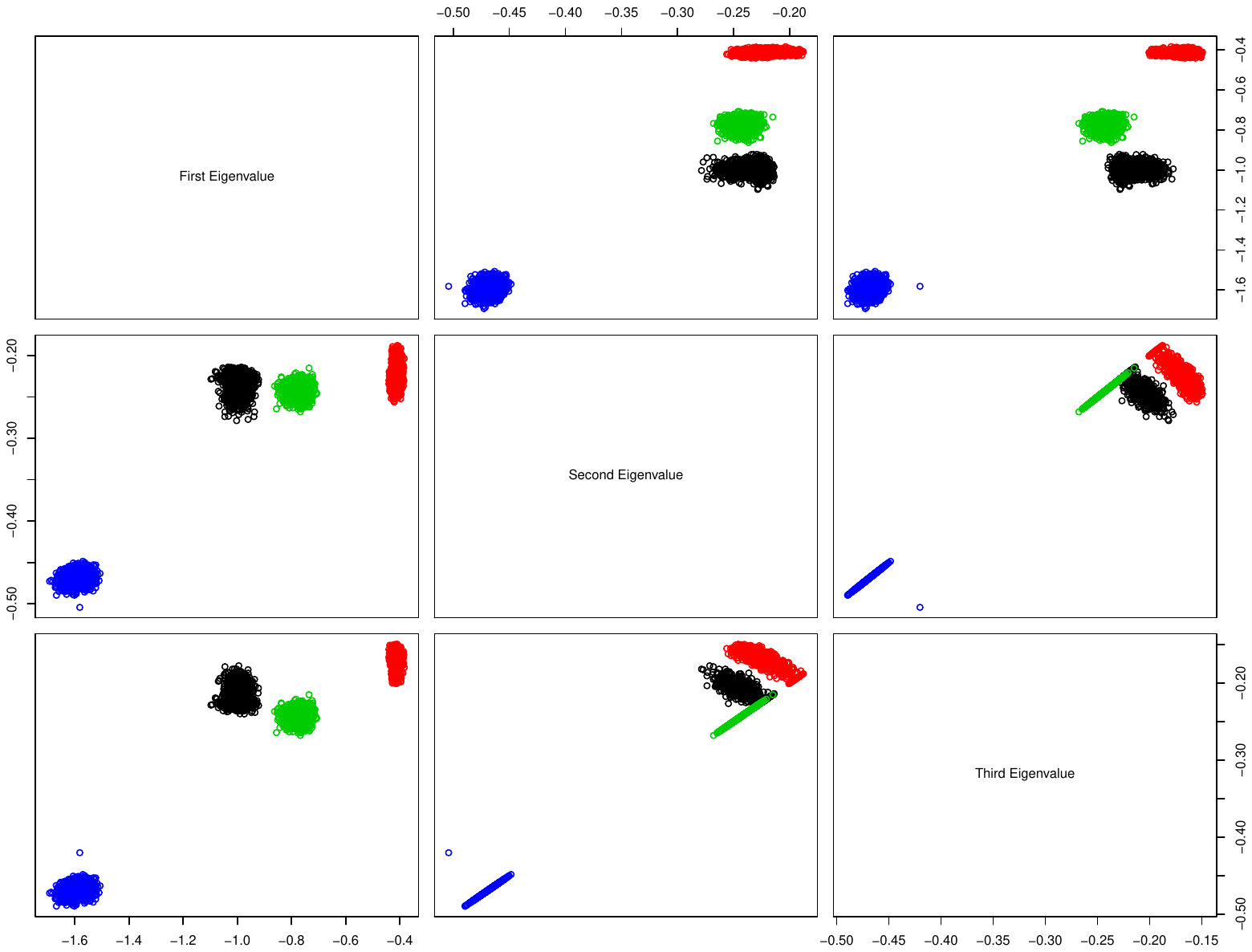}
		%\caption*{Gamma$\left(20,500\right)$}
	\end{minipage}
	\hfill
	\begin{minipage}[b]{0.45\textwidth}
		\includegraphics[width=\textwidth]{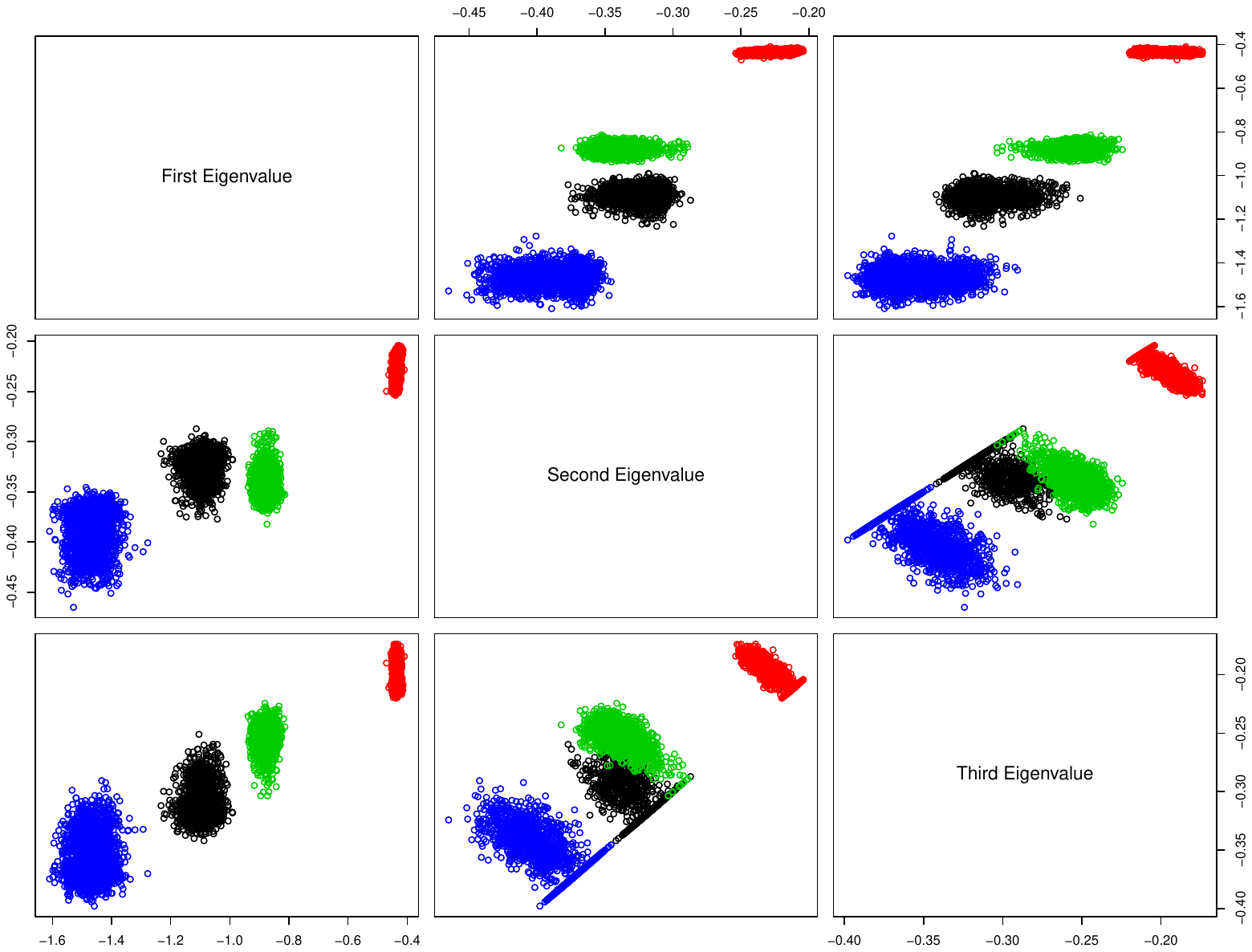}
		%\caption*{Gamma$\left(20,500\right)$}
	\end{minipage}
	\hfill	\label{eigs1}
\end{figure}

Due to the data structure, with different start times and observations times for each subject, and the latent nature of the modeled process, it is challenging to demonstrate the clustering of individual trajectories by using conventional plotting of longitudinal curves so other summaries must suffice.  The transition probabilities Figures discussed previously reveal some but not all of the differences between clusters. To examine further the distinction between clusters, Figure \ref{eigs1} display the real component of the three (non-trivial) eigenvalues of the posterior sampled $Q$ matrices in the four cluster realizations.  In both plots, it clear that although there is some overlap for the second and third eigenvalues between Clusters 1 and 3, they are distinct.   This evidence, coupled with the differences in outcome response levels indicated by the results in Tables of the posterior mean of the exponential of the coefficients --  Cluster 1 has on average a greater number of drugs prescribed, indicating that patients in Cluster 1 have more severe health condition than those in Cluster 3 -- confirms the presence of population substructure.

Table \ref{comatrix20k} displays the co-clustering matrix for two methods. The two methods are consistent in grouping the majority of patients into the same cluster. Clusters 1 and 3 showed some overlap in Figure \ref{eigs1},  while the Dirichlet mixture model has a more distinct separation. This is because a high percentage of subjects in Cluster 1 from the Dirichlet mixture model is grouped in Cluster 3 in the finite mixture model, which brings the separation less distinct in the left panel of Figure \ref{eigs1}. As discussed in the simulation example, the discrepancy in cluster memberships of two methods mainly comes from the updating mechanism of  the number of clusters and the cluster membership of the two algorithms. Reversible-jump MCMC updates the cluster membership each iteration after the split/combine move of the number of clusters. However, split-merge MCMC updates both the cluster membership and the number of clusters simultaneously, which results in  the cluster membership changing less often as reversible-jump MCMC.

\begin{table}[ht]
	\caption{\label{comatrix20k} COPD data analysis: Co-clustering matrix for the two clustering methods. Cluster membership determined by the posterior modal cluster.}
	{
		\begin{tabular*}{40pc}{@{\hskip5pt}@{\extracolsep{\fill}}c@{}c@{}c@{}c@{}c@{}c@{\hskip5pt}}
			\hline
			&	\multicolumn{5}{c}{Finite mixture model} \\
			\hline
			&	& Cluster 1 & Cluster 2 & Cluster 3 & Cluster 4\\
			\hline
			\multirow{3}{*}{Dirichlet mixture model} & Cluster 1  & 1002  & 563& 584& 697  \\
			& Cluster 2  &322 & 9950 & 2421 & 1005  \\
			& Cluster 3     &535  &1077 &2903 & 847  \\
			&	Cluster 4 & 249 &654 & 613 &1290\\
			\hline
	\end{tabular*}}
\end{table}

\section{Discussion}
\label{sec:dis}
We implemented model-based clustering procedures for health trajectories based on continuous-time hidden Markov models based on finite and infinite mixture models. The methodology was applied to simulated examples, where the Markov transition rate matrices dictate that each cluster has its own transition characteristics and observation process.  The posterior distribution for the cluster labels was computed by reversible-jump MCMC \citep{green1995reversible} for finite mixture models and via P\'{o}lya urn schemes \citep{neal2000markov} and more efficient split-merge updates \citep{jain2004split} for Dirichlet mixture models. Simulation studies demonstrated that both mixture models can identify the correct number of clusters and sample the target posterior distribution.  Only the conjugate DP mixture case was explored in this paper; for handling the non-conjugate case, an auxiliary variable method (\cite{neal2000markov}, Algorithm 8) or non-conjugate split-merge proposals \citep{jain2007splitting} could be used.  This would allow us to incorporate continuous time-varying covariates in the observation process and baseline covariates of general form in $Q$.  Alternatively, if log-linear specifications are used in latent and observation models, an approximate conjugate analysis can be carried out using Gaussian approximations.  In our illustrative analyses, we have examined a cohort of over 24,000 patients which comprised around one million records.  This demonstrates the computing capability of our proposed algorithm.  The main computational obstacle is the imputation of the latent process which is needed to facilitate all aspects of the likelihood calculation, and this may be in part overcome by parallelization of the individual likelihood calculations.

Focusing on the number of clusters, the Dirichlet precision parameter $\alpha$ and the prior distribution on the component specific parameters both influence the number of inferred clusters -- this is inevitable in the context of the unsupervised learning problem to which model-based clustering corresponds.  For example, a more precise prior on the components of $Q$ will encourage more clusters for the same observed data.  Similarly, larger values of $\alpha$ will encourage more clusters; it is also straightforward to treat $\alpha$ as an unknown parameter in the MCMC algorithm, although we did not pursue that approach here.

In this paper, we have not explored an approach that allows the number of states to vary, but that is also possible. An efficient construction of proposal distributions is required in order to allow the sampler to move around and explore the parameter space  for both the number of clusters and states \citep{brooks2003efficient}.  We have extended our algorithm to allow the number of states to be inferred using the reversible jump approach, and will report on this elsewhere.

In \cite{luocthmm2018a}, other potential extensions or the model are discussed, and we recap them here. We have not addressed issues of informative dropout; in principle, dropout can be handled using standard Bayesian missing data procedures once a suitable missingness mechanism has been proposed, although the specification of a realistic model may be challenging to construct, and for the COPD data, the influence of informative dropout is likely to be minimal, as the principal cause of dropout is subjects leaving the province, which mainly affects younger subjects.  Secondly, censoring due to death of the subject can be handled within the framework of the latent HMM by the inclusion of an absorbing state: however, information on time of death is not available in our real data set as the cohort contains only data on interactions with the healthcare system.  This could be addressed by linking the RAMQ data to provincial death records. We assume here that the sequence of observation times $\tau_t, t=1,\ldots,T$ is not informative about the underlying Markov chain or the outcome process, but the model could be extended to account for informative observation times by introducing a further stochastic process, for example an inhomogeneous Poisson process with rate dependent on the latent HMM -- this would complicate the computation and be inhibitive for large data sets, but would still be feasible in principle.

\bibliographystyle{chicago}
\bibliography{sample.bib}

\clearpage

\appendix

 \begin{center}
	{\LARGE\bf Supplementary Materials for ``Bayesian Clustering for Continuous-Time Hidden Markov Models"}
\end{center}

\section{The CTHMM-GLM Model}

This section outlines the formulation of \cite{luocthmm2018a}.  To recap the model for a single individual, a sequence $\{O_1,\ldots,O_T\}$ of variables is observed at time points $\{\tau_1,\dots,\tau_T\}$.  A latent process $\{X_s\}$ ($s \in \mathbb{R}^+$), representing the health status for a condition of interest, is assumed to be a continuous-time Markov chain (CTMC) with parameters $\left(\pi ,Q \right)$, where $\pi$ is the initial distribution and $Q$ is the infinitesimal generator, taking values on the finite state space $\{1,2,\ldots,K\}$.  The transition probability from state $i$ to $j$ in the time interval of length $\Delta_t = \tau_{t+1}-\tau_t$ between observation times $t$ and $t+1$ takes the form
\[
p_{ij}\left(\Delta_t\right)=\mathbb{P}\left(X_{\tau_{t+1}}=j \left | X_{\tau_{t}}=i,\tau_{t+1}-\tau_{t}=\Delta_t \right. \right)=\text{expm}\left(\Delta_t Q\right)_{\left(ij\right)}
\]
where $t=1,\ldots,T-1$, $Q=\left(q_{ij}\right)$ for $1\leq i,j\leq K$ is the infinitesimal generator for the continuous-time Markov process $\{X_s\}$, and
$\text{expm}\left(A\right)$ is the matrix exponential of matrix $A$. The initial state distribution $\pi$ for $\{X_s\}$ is $\pi_i = \mathbb{P}\left(X_{0}=i\right)$ for $i=1,\ldots,K$. The model is parameterized by $\Theta=\left\{q_{ij}, 1\le i,j\le K ,\pi,B, \phi\right\}$. The following diagram provides a schematic of the presumed data generating structure for one subject.

\tikzset{filled/.style={fill=circle area, draw=circle edge, thick},
	outline/.style={draw=circle edge, thick}}

\begin{figure}[ht]
	\centering
	\includegraphics[scale=0.3]{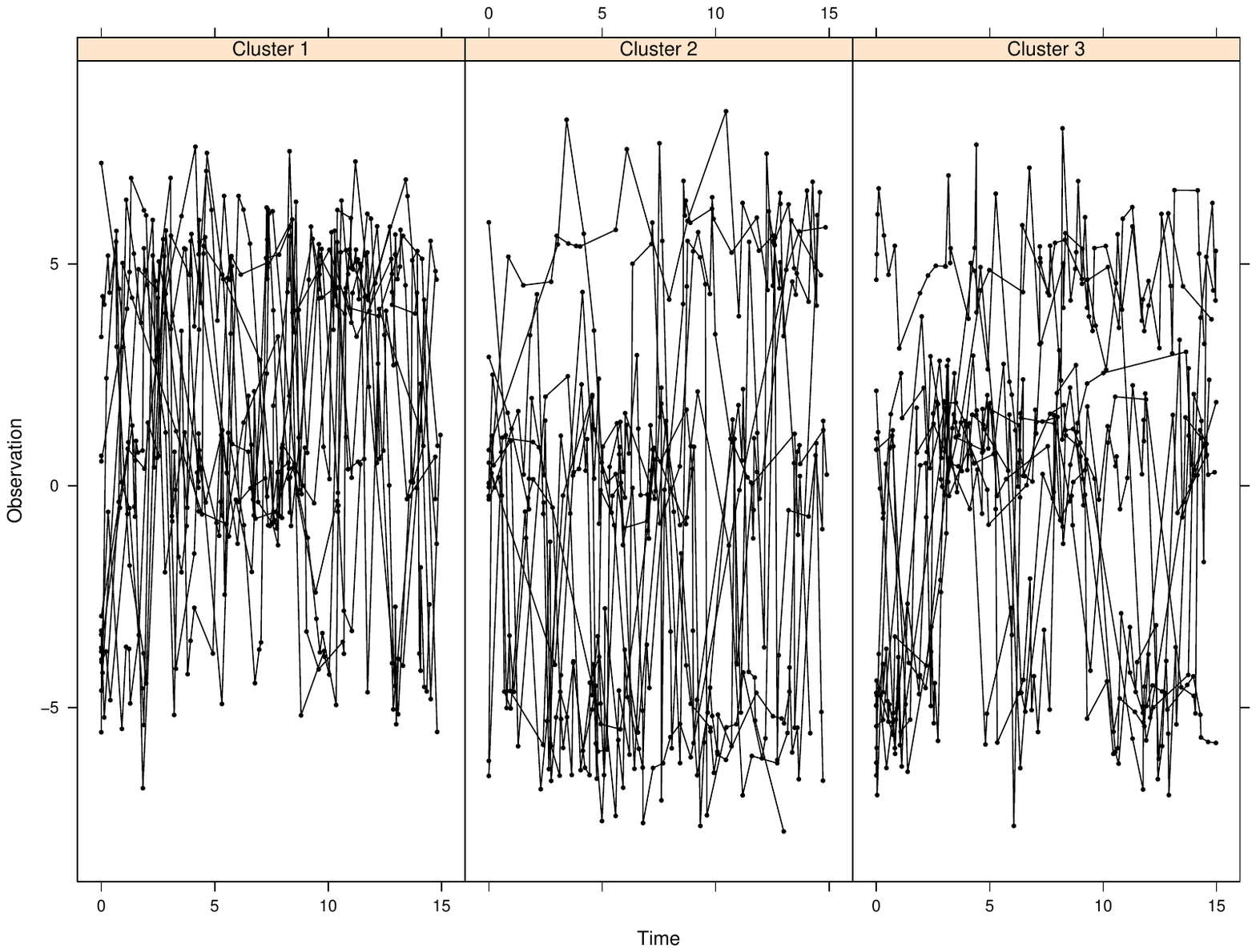}
	\caption{Observed data trajectories from three clusters from simulated Example 1, with different Markov and outcome parameters. Panels represent different clusters.}
	\label{traj}
\end{figure}

\tikzset{cross/.style={cross out, draw=black, minimum size=2*(#1-\pgflinewidth), inner sep=0pt, outer sep=0pt},
	%default radius will be 1pt.
	cross/.default={1pt}}
%\begin{figure}[ht]
\begin{center}
	\begin{tikzpicture}
	
	\draw[decorate,
	decoration={zigzag,
		pre length=5.5cm,
		post length=2cm,
		amplitude=2mm
	}] (0,0) -- (10,0);
	\draw[decorate,
	decoration={zigzag,
		pre length=5.5cm,
		post length=2cm,
		amplitude=2mm
	}] (0,1) -- (10,1);
	\draw[decorate,
	decoration={zigzag,
		pre length=5.5cm,
		post length=2cm,
		amplitude=2mm
	}] (0,2) -- (10,2);
	
	\draw (0,0)  node[cross=3pt]{};
	\draw (0,0.1)  node[above] {1};
	\draw (2,0)  node[cross=3pt]{};
	\draw (2,0.1)  node[above] {2};
	\draw (2.5,0)  node[cross=3pt]{};
	\draw (2.5,0.1)  node[above] {3};
	\draw (4.1,0)  node[cross=3pt]{};
	\draw (4.1,0.1)  node[above] {2};
	\draw (9,0)  node[cross=3pt]{};
	\draw (9,0.1)  node[above] {1};

	\fill (0,2)  circle[radius=2pt] node[above] {$\tau_1=0$};
	\fill (1,2)  circle[radius=2pt] node[above] {$\tau_2$};
	\fill (1.5,2)  circle[radius=2pt] node[above] {$\tau_3$};
	\fill (2.4,2)  circle[radius=2pt] node[above] {$\tau_4$};
	\fill (4.5,2)  circle[radius=2pt] node[above] {$\tau_5$};
	%\fill (6.0,2)  circle[radius=2pt] node[above] {$\tau_6$};
	%\fill (7.5,2)  circle[radius=2pt] node[above] {$\tau_7$};
	\fill (8.5,2)  circle[radius=2pt] node[above] {$\tau_T$};
	%\fill (10.0,2) square[radius=2pt] node[above] {$10$};
	%\draw [black] plot [only marks, mark=square*] coordinates {(10,2)} node[above] {$s=10$};
	
	\draw (0,1)  circle[radius=2pt] node[above] {$O_1$};
	\draw (1,1)  circle[radius=2pt] node[above] {$O_2$};
	\draw (1.5,1)  circle[radius=2pt] node[above] {$O_3$};
	\draw (2.4,1)  circle[radius=2pt] node[above] {$O_4$};
	\draw (4.5,1)  circle[radius=2pt] node[above] {$O_5$};
	%\draw (6.0,1)  circle[radius=2pt] node[above] {$O_6$};
	%\draw (7.5,1)  circle[radius=2pt] node[above] {$O_7$};
	\draw (8.5,1)  circle[radius=2pt] node[above] {$O_T$};
	
	\node[text width=1cm] at (-0.5,2) {$s$};
	\node[text width=1cm] at (-1,1) {$O_s\left|C\right.$};
	\node[text width=1cm] at (-1,0) {$X_s\left|C\right.$};
	
	\end{tikzpicture}
	%	\caption{Schematic of the presumed data generating mechanism for one subject.  \label{tikzfig:data}}
	%\end{figure}
\end{center}
In this diagram, outcome $O_s$ is observed at $s = \tau_t, t=1,\ldots,T$.  The underlying trajectory determines that the subject begins in state 1, then progresses through states 2,3,2 etc. until finally reverting to state 1 after the final observation.  Observations are made at times that do not coincide with the transition times between states, and $\Delta _{t} = \tau_{t+1}-\tau_{t}$ records the interval between observations $O_{{t+1}}$ and  $O_{t}$. $C$ represents the cluster label of this individual. In one cluster case, $C$ will be the same across all the subjects. In this paper, we assume that the measurement process itself (that is, the collection of times $\tau_t, t=1,\ldots,T$) is not informative about the system either in its hidden or observed components.	 To illustrate the challenging nature of clustering trajectories, Figure \ref{traj} shows simulations from three groups with different parameters (taken from the Gaussian example in Example 1); Cluster 1 has in general smaller variation, but it is not easy to distinguish between Clusters 2 and 3.  As differences among clusters are also coming from the underlying continuous-time Markov process $X_s$, it is necessary to extend the basic model to allow for heterogeneity in observation and latent processes.

\noindent\textbf{Likelihood:} Using standard arguments, the transition probability from state $i$ to $j$ in the time interval of length $\Delta_t = \tau_{t+1}-\tau_t$ between observation times $t$ and $t+1$ takes the form
\[
p_{ij}\left(\Delta_t\right)=\mathbb{P}\left(X_{\tau_{t+1}}=j \left | X_{\tau_{t}}=i,\tau_{t+1}-\tau_{t}=\Delta_t \right. \right)=\exp\left(\Delta_t Q\right)_{\left(ij\right)}
\]
where $t=1,\ldots,T-1$, $Q=\left(q_{ij}\right)$ for $1\leq i,j\leq K$ is the infinitesimal generator for the continuous-time Markov process $\{X_s\}$ with $\sum_{j\ne i}q_{ij}=-q_{ii}>0$, and
$\exp\left(A\right)$ is the matrix exponential of matrix $A$. The initial state distribution $\pi$ for $\{X_s\}$ is $\pi_i = \mathbb{P}\left(X_{0}=i\right)$ for $i=1,\ldots,K$. The model is parameterized by $\Theta=\left\{q_{ij}, 1\le i,j\le K ,\pi,B, \phi\right\}$. For any individual, as $\{X_s\}$ is a Markov process with infinitesimal generator $Q$,  if $\{X_s\}$ has been observed continuously in the time interval $\left[0,\tau \right]$, the likelihood function of $Q$ is
\begin{equation*}
\prod\limits_{l = 1}^K {\prod\limits_{m \ne l} {{q_{l,m}}^{{N_{l,m}}\left( \tau  \right)}\exp \left( { - {q_{l,m}}{R_l}\left( \tau  \right)} \right)} }
\end{equation*}
where $N_{l,m} ( \tau )$ is the number of transitions from state $l$ to state $m$ in the time interval $\left[0,\tau\right]$ and $R_l\left( {{\tau}} \right)$ is the total time that the process has spent in state $l$ in $\left[0,\tau\right]$,
\begin{equation*}
R_l ( \tau ) = \int_{0}^{\tau} \mathbbm{1} (X_s = l) \: ds
\end{equation*}
Note that the quantities, $N_{l,m} \left( \tau  \right)$ and $R_l\left( {{\tau}} \right)$, are unobserved, but can be computed given a realization of the latent process on $[0,\tau]$.

For a random sample of $N$ subjects, let $O_{n,t}$ $\left(t=1,\ldots,T_n\right)$  be the $t^{\text{th}}$ observation for subject $n$ with the associated observation time $\tau_{n,t}$.  The complete data likelihood derived from $\left\{O_{n}\right\}$ and $\left\{X_{n,\tau_n}\right\}$ can be factorized $\mathcal{L}(\Theta) \equiv \mathcal{L}(\bO, \bX | \Theta ) = \mathcal{L} (\bX | \Theta )\mathcal{L} (\bO | \bX, \Theta )$ where $\bO = \{O_{n,t}\}$ and $\bX = \{X_{n,\tau_{n,t}}\}$ for $n=1,\ldots,N, t=1,\ldots,T_n$ and
\begin{align*}
&\mathcal{L} (\bO  |  \bX, \Theta )
=\prod\limits_{n=1}^N \prod\limits_{t = 1}^{T_n} {f\left( {{O_{n,t}}\left| {{X_{n,\tau_{n,t}}}} \right.} \right)}\\
&\mathcal{L} (\bX | \Theta )  = \prod\limits_{n=1}^N {\pi_{X_{n,0}}} \left( \prod\limits_{t = 1}^{{T_n}  - 1} {\prod\limits_{l = 1}^K {\prod\limits_{m \ne l} q_{lm}^{N_{n,l,m}\left( {{\Delta _{n,t}}} \right)}\exp \left( { - {q_{lm}} R_{n,l}\left( {{\Delta _{n,t}}} \right)} \right)} } \right)
\end{align*}
The log-likelihood written in terms of the latent state indicator random vectors $\left\{S_k\right\}_{k=1}^{K}$ is
\begin{equation} \label{loglik}
\scriptstyle
\sum\limits_{n=1}^N {\sum\limits_{t = 1}^{T_n}  {\sum \limits_{k=1}^K{S_{n,t,k} \log f\left( {{O_{n,t}}\left| {{S_{n,t,k}}} \right.} \right)}}} + \sum\limits_{n=1}^N { \sum \limits_{k=1}^K S_{n,1,k} \log\left( \pi_{k}\right)} + \sum\limits_{n=1}^N \sum\limits_{t = 1}^{{T_n} -1}{\sum \limits_{k=1}^K  \sum \limits_{j=1}^K S_{n,t,k}S_{n,{t+1},j}  p_{n,t}^{k,j} } \nonumber
\end{equation}
where
\begin{equation} \label{eq:transprob}
p_{n,t}^{k,j} = \sum\limits_{l = 1}^K \sum\limits_{m \ne l} N_{n,l,m}^{k,j}\left( \Delta _{n,t} \right)\log(q_{lm}) -   q_{lm} R_{n,l}^{k,j} \left( \Delta _{n,t} \right).
\end{equation}
records the probability of transition from state $k$ to state $j$ in the interval $\Delta _{n,t} = \tau_{n,t+1}-\tau_{n,t}$. $N_{n,l,m}^{j,k}\left( \Delta _{n,t} \right)$ and $R_{n,l}^{j,k} \left( \Delta _{n,t} \right)$ are the amended versions of $N_{l,m}$ and $R_l$ computed conditional on starting in state $j$ and ending in state $k$ over the interval $\Delta _{n,t}$.
\subsection{Definition of $a_{n,t,k}$ and $b_{n,t,k,j}$}
For $1 \leq k \leq K$, let
\[
a_{n,t,k} = \mathbb{E} [S_{n,t,k} |O;\Theta] = \mathbb{P} (S_{n,t,k}=1 | O;\Theta) = \sum\limits_{j=1}^K { b_{n,t,k,j}}
\]
say, where $b_{n,t,k,j} = \mathbb{P} (S_{n,t,k}=S_{n,{t+1},j}=1 |O;\Theta )$. The $\left(a_{n,t,k},b_{n,t,k,j}\right)$ can be obtained by the forward-backward algorithm \citep{baum1967inequality,baum1968growth}.  The forward variable $\alpha_{n,t,k}$ is defined as $\alpha_{n,t,k} = \mathbb{P}\left(O_{n,1} , \ldots,O_{n,t},X_{n,\tau_{n,t}}=k \right)$. Let $O_{n}^{(t)} = (O_{n,1},\ldots,O_{n,t})$.  Then
{\small
	\begin{equation*}
	\begin{aligned}
	\mathbb{P}\left(O_{n,1} \right.&,\left.\ldots,O_{n,t},X_{n,\tau_{n,t}}=k \right) = \mathbb{P}\left(O_{n,t} \left| X_{n,\tau_{n,t}}=k,O_{n}^{(t-1)} \right. \right) \mathbb{P}\left(O_{n}^{(t-1)},X_{n,\tau_{n,t}}=k \right)\\
	&= f\left(O_{n,t} \left| X_{n,\tau_{n,t}}=k \right. \right) \times \sum \limits_{i=1}^K{\mathbb{P}\left(O_{n}^{(t-1)},X_{n,\tau_{n,t}}=k, X_{{n,\tau_{n,t-1}}}=i \right)}\\
	&= f\left(O_{n,t} \left| X_{n,\tau_{n,t}}=k \right. \right) \sum \limits_{i=1}^K{\mathbb{P}\left(O_{n}^{(t-1)}, X_{{n,\tau_{n,t-1}}}=i \right)\mathbb{P}\left(X_{n,\tau_{n,t}}=k \left | X_{{n,\tau_{n,t-1}}}=i\right.\right)}\\
	&= f\left(O_{n,t} \left| X_{n,\tau_{n,t}}=k \right. \right) \times
	\sum \limits_{i=1}^K{\alpha_{n,t-1,i} p_{n,ik}\left(\Delta_{n,t-1}\right)}
	\end{aligned}
	\end{equation*}}The initial value $\alpha_{n,1,k}=\pi_k \times  f\left(O_{n,1} \left| X_{n,\tau_{n,1}}=k \right. \right)$.   The backward variable $\gamma_{n,t,k}$ is defined as $\gamma_{n,t,k} = \mathbb{P}\left(O_{n,T},\ldots,O_{n,t+1}\left|X_{\tau_{n,t}}=k \right. \right)$. Let $\overleftarrow{O}_{n}^{(t)} = (O_{n,T},\ldots,O_{n,t+1})$.  Then
{\small
	\begin{align*}
	\mathbb{P}\left(O_{n,T}\right.&,\left.\ldots,O_{n,t+1}\left|X_{\tau_{n,t}}=k \right. \right)=\sum \limits_{i=1}^K{\mathbb{P}\left(\overleftarrow{O}_{n}^{(t)} ,X_{{\tau_{n,t+1}}}=i\left|X_{\tau_{n,t}}=k \right. \right)}\\
	&=\sum \limits_{i=1}^K{\mathbb{P}\left(\overleftarrow{O}_{n}^{(t)}\left|X_{{\tau_{n,t+1}}}=i,X_{\tau_{n,t}}=k \right. \right) \times \mathbb{P}\left(X_{{\tau_{n,t+1}}}=i \left | X_{\tau_{n,t}}=k \right.\right)}\\
	&=\sum \limits_{i=1}^K{\mathbb{P}\left(O_{n,t+1}\left|X_{\tau_{n,t+1}}=i \right. \right)\mathbb{P}\left(\overleftarrow{O}_{n}^{(t+1)}\left|X_{{\tau_{n,t+1}}}=i\right. \right) \times p_{n,ki}\left(\Delta_{n,t}\right)}\\
	&=\sum \limits_{i=1}^K{f\left(O_{n,t+1} \left| X_{{\tau_{n,t+1}}}=i \right. \right) \times \gamma_{n,t+1,i}  p_{n,ki}\left(\Delta_{n,t}\right)}
	\end{align*}
}The first backward value $\gamma_{n,T,k}$ is initialized to 1 for all $n$ and $k$. Then, for $t=1,\ldots,T_n-1$,
$b_{n,t,k,j}=\mathbb{P}\left(S_{n,t,k}=S_{n,{t+1},j}=1 \left|O \right.\right)$. Define $\overrightarrow{O}_{n}^{(t)} = (O_{n,t},\ldots,O_{n,T})$.  Then
{\small
	\[
	\begin{aligned}
	b_{n,t,k,j}&=\mathbb{P}\left(X_{{n,\tau_{n,t}}}=k,X_{{n,\tau_{n,t+1}}}=j \left|O \right.\right)= \frac{\mathbb{P}\left(X_{{n,\tau_{n,t}}}=k,X_{{n,\tau_{n,t+1}}}=j, \overrightarrow{O}_{n}^{(1)}\right)}{\sum \limits_{j=1}^K{\sum \limits_{k=1}^K{\mathbb{P}\left(X_{{n,\tau_{n,t}}}=k,X_{{n,\tau_{n,t+1}}}=j, \overrightarrow{O}_{n}^{(1)}\right)}}}\\[6pt]
	&=
	\frac{\mathbb{P}\left(\overrightarrow{O}_{n}^{(t+1)},X_{{n,\tau_{n,t+1}}}=j\left| X_{n,\tau_{n,t}}=k,{O}_{n}^{(t)} \right.\right)\mathbb{P}\left(X_{n,\tau_{n,t}}=k, {O}_{n}^{(t)} \right)}{\sum \limits_{l=1}^K{\sum \limits_{m=1}^K{\mathbb{P}\left(\overrightarrow{O}_{n}^{(t+1)},X_{{n,\tau_{n,t+1}}}=l\left| X_{n,\tau_{n,t}}=m,{O}_{n}^{(t)} \right.\right)\mathbb{P}\left(X_{n,\tau_{n,t}}=m, {O}_{n}^{(t)}\right)}}}\\[12pt]
	&= \frac{f\left(O_{n,t+1} \left| X_{{n,\tau_{n,t+1}}}=j \right. \right) \times \gamma_{n,t+1,j} \times p_{n,kj}\left(\Delta_{n,t}\right)\alpha_{n,t,k}}{\sum \limits_{l=1}^K{\sum \limits_{m=1}^K{f\left(O_{n,t+1} \left| X_{{n,\tau_{n,t+1}}}=l \right. \right) \times \gamma_{n,t+1,l} \times p_{n,mj}\left(\Delta_{n,t}\right)\alpha_{n,t,m}}}}\\
	\end{aligned}
	\]
	with $	a_{n,t,k}=\sum \limits_{j=1}^K {b_{n,t,k,j}}$.
}
For $t=T_n$,
\[
a_{n,T_n,k}=\mathbb{P}\left(X_{\tau_{n,T_n}}=k \left|O \right. \right)=\frac{\mathbb{P}\left(X_{\tau_{n,T_n}}=k ,{O}_{n}^{(T_n)}\right)}{\sum \limits_{j=1}^K {\mathbb{P}\left(X_{\tau_{n,T_n}}=j ,{O}_{n}^{(T_n)}\right)}} =\frac {\alpha_{n,T_n,k}}{\sum \limits_{j=1}^K{\alpha_{n,T_n,j}}}.
\]

\section{Update $\mathbf{C}$ by split-merge by  Jain and Neal (2004)}
We layout the split-merge update for the cluster membership by \cite{jain2004split} in this section. Denote by $M^{i-1}$ the number of components in the current label configuration.
Select two distinct subjects $d$ and $e$ and denote their cluster labels $j_d = C_{d}^{i-1}$ and $j_e = C_{e}^{i-1}$.
\begin{enumerate}
	\item  Let $\mathcal{M} = \{f : f \ne d,e , \textrm{but } C_f^{i-1}=j_d \textrm{ or } C_f^{i-1}=j_e\} \subseteq \{1,2,\ldots,N\}$.
	
	\item Define the \textit{launch} state, $\mathbf{C}^l$ by running a Gibbs sampler scan restricted to the labels of subjects $f \in \mathcal{M}$.
	\begin{enumerate}[(i)]
		
		\item If $j_d \ne j_e$: for $f \in \mathcal{M}$, $\mathbb{P}\left(C_f= j \left|C_{-f}\right.\right)$ for $j \in \{j_d,j_e\}$ is given by
		\begin{equation}
		\label{dmsp22s}
		\dfrac{N_{-f,j}\displaystyle \int \mathcal{L}_f\left(\Theta\right) dH_{-f,j}\left(\Theta\right)}{N_{-f,j_d}\displaystyle\int \mathcal{L}_f\left(\Theta\right)dH_{-f,j_d}\left(\Theta\right)+N_{-f,j_e}\displaystyle\int \mathcal{L}_f\left(\Theta\right)dH_{-f,j_e}\left(\Theta\right)
		}
		\end{equation}
		where
		
		\begin{itemize}
			
			\item $N_{-f,j} = \sum\limits_{m\ne f} {\mathbbm{1}\left( {{C_m} = j} \right)}$. For MFM, $N_{-f,j}$ will be replaced by $N_{-f,j}+\delta$;
			
			\item $\mathcal{L}_f\left(\Theta\right)$ is defined in the main paper Section 5.2;
			
			\item $H_{-f,j}\left(\Theta\right)$ is the posterior distribution of $\Theta$ based on the prior $G_0$ and subjects $g \in \mathcal{M} \cup \left\{d,e\right\}$ but $g \ne f$, such that their labels $C_g=j$.
			
		\end{itemize}
		
		\item If $j_d = j_e$: for $f \in \mathcal{M}$, the conditional probability $\mathbb{P}\left(C_f= j \left|C_{-f}\right.\right)$ for $j \in \{j_e,M^{i-1}+1\}$ is calculated in the same fashion in \eqref{dmsp22s}.  The value $M^{i-1}+1$ is a new component label, not represented in the current label set.
		
	\end{enumerate}

	\item Split-merge step:
	
	\begin{enumerate}[(i)]
		\item If $j_d=j_e$, propose a split configuration, $\mathbf{C}^{\text{split}}$.
		\begin{itemize}
			
			\item Set $C_d^{\text{split}} = M^{i-1}+1$ and $C_e^{\text{split}}= C_e^{i-1}=j_e$.
			
			\item For $f\notin \mathcal{M}$, let $C_f^{\text{split}}=C_f^{i-1}$; for $f\in \mathcal{M}$, set $C_f^{\text{split}}$ by performing one more P\'{o}lya urn scan from the launch state.
			
			\item Perform a MH update with acceptance probability $a\left(\mathbf{C}^{\text{split}},\mathbf{C}^{i-1}\right)$. If the proposal is accepted, then set $\mathbf{C}^{i}=\mathbf{C}^{\text{split}}$; if rejected, set $\mathbf{C}^{i}=\mathbf{C}^{i-1}$.
			
		\end{itemize}
		\item
		If $j_d \neq j_e$, propose the merge configuration, $\mathbf{C}^{\text{merge}}$.
		\begin{itemize}
			\item Set $C_d^{\text{merge}}=C_e^{\text{merge}}= C_e^{i-1}=j_e$.
			\item For $f\in \mathcal{M}$, set $C_f^{\text{merge}}= j_e$; for $f\notin \mathcal{M}$, let $C_f^{\text{merge}}=C_f^{i-1}$.
			\item Perform a MH update with acceptance probability $a\left(\mathbf{C}^{\text{merge}},\mathbf{C}^{i-1}\right)$. If the proposal is accepted, then set $\mathbf{C}^{i}=\mathbf{C}^{\text{merge}}$; if rejected, set $\mathbf{C}^{i}=\mathbf{C}^{i-1}$.
		\end{itemize}
	\end{enumerate}
	
\end{enumerate}
The acceptance probability for this proposal is computed in the following subsection.

\section{Split-Merge Proposals for the Conjugate Dirichlet Process Mixture Model with Component Parameter $Q$}
\label{dirmixQ}
We now propose an algorithm to cluster individual trajectories via the DP mixture for the CTHMM-GLM with component parameters $Q$.

\begin{itemize}
	
	\item \textbf{Initialization:} Randomly sample cluster label $\mathbf{C}^{0}$ from  $\{1,\dots,M^{0}\}$ for each subject, where $M^{0}$ is an arbitrary positive integer with $\left|\mathbf{C}^{0}\right|=M^{0}$. Starting with initial values $\pi^{0}$, $Q^{0}_{C^0}$, $\theta^{0}$ and $\phi^{0}$, compute $a_{n,t,k}$  and $b_{n,t,k,j}$ using the forward-backward algorithm.
	
	\item \textbf{Update latent state indicators:} For each $n$ and $t$, generate the random vector $S_{n,t}^{i}$ from the multinomial distribution with the parameter set $a_{n,t}^{i}=(a_{n,t,1}^{i},\ldots,a_{n,t,K}^{i})$ where $S_{n,t}=\left(S_{n,t,1},\ldots,S_{n,t,K}\right)^\top$ is an indicator random vector with $S_{n,t,k}=1$ if $X_{\tau_t}=k$ and 0 otherwise.
	
	\item \textbf{Update $B$ and $\phi$:} Sample coefficient matrix $B^{i}$ and scale parameter $\phi^{i}$ given $S_{n,t}^{i}$ via the MH algorithm as their conditional posterior distributions are not of standard form.
	
	\item \textbf{Update $\pi$:} Under prior $Dirichlet\left(\alpha_1,\ldots,\alpha_K\right)$, sample $\pi^{i}$ from a Dirichlet distribution with parameters
	\[
	\left(\sum \limits_{n=1}^{N}{S_{n,1,1}^{i}}+\alpha_1,\ldots,\sum \limits_{n=1}^{N}{S_{n,1,K}^{i}}+\alpha_K\right)
	\]

	\item \textbf{Simulate the path for the latent process:}
	\begin{itemize}
		\item 	Sample the current state and next state ($X_{n,\tau_{n,t}}$ and $X_{n,\tau_{n,t+1}}$) from a multinomial distribution with the parameter matrix $b_{n,t,k,j}$.
		
		\item Simulate ${N_{n,l,m}}\left( {{\Delta _{n,t}}}\right)$ and ${R_{n,l}}\left( {{\Delta _{n,t}}}\right)$ from the Markov jump processes step-by-step with infinitesimal generator $Q_{C_n}^{i-1}$ through the intervals $\left[\tau_{n,t},\tau_{n,t+1}\right)$ initiated at $X_{n,\tau_{n,t}}$ and end point $X_{n,\tau_{n,t+1}}$ sampled previously with current label $C^{i-1}$.
	\end{itemize}
	\item \textbf{Update label $\mathbf{C}$ by split-merge}:
	Denote by $M^{i-1}$ the number of clusters in the current label configuration.
	Select two distinct subjects $d$ and $e$ and denote their cluster labels $j_d = C_{d}^{i-1}$ and $j_e = C_{e}^{i-1}$.
	\begin{enumerate}
		\item  Let $\mathcal{M} = \{f : f \ne d,e , \textrm{but } C_f^{i-1}=j_d \textrm{ or } C_f^{i-1}=j_e\} \subseteq \{1,2,\ldots,N\}$.
		
		\item Define the \textit{launch} state, $\mathbf{C}^l$ by running a Gibbs sampler scan restricted to the labels of subjects $f \in \mathcal{M}$.
		\begin{enumerate}[(i)]
			
			\item If $j_d \ne j_e$: for $f \in \mathcal{M}$, the conditional probability $\mathbb{P}\left(C_f= j \left|C_{-f}\right.\right)$ for $j \in \{j_d,j_e\}$ is given by
			{\small \[
				\frac{N_{-f,j} \displaystyle \int\prod\limits_{l\ne m }\prod\limits_{t=1}^{T_f}\mathcal{L}\left(q_{lm}\left|\Delta_{f,t}\right.\right)dH_{-f,j}\left(q_{lm}\right)}{N_{-f,j_d} \displaystyle \int\prod\limits_{l\ne m }\prod\limits_{t=1}^{T_f}\mathcal{L}\left(q_{lm}\left|\Delta_{f,t}\right.\right)dH_{-f,j_d}\left(q_{lm}\right)+N_{-f,j_e} \displaystyle \int\prod\limits_{l\ne m }\prod\limits_{t=1}^{T_f}\mathcal{L}\left(q_{lm}\left|\Delta_{f,t}\right.\right)dH_{-f,j_e}\left(q_{lm}\right)
				}
				\]}
			where 
			\[
			N_{-f,j} = \sum\limits_{m\ne f} {\mathbbm{1}\left( {{C_m} = j} \right)} \qquad \mathcal{L}\left(q_{lm}\left|\Delta_{f,t}\right.\right)={{q_{l,m}}^{N_{l,m}\left( {{\Delta _{f,t}}} \right)}e^ {\left( { - {q_{l,m}}R_l\left( {{\Delta _{n,t}}} \right)} \right)}}
			\]
			and where $H_{-f,j}\left(q_{lm}\right)$ denotes the posterior distribution of $q_{lm}$ based on the prior $G_0$ and subjects $g \in \mathcal{M} \cup \left\{d,e\right\}$ but $g \ne f$ such that $C_g=j$.
			
			\item If $j_d = j_e$: for $f \in \mathcal{M}$, the conditional probability $\mathbb{P}\left(C_f= j \left|C_{-f}\right.\right)$ for $j \in \{j_e,M^{i-1}+1\}$ is calculated in the same fashion.  The value $M^{i-1}+1$ is a new component label, not represented in the current label set.
			
		\end{enumerate}	
		
		\item Split-merge step:
		
		\begin{enumerate}[(i)]
			\item If subjects $d$ and $e$ are in the \textbf{same} mixture component, i.e., $C_d^{i-1}=C_e^{i-1}$, propose the split procedure, $\mathbf{C}^{\text{split}}$.
			\begin{itemize}
				
				\item Set $C_d^{\text{split}} = M^{i-1}+1$ and $C_e^{\text{split}}= C_e^{i-1}=j_e$.
				\item For $f\notin \mathcal{M}$, let $C_f^{\text{split}}=C_f^{i-1}$; for $f\in \mathcal{M}$, modify $C_f^{\text{split}}$ by performing one more P\'{o}lya urn scan from the launch state label $C^l_f$.
				\item Perform the MH update with acceptance probability $a\left(\mathbf{C}^{\text{split}},\mathbf{C}^{i-1}\right)$. If the proposal is accepted, then set $\mathbf{C}^{i}=\mathbf{C}^{\text{split}}$; if rejected, set $\mathbf{C}^{i}=\mathbf{C}^{i-1}$.
				
			\end{itemize}
			\item
			If subjects $d$ and $e$ are in \textbf{different} mixture components, i.e., $C_d^{i-1}\ne C_e^{i-1}$, propose the merge procedure, $\mathbf{C}^{\text{merge}}$.
			\begin{itemize}
				\item Set $C_d^{\text{merge}}=C_e^{\text{merge}}= C_e^{i-1}=j_e$.
				\item For $f\in \mathcal{M}$, let $C_f^{\text{merge}}=C_e^{i-1}$; for $f\notin \mathcal{M}$, let $C_f^{\text{merge}}=C_f^{i-1}$.
				\item Perform the MH update with acceptance probability $a\left(\mathbf{C}^{\text{merge}},\mathbf{C}^{i-1}\right)$. If the proposal is accepted, then set $\mathbf{C}^{i}=\mathbf{C}^{\text{merge}}$; if rejected, set $\mathbf{C}^{i}=\mathbf{C}^{i-1}$.
			\end{itemize}
		\end{enumerate}
		
	\end{enumerate}
	
	\item \textbf{Update the component parameter $Q$}:
	For all $C$ in $\mathbf{C}^{i}=\left\{C_1^{i},\ldots,C_N^{i}\right\}$, update ${N_{l,m}}\left( {{\Delta _{n,t}}}\right)$ and ${R_{l}}\left( {{\Delta _{n,t}}}\right)$ from the updated label component generator $Q_C$ (obtained when calculating $a\left(\mathbf{C}^{*},\mathbf{C}^{i-1}\right)$). Then $q_{l,m\left|C\right.}^{i}$ associated with component $C$ is generated from a Gamma distribution with shape parameter $\Lambda_{l,m}^{C}$ and rate parameter $\Upsilon_{l}^{C}$.
\end{itemize}
In terms of calculating the acceptance probability, the procedure is the same as the general algorithm except for the likelihood ratio. In the general algorithm, the likelihood of the component parameter $Q$ is based on the unobserved Markov process, namely,  ${N_{l,m}}\left( {{\Delta _{n,t}}}\right)$ and ${R_{l}}\left( {{\Delta _{n,t}}}\right)$. Therefore, when calculating the likelihood ratio for split-merge procedures, the unobserved Markov process should be modified as well. The likelihood of the current label for $f\in \mathcal{M}\cup \left\{d,e\right\}$ is
\begin{equation*}\scriptstyle
{\mathcal{L}\left(C_f\right)}={\prod_{f:C_f=j_d}\int\prod\limits_{l\ne m }\prod\limits_{t=1}^{T_f}\mathcal{L}\left(q_{lm}\left|\Delta_{f,t}\right.\right))dH_{-f,j_d}\left(q_{lm}\right)} \times {\prod_{f:C_f=j_e}\int\prod\limits_{l\ne m }\prod\limits_{t=1}^{T_f}\mathcal{L}\left(q_{lm}\left|\Delta_{f,t}\right.\right)dH_{-f,j_e}\left(q_{lm}\right)}
\end{equation*}
where ${N_{l,m}}\left( {{\Delta _{f,t}}}\right)$ and ${R_{l}}\left( {{\Delta _{f,t}}}\right)$ are generated from the previous step. In the split step, the likelihood of the split label is calculated by
\begin{equation*}
\begin{aligned}
{\mathcal{L}\left(C_f^{\text{split}}\right)}&=\prod_{f:C_f^{\text{split}}=C_d^{\text{split}}}\int{\prod\limits_{l\ne m }\prod\limits_{t=1}^{T_f}\mathcal{L}^{d,\text{split}}\left(q_{lm}\left|\Delta_{f,t}\right.\right)dH_{-f,C_d^{\text{split}}}\left(q_{lm}\right)} \\
&\times \prod_{f:C_f^{\text{split}}=j_e}\int\prod\limits_{l\ne m }\prod\limits_{t=1}^{T_f}\mathcal{L}^{e,\text{split}}\left(q_{lm}\left|\Delta_{f,t}\right.\right)dH_{-f,j_e}\left(q_{lm}\right)
\end{aligned}
\end{equation*}
where
\[
\mathcal{L}^{s,\text{split}}\left(q_{lm}\left|\Delta_{f,t}\right.\right)={{q_{l,m}}^{{N^{s,\text{split}}_{l,m}}\left( {{\Delta _{f,t}}} \right)}e^ {\left( { - {q_{l,m}}{R^{s,\text{split}}_l}\left( {{\Delta _{f,t}}} \right)} \right)}} \qquad s=d,e
\]
and ${N^{e,\text{split}}_{l,m}}\left( {{\Delta _{f,t}}} \right)$, ${R^{e,\text{split}}_{l}}\left({{\Delta _{f,t}}}\right)$  are simulated from the Markov jump processes step-by-step with infinitesimal generator $Q^{e,\text{split}}$ through the intervals $\left[\tau_{n,t},\tau_{n,t+1}\right)$ initiated at $X_{\tau_{n,t}}$ and end point $X_{\tau_{n,t+1}}$ respectively and each entry in $Q^{e,\text{split}}$ is generated from a Gamma distribution with shape parameter $\Lambda_{l,m}^{j_e}$ and rate parameter $\Upsilon_{l}^{j_e}$. Similarly, ${N^{d,\text{split}}_{l,m}}\left( {{\Delta _{f,t}}} \right), {R^{d,\text{split}}_{l}}\left({{\Delta _{f,t}}}\right)$ are sampled associated with $Q^{d,\text{split}}$ where each entry in  $Q^{d,\text{split}}$ is sampled from a Gamma distribution with
shape parameter $\Lambda_{l,m}^{C_d^{\text{split}}}$ and rate parameter $\Upsilon_{l}^{C_d^{\text{split}}}$.

If the split step is accepted, $Q^{d,\text{split}}$ and $Q^{e,\text{split}}$ will be carried into the next step to update $Q$. Similarly in the merge step, the likelihood of the merge label is calculated by
\begin{equation*}
\mathcal{L}\left(C_f^{\text{merge}}\right)=\prod_{f:C_f^{\text{merge}}=j_e}\int{\prod\limits_{l\ne m }\prod\limits_{t=1}^{T_f}\mathcal{L}^{e,\text{merge}}\left(q_{lm}\left|\Delta_{f,t}\right.\right)dH_{-f,j_e}\left(q_{lm}\right)}
\end{equation*}
where $\mathcal{L}^{e,\text{merge}}\left(q_{lm}\left|\Delta_{f,t}\right.\right)={{q_{l,m}}^{{N^{e,\text{merge}}_{l,m}}\left( {{\Delta _{f,t}}} \right)}e^ {\left( { - {q_{l,m}}{R^{e,\text{merge}}_l}\left( {{\Delta _{n,t}}} \right)} \right)}}$ and ${N^{e,\text{merge}}_{l,m}}\left( {{\Delta _{f,t}}} \right)$, ${R^{e,\text{merge}}_{l}}\left({{\Delta _{f,t}}}\right)$  are simulated associated with $Q^{e,\text{merge}}$ where each entry in  $Q^{e,\text{merge}}$ is sampled from a Gamma distribution with  shape parameter $\Lambda_{l,m}^{j_e}$ and rate parameter $\Upsilon_{l}^{j_e}$.

If the merge step is accepted, $Q^{e,\text{merge}}$ will be carried into the next step when updating $Q$.

\section{Split-merge illustration: Poisson model}
Suppose $O_t\left|X_{\tau_t}=k\right. \sim \text{Poisson}\left(\theta_k\right)$ and that there are no covariates in outcome model or latent model.  Data $\mathbf{o}=\left\{o_{n,t}\right\}$ for $n=1,\ldots,N,t=1,\ldots,n$, have the complete data log-likelihood
\begin{align*}
\ell (\Theta ) & =\sum\limits_{n=1}^N {\sum\limits_{t = 1}^{T_n}  {\sum \limits_{k=1}^K{ S_{n,t,k}\left( o_{n,t}\log\theta_k-\log o_{n,t}!-\theta_k \right)}}} + \sum\limits_{n=1}^N { \sum \limits_{k=1}^K S_{n,1,k} \log\left( \pi_{k}\right)} \\
& \qquad \qquad \qquad + \sum\limits_{n=1}^N \sum\limits_{t = 1}^{{T_n} -1}{\sum \limits_{k=1}^K  \sum \limits_{j=1}^K S_{n,t,k}S_{n,{t+1},j}  p_{n,t}^{k,j} } \nonumber
\end{align*}
where $p_{n,t}^{k,j}$ is the probability of transition from state $k$ to state $j$, defined in \eqref{eq:transprob}.  If \textit{a priori} $ \pi \sim  Dirichlet \left(\alpha_1,\ldots,\alpha_K\right)$, $ q_{lm} \sim Gamma\left(a_{lm},b_{l}\right)$ for $1\le l\ne m \le K$  and $ \theta_k \sim Gamma\left(a_k,b_k\right)$ independently, then we have a conjugate model, and the integrals defining the split and merge probabilities are analytically tractable; we can integrate out the model parameters separately.  To update $C_f$ via intermediate Gibbs sampling, a new value of $C_f$ is drawn from \eqref{dmsp22s}.

\medskip

\noindent \textbf{Outcome model parameters: } For $\theta_k$ ($k=1,\ldots,K$), given latent state indicators, since this is a conjugate Poisson-Gamma model, $H_{-f,C}\left(\theta_k\right) \sim Gamma(a_{k}^{\prime},b_{k}^{\prime})$ where
\[
a_{k}^{\prime}=a_k+\sum\limits_{n\ne f:C_n=C}\sum\limits_{t: S_{n,t,k}=1}o_{n,t} \qquad \qquad b_{k}^{\prime}=b_k+\sum\limits_{n\ne f:C_n=C}\sum\limits_{t=1}^{T_n} S_{n,t,k}.
\]
Therefore, as
\[
\mathcal{L}_{\theta_k}\left(C_f=j\left|C_{-f}\right.\right)=\int \mathcal{L}\left(\theta_k\right)dH_{-f,j}\left(\theta_k\right)
\]
we have that
\[
\mathcal{L}_{\theta_k}\left(C_f=j\left|C_{-f}\right.\right) =  \frac{\Gamma\left(\sum\limits_{t: S_{f,t,k}=1}o_{f,t}+a_{k}^{\prime}\right)}{\left(\sum\limits_{t=1}^{T_f} S_{f,t,k}+b_{k}^{\prime}\right)^{\sum\limits_{t: S_{f,t,k}=1} o_{f,t} + a_k^{\prime} }}\times\frac{\left(b_{k}^{\prime}\right)^{a_k^{\prime} }}{\Gamma\left(a_{k}^{\prime} \right)\times\prod\limits_{t: S_{f,t,k}=1}o_{f,t}!}
\]

\medskip
\noindent\textbf{Initial state parameters:} Within each cluster, $H_{-f,j}\left(\pi\right) \sim Dirichlet \left(\alpha_1^{\prime},\ldots,\alpha_K^{\prime}\right)$, where $\alpha_k^{\prime}=\alpha_k+\sum\limits_{n\ne f:C_n=j}S_{n,1,k} $. Then,
\[
\mathcal{L}_{\pi}\left(C_f=j\left|C_{-f}\right.\right)=\int \mathcal{L}\left(\pi\right)dH_{-f,j}\left(\pi\right)= \frac{\prod\limits_{k=1}^{K}\Gamma \left(S_{n,1,k}+\alpha_k^{\prime}\right)}{\Gamma\left(\sum\limits_{k=1}^{K} S_{n,1,k}+\alpha_k^{\prime}\right)} \times \frac{\Gamma\left(\sum\limits_{k=1}^{K}\alpha_k^{\prime}\right)} {\prod\limits_{k=1}^{K}\Gamma \left(\alpha_k^{\prime}\right)}
\]

\medskip
\noindent\textbf{Transition model parameters:} For the components $q_{lm}$ of $Q$, given $N_{lm}\left(\tau\right)$ and $R_l\left(\tau\right)$ as defined previously, $H_{-f,C}\left(q_{lm}\right) \sim Gamma(a_{lm}^{\prime},b_{lm}^{\prime})$, where
\[
a_{lm}^{\prime}=\sum\limits_{n\ne f:C_n=C} {\sum\limits_{t = 1}^{T_n} { {N_{l,m}}\left( {{\Delta _{n,t}}} \right)} }+a_{lm} \qquad \qquad b_{l}^{\prime}=\sum\limits_{n\ne f:C_n=C} {\sum\limits_{t = 1}^{T_n} { {R_{l}}\left( {{\Delta _{n,t}}} \right)} }+b_l.
\]
Then,
\small
\[
\mathcal{L}_{q_{lm}}\left(C_f=j\left|C_{-f}\right.\right)=\int \mathcal{L}\left(q_{lm}\right)dH_{-f,j}\left(q_{lm}\right)=  \frac{\Gamma\left(\sum\limits_{t=1}^{T_f}  {N_{l,m}}\left( {{\Delta _{f,t}}} \right)+a_{lm}^{\prime}\right)}{\left(\sum\limits_{t=1}^{T_f} {R_{l}}\left( {{\Delta _{f,t}}} \right)+b_{k}^{\prime}\right)^{\sum\limits_{t=1}^{T_f}  {N_{l,m}}\left( {{\Delta _{f,t}}} \right)+a_{lm}^{\prime}}}\times\frac{\left(b_{l}^{\prime}\right)^{a_{lm}^{\prime} }}{\Gamma\left(a_{lm}^{\prime} \right)}.
\]\normalsize
Therefore,
\small
\[
\scriptsize
\int \mathcal{L}_f\left(\Theta\right)dH_{-f,j}\left(\Theta\right)=\prod\limits_{k=1}^{K}\mathcal{L}_{\theta_k}\left(C_f=j\left|C_{-f}\right.\right) \times \mathcal{L}_{\pi}\left(C_f=j\left|C_{-f}\right.\right) \times \prod\limits_{1\le l \ne m \le K} \mathcal{L}_{q_{lm}}\left(C_f=j\left|C_{-f}\right.\right)
\]\normalsize
and the probability in \eqref{dmsp22s} readily computed.

The MH acceptance probability %in \eqref{eq:Cupdate} 
also involves calculating the marginal likelihood of the cluster membership, $\mathcal{L}\left(\mathbf{C}\right)$ defined in the main paper Section 5.2.
%in \eqref{likc}.
The integrals can be calculated in the same fashion above and the only difference is to change $H_{-f,j}$ to $H_{f,j}$. Specifically,
\begin{align*}
H_{f,j}\left(\theta_k\right) & \sim Gamma \left(a_k+\sum\limits_{n < f :C_n=j}\sum\limits_{t: S_{n,t,k}=1}o_{n,t}, b_k+\sum\limits_{n<f:C_n=j}\sum\limits_{t=1}^{T_n} S_{n,t,k} \right)\\[6pt]
H_{f,j}\left(\pi\right) & \sim Dirichlet \left(\alpha_1+\sum\limits_{n<f :C_n=j}S_{n,1,1},\ldots, \alpha_K+\sum\limits_{n< f:C_n=j}S_{n,1,K} \right)\\[6pt]
H_{f,j}\left(q_{lm}\right) & \sim Gamma \left(\sum\limits_{n< f:C_n=j} {\sum\limits_{t = 1}^{T_n} { {N_{l,m}}\left( {{\Delta _{n,t}}} \right)} }+a_{lm}, \sum\limits_{n< f:C_n=j} {\sum\limits_{t = 1}^{T_n} { {R_{l}}\left( {{\Delta _{n,t}}} \right)} }+b_l \right)
\end{align*}
A similar conjugate approach can be used if factor predictors are incorporated into the models for the outcome or the transition infinitesimal generator.

\section{Example Results}

The following results demonstrate that the algorithm successfully recovers the simulating parameters in the posterior distributions.  In each case, the posterior mean was used to calculate the norm difference from the true values.

\subsection{Example 1}

\begin{table}[ht]
	\caption{\label{sim101}Example 1: Inference for simulated data with three clusters and three latent states via a DP mixture model.}
	
	{\centering
		\begin{tabular*}{40pc}{@{\hskip5pt}@{\extracolsep{\fill}}c@{}c@{}c@{}c@{}c@{}c@{}c@{}@{\hskip5pt}}
			
			\hline 
			%\multicolumn{6}{c}{EM}\\\cline{1-6}
			& \multicolumn{3}{c}{Gaussian} &\multicolumn{3}{c}{Poisson}\\ \cline{2-7}
			%\midrule[2pt]
			%\multicolumn{8}{c}{EM algorithm with tolerance 0.05 }\\\cline{1-8}
			&$T=30$ &$T=50$  & $T=100$ &$T=30$& $T=50$ & $T=100$\\
			\hline 
			$\left\| {\pi_1-\hat \pi_1} \right\|$  & 0.04& 0.02 & 0.02& 0.04&0.02&0.04\\
			$\left\| {B_1-\hat B_1} \right\|$ & 0.47 & 0.43 &0.41&0.22&0.08& 0.04\\
			$\left\| {Q_1-\hat Q_1} \right\|$ &0.79 & 0.60&0.37&1.37 &0.44&0.22\\
			$\left\| {\pi_2-\hat \pi_2} \right\|$  & 0.04 & 0.03& 0.02& 0.05&0.04&0.01\\
			$\left\| {B_2-\hat B_2} \right\|$ & 0.37& 0.44 &0.28&0.02 &0.01&0.02\\
			$\left\| {Q_2-\hat Q_2} \right\|$ &0.59 & 0.58&0.27&0.88&0.45&0.17\\
			$\left\| {\pi_3-\hat \pi_3} \right\|$  & 0.02 & 0.01 & 0.02& 0.03&0.03&0.03\\
			$\left\| {B_3-\hat B_3} \right\|$ & 0.58 & 0.31 &0.25&0.12 &0.03&0.02\\
			$\left\| {Q_3-\hat Q_3} \right\|$ &0.34 & 0.25&0.09&0.30&0.12& 0.16\\
			\% of 3-cluster iterations& 59.65\%&  99.25\% & 99.75\% & 56.40\% &93.05\% &98.25\%\\
			Misclassification rate&20.97\% &19.16\% &12.9\%& 14.10\% &6.53\%&2.40\%\\
			\hline 
		\end{tabular*}
	}
\end{table}

\begin{table}[ht]
	\caption{\label{simfin1}Example 1: Inference for simulated data with three clusters and three latent states via a mixture of finite mixture models  using reversible-jump MCMC.}
	
	{\centering
		\begin{tabular*}{40pc}{@{\hskip5pt}@{\extracolsep{\fill}}c@{}c@{}c@{}c@{}c@{}c@{}c@{}@{\hskip5pt}}
			
			\hline 
			%\multicolumn{6}{c}{EM}\\\cline{1-6}
			& \multicolumn{3}{c}{Gaussian} &\multicolumn{3}{c}{Poisson}\\ \cline{2-7}
			%\midrule[2pt]
			%\multicolumn{8}{c}{EM algorithm with tolerance 0.05 }\\\cline{1-8}
			&$T=30$ &$T=50$  & $T=100$ &$T=30$& $T=50$ & $T=100$\\
			\hline 
			$\left\| {\pi_1-\hat \pi_1} \right\|$  & 0.03& 0.02 & 0.04& 0.04&0.03&0.03\\
			$\left\| {B_1-\hat B_1} \right\|$ & 0.10 & 0.03 &0.04&0.13&0.20& 0.13\\
			$\left\| {Q_1-\hat Q_1} \right\|$ &0.21 & 0.19&0.11&0.38 &0.32&0.20\\
			$\left\| {\pi_2-\hat \pi_2} \right\|$  & 0.03 & 0.03& 0.02& 0.04&0.03&0.04\\
			$\left\| {B_2-\hat B_2} \right\|$ & 0.10& 0.02 &0.03&0.04 &0.02&0.04\\
			$\left\| {Q_2-\hat Q_2} \right\|$ &0.09 & 0.06&0.03&0.10&0.21&0.09\\
			$\left\| {\pi_3-\hat \pi_3} \right\|$  & 0.03 & 0.01 & 0.05& 0.06&0.03&0.03\\
			$\left\| {B_3-\hat B_3} \right\|$ & 0.09 & 0.03 &0.02&0.05 &0.05&0.02\\
			$\left\| {Q_3-\hat Q_3} \right\|$ &0.13 & 0.04&0.04&0.06&0.05& 0.04\\
			\% of 3-cluster iterations& 62.35\%& 54.40\% & 81.05\%  & 66.85\%  &80.65\%  & 89.95\%\\
			Misclassification rate&1.20\% & 0.60\% &0.00\% & 4.00\%&2.30\%& 0.70\%\\
			\hline 
		\end{tabular*}
	}
\end{table}

\begin{table}[ht]
	\caption{\label{simfin2}Example 1: Inference for simulated data with three clusters and three latent states via a mixture of finite mixture models using a split-merge algorithm.}
	
	{\centering
		\begin{tabular*}{40pc}{@{\hskip5pt}@{\extracolsep{\fill}}c@{}c@{}c@{}c@{}c@{}c@{}c@{}@{\hskip5pt}}
			
			\hline 
			%\multicolumn{6}{c}{EM}\\\cline{1-6}
			& \multicolumn{3}{c}{Gaussian} &\multicolumn{3}{c}{Poisson}\\ \cline{2-7}
			%\midrule[2pt]
			%\multicolumn{8}{c}{EM algorithm with tolerance 0.05 }\\\cline{1-8}
			&$T=30$ &$T=50$  & $T=100$ &$T=30$& $T=50$ & $T=100$\\
			\hline 
			$\left\| {\pi_1-\hat \pi_1} \right\|$  & 0.04& 0.02 & 0.04& 0.07&0.03&0.03\\
			$\left\| {B_1-\hat B_1} \right\|$ & 0.59& 0.33 &0.06&0.08&0.18& 0.14\\
			$\left\| {Q_1-\hat Q_1} \right\|$ &0.86 & 0.45&0.10&0.94 &0.65&0.31\\
			$\left\| {\pi_2-\hat \pi_2} \right\|$  & 0.04 & 0.04& 0.04& 0.09&0.04&0.04\\
			$\left\| {B_2-\hat B_2} \right\|$ & 0.38& 0.23 &0.17&0.43 &0.03&0.04\\
			$\left\| {Q_2-\hat Q_2} \right\|$ &0.35 & 0.21&0.19&0.88&0.45&0.06\\
			$\left\| {\pi_3-\hat \pi_3} \right\|$  & 0.08 & 0.02 & 0.03& 0.12&0.05&0.03\\
			$\left\| {B_3-\hat B_3} \right\|$ & 0.98 & 0.72 &0.16&0.26 &0.02&0.04\\
			$\left\| {Q_3-\hat Q_3} \right\|$ &0.64 & 0.45&0.05&0.54&0.12& 0.08\\
			\% of 3-cluster iterations&45.25\% &46.85\%  & 50.25\%   & 48.65\%  &56.60\%   & 58.15\% \\
			Misclassification rate&15.60\% & 12.70\%  &9.10\%  &12.10\%  & 7.30\% & 2.40\%\\
			\hline 
		\end{tabular*}
	}
\end{table}

\begin{figure}[ht]
	\centering
	\caption{Example 1: Trace plots of the number of clusters via a DP mixture model for Normal (left) and Poisson (right) models and for $T=30, 50$ and $100$ observations per subject (top, middle, bottom rows respectively).  The plot reveals that as the amount of data increases, the posterior distribution on the number of clusters becomes more concentrated on the three cluster model.}
	\begin{minipage}[b]{0.425\textwidth}
		\includegraphics[width=\textwidth]{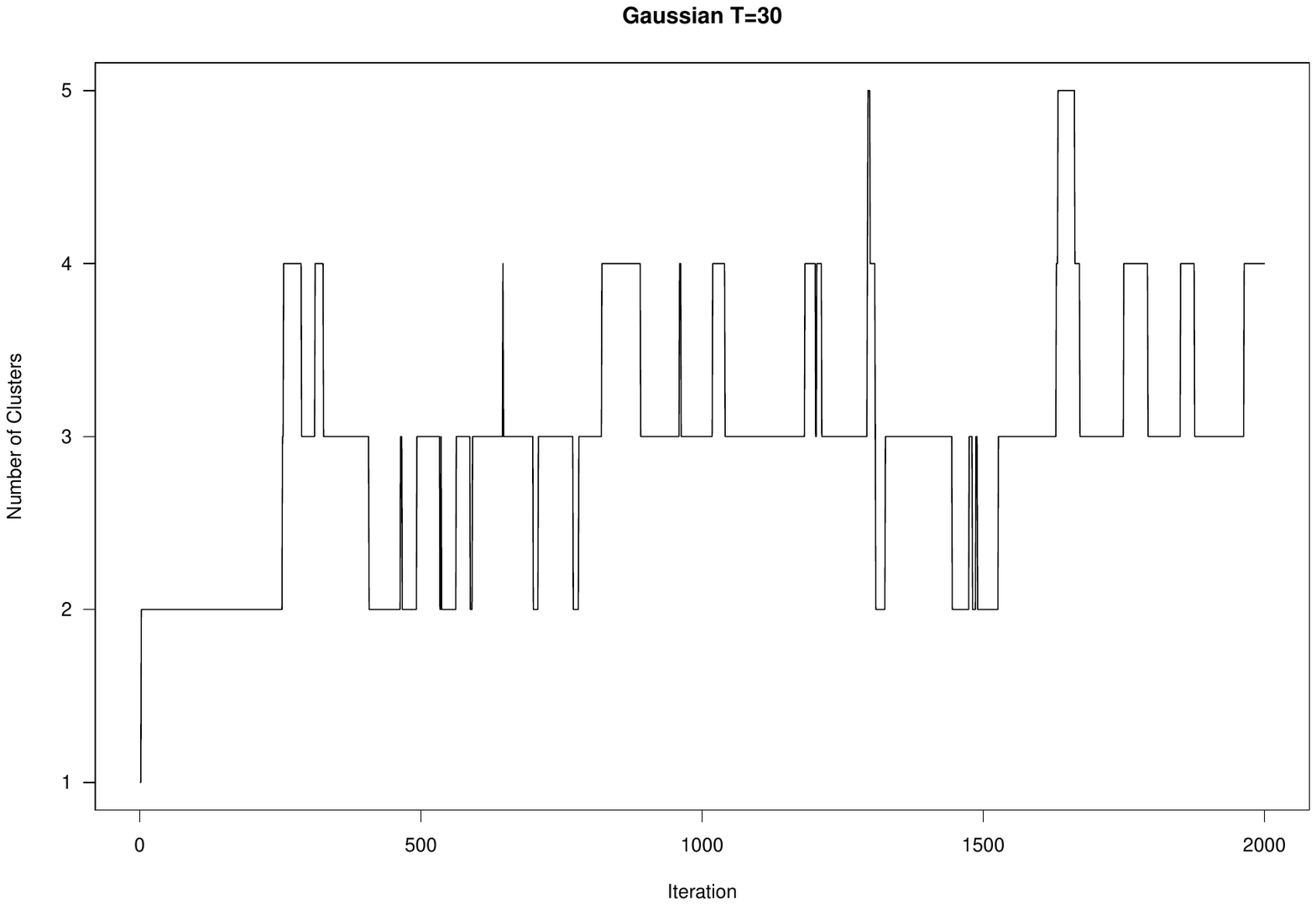}
		%\caption*{Gamma$\left(20,500\right)$}
	\end{minipage}
	\hfill
	\begin{minipage}[b]{0.425\textwidth}
		\includegraphics[width=\textwidth]{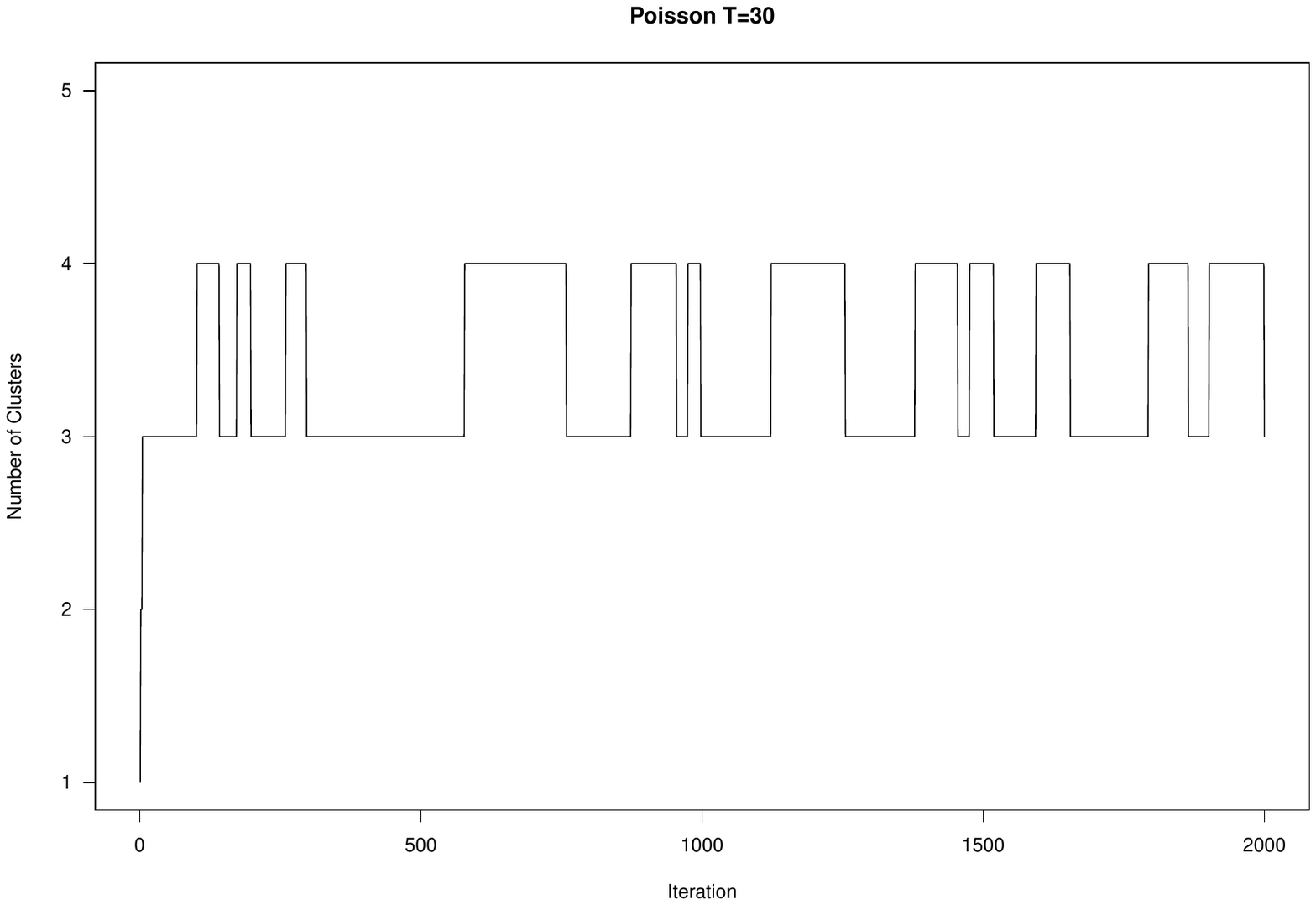}
		%\caption*{Gamma$\left(20,500\right)$}
	\end{minipage}
	\begin{minipage}[b]{0.425\textwidth}
		\includegraphics[width=\textwidth]{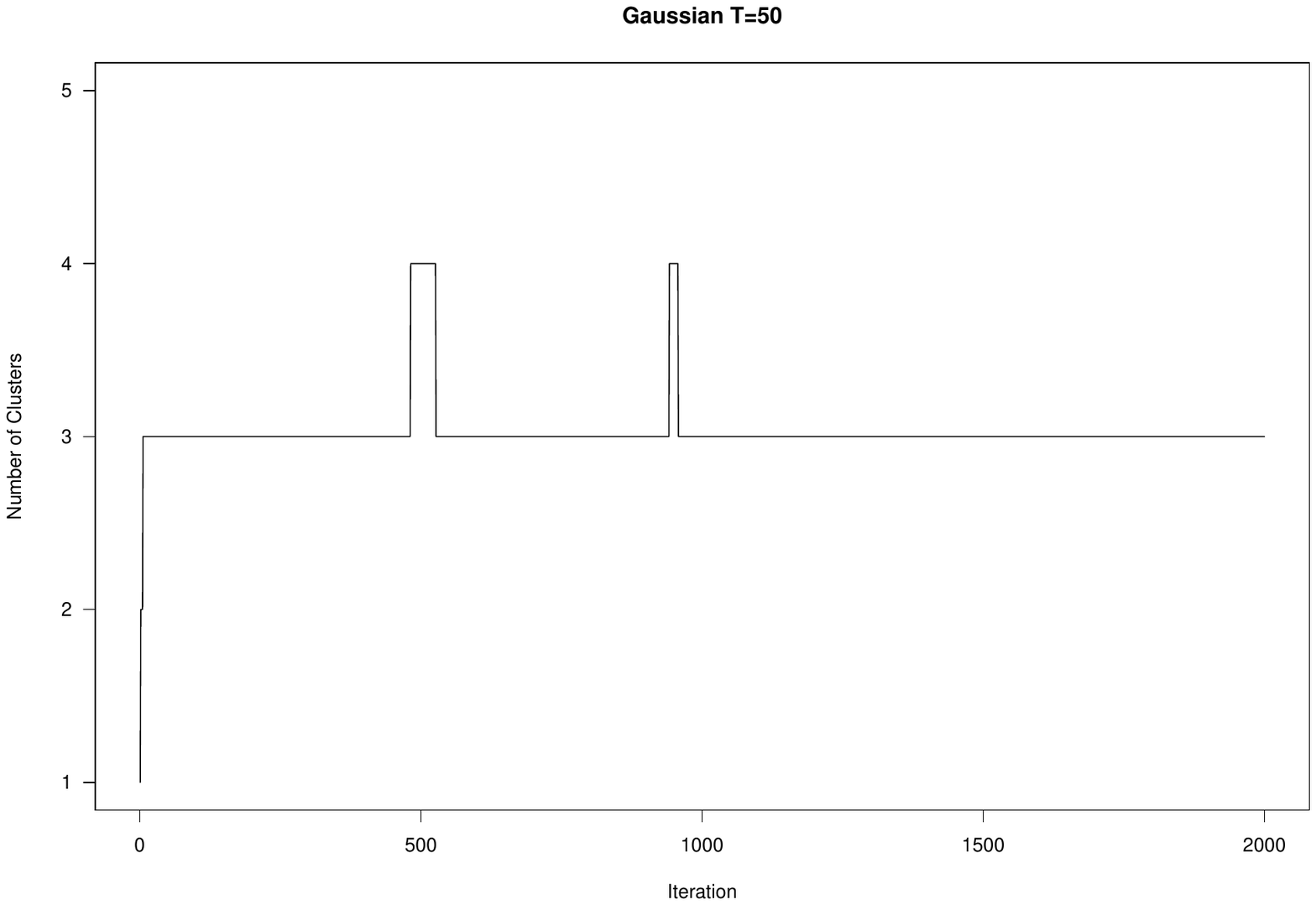}
		%\caption*{Gamma$\left(2,50\right)$}
	\end{minipage}
	\hfill
	\begin{minipage}[b]{0.425\textwidth}
		\includegraphics[width=\textwidth]{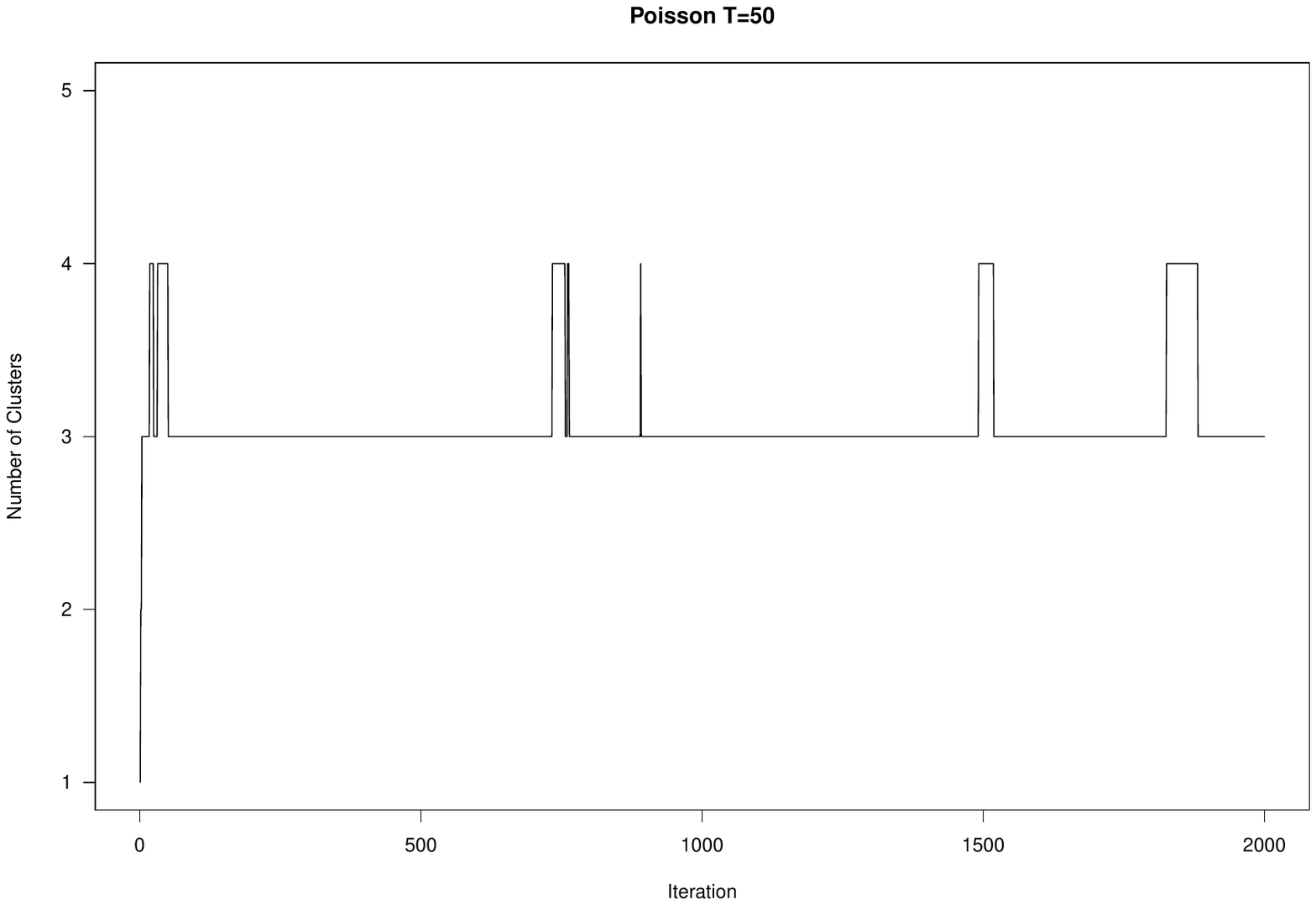}
		%\caption*{Gamma$\left(20,500\right)$}
	\end{minipage}
	\hfill
	\begin{minipage}[b]{0.425\textwidth}
		\includegraphics[width=\textwidth]{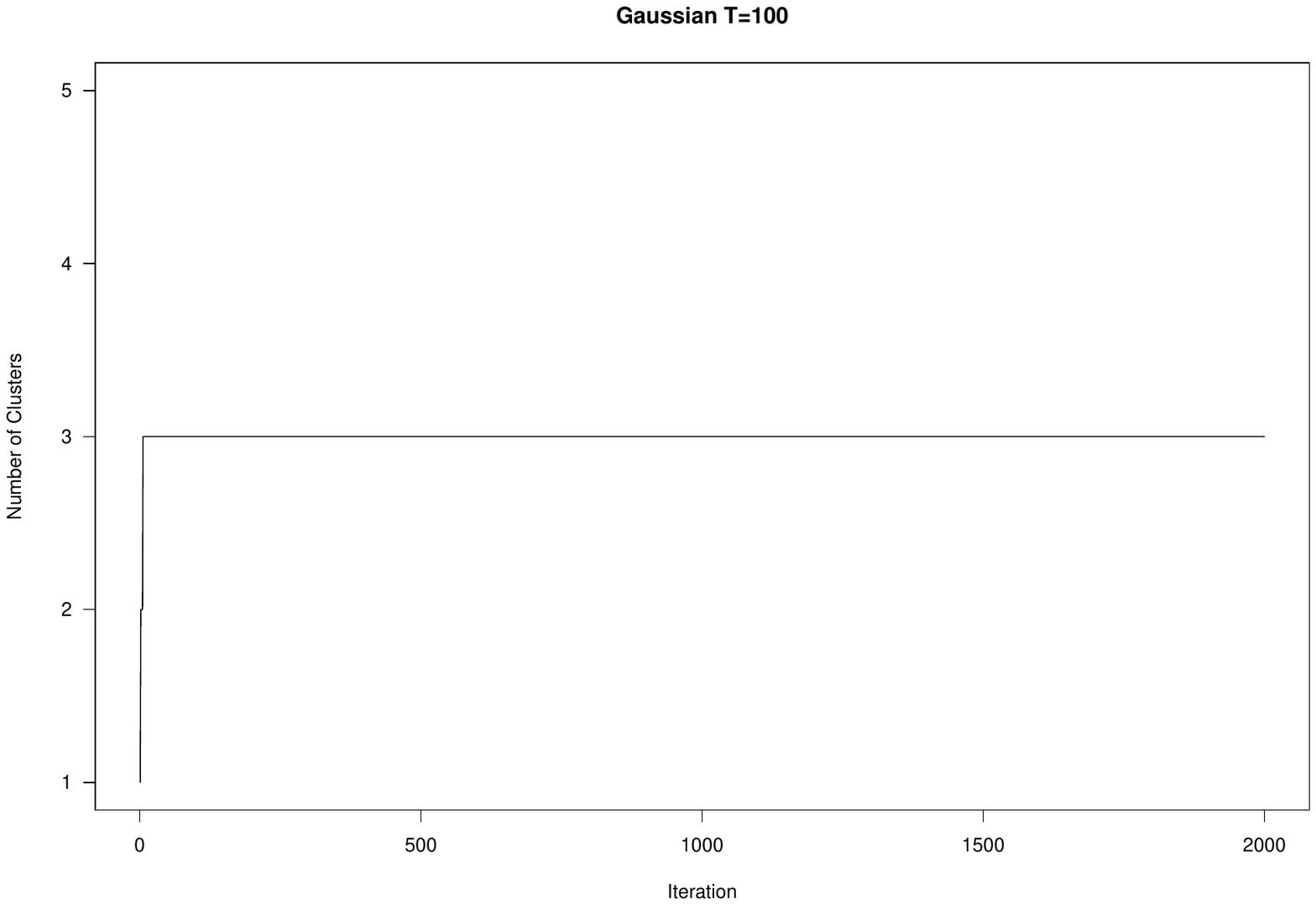}
		%\caption*{Gamma$\left(20,500\right)$}
	\end{minipage}
	\hfill
	\begin{minipage}[b]{0.425\textwidth}
		\includegraphics[width=\textwidth]{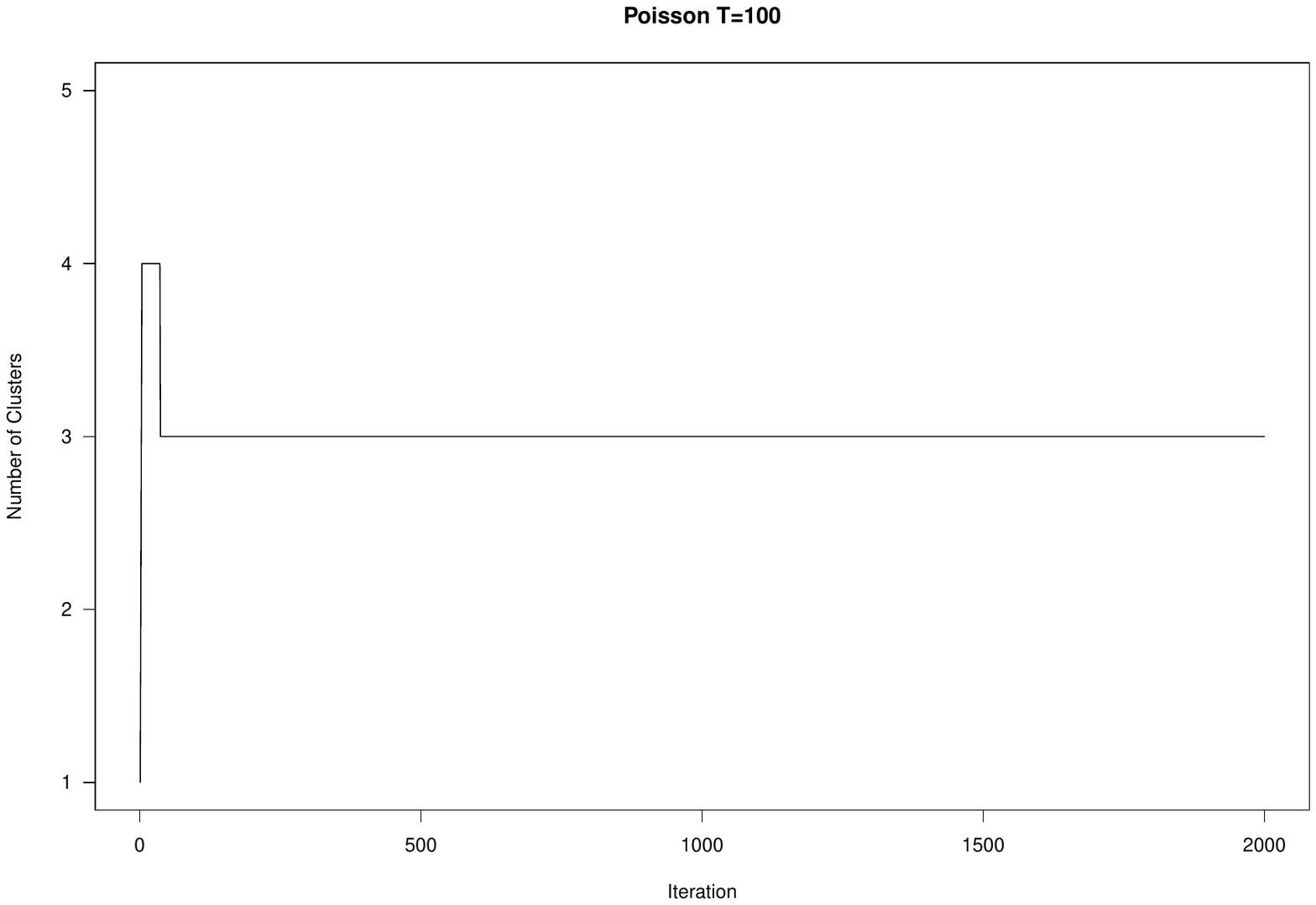}
		%\caption*{Gamma$\left(20,500\right)$}
	\end{minipage}
	\label{tracesim2}
\end{figure}

\begin{figure}[ht]
	\centering
	\caption{Example 1: Trace plots of the number of clusters via a mixture of finite mixture models using reversible-jump MCMC for Normal (left) and Poisson (right) models and for $T=30, 50$ and $100$ observations per subject (top, middle, bottom rows respectively).  The plot reveals that as the amount of data increases, the posterior distribution on the number of clusters becomes more concentrated on the three cluster model.}
	\begin{minipage}[b]{0.425\textwidth}
		\includegraphics[width=\textwidth]{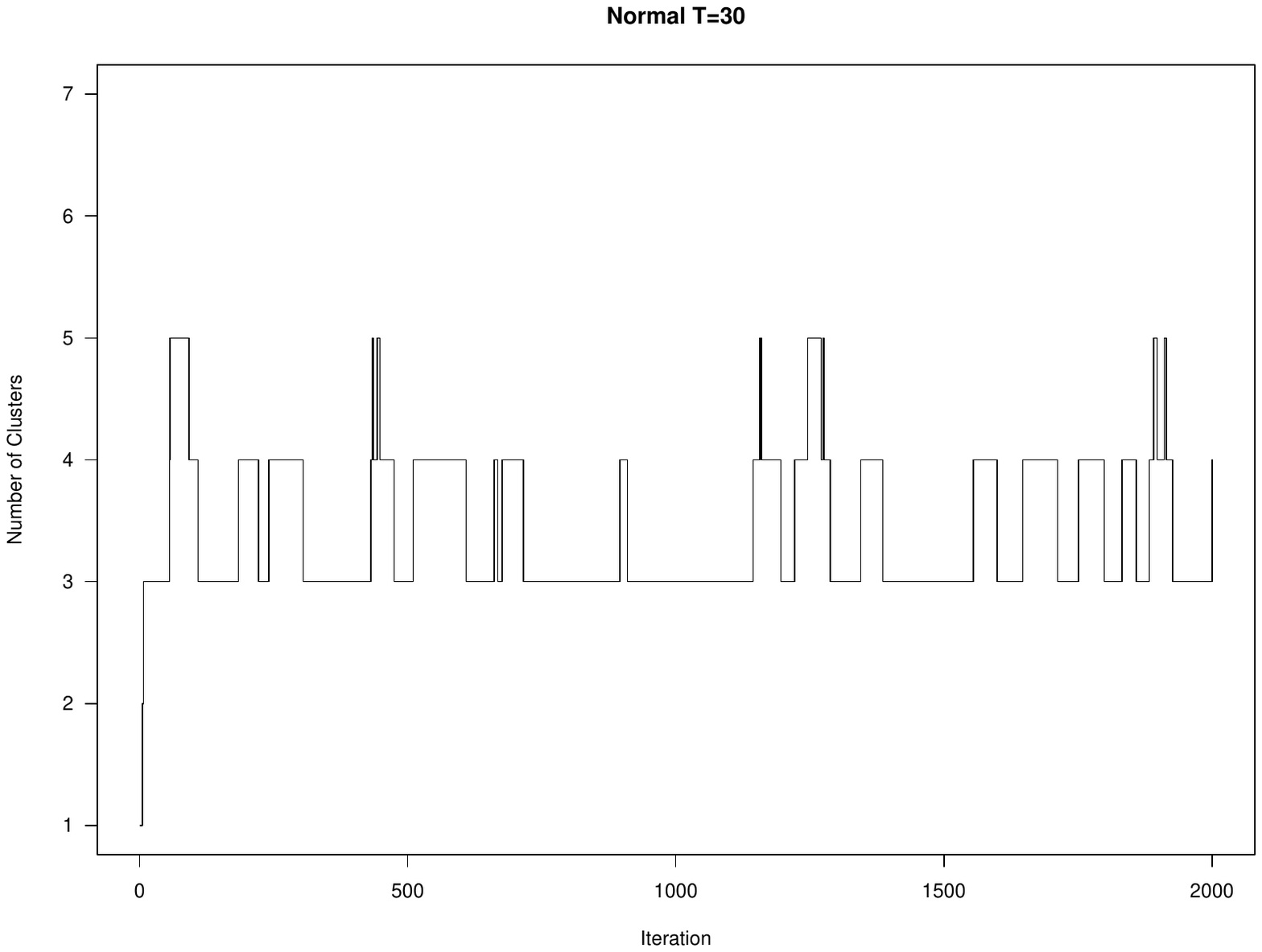}
		%\caption*{Gamma$\left(20,500\right)$}
	\end{minipage}
	\hfill
	\begin{minipage}[b]{0.425\textwidth}
		\includegraphics[width=\textwidth]{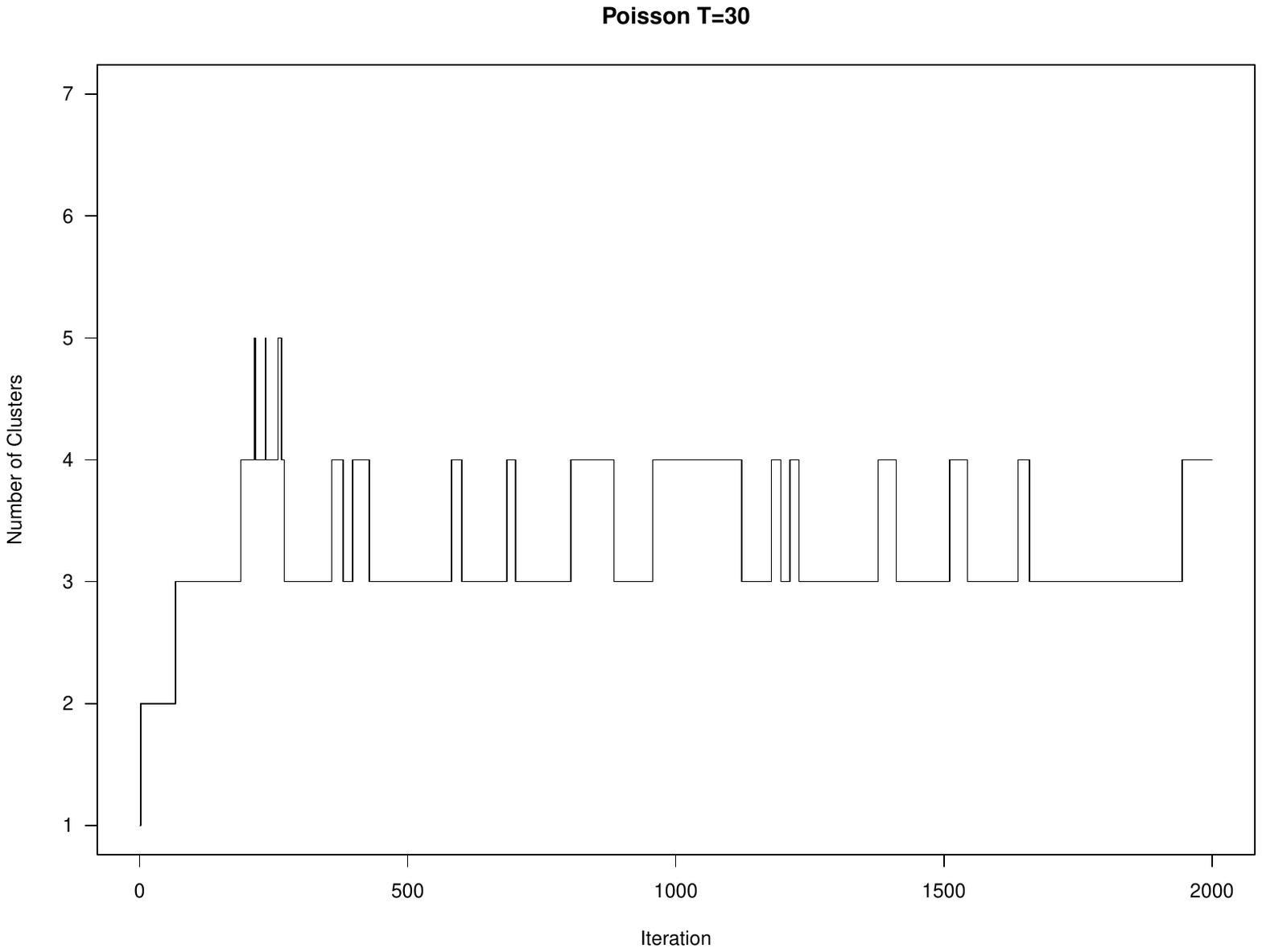}
		%\caption*{Gamma$\left(20,500\right)$}
	\end{minipage}
	\begin{minipage}[b]{0.425\textwidth}
		\includegraphics[width=\textwidth]{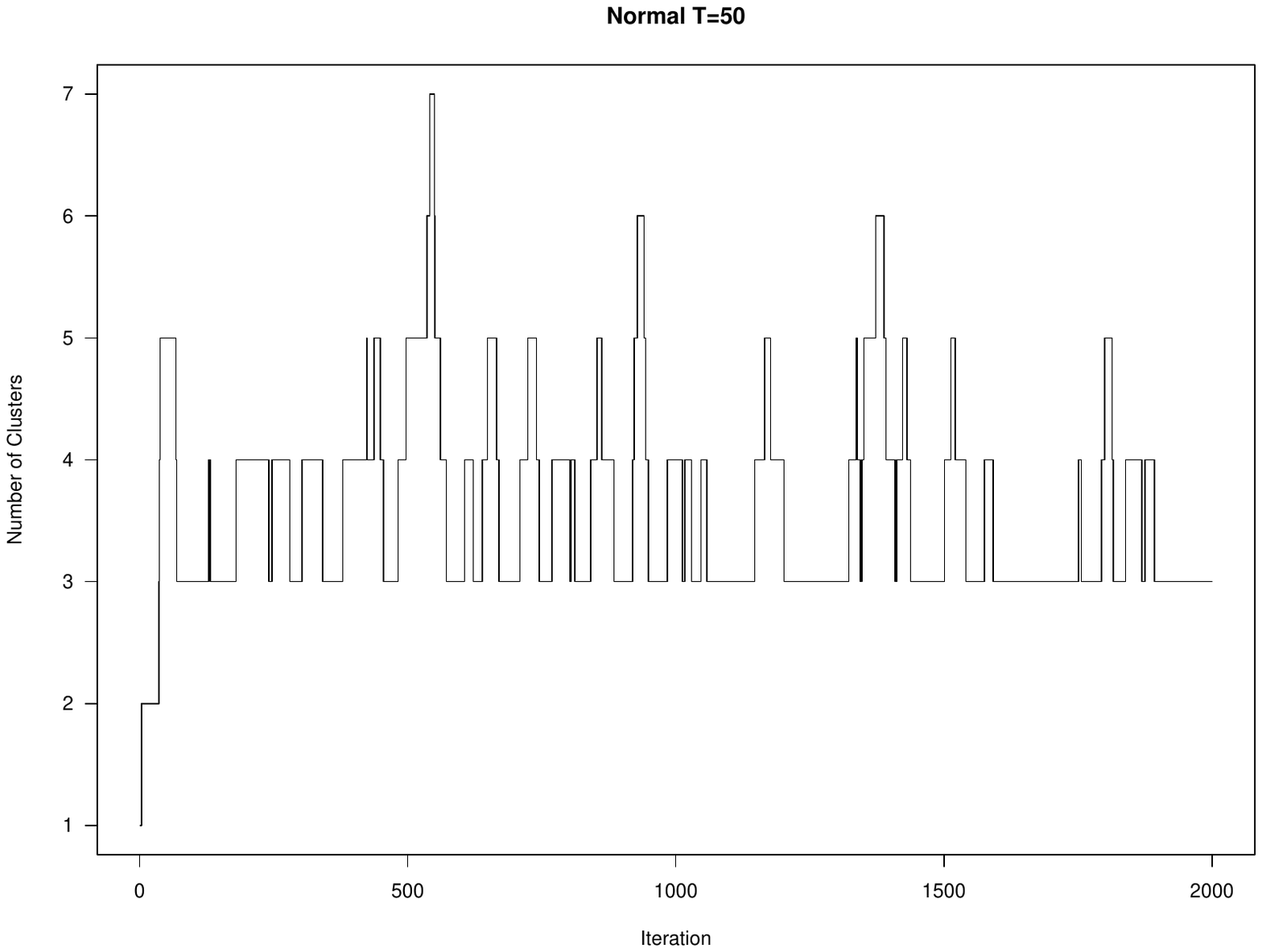}
		%\caption*{Gamma$\left(2,50\right)$}
	\end{minipage}
	\hfill
	\begin{minipage}[b]{0.425\textwidth}
		\includegraphics[width=\textwidth]{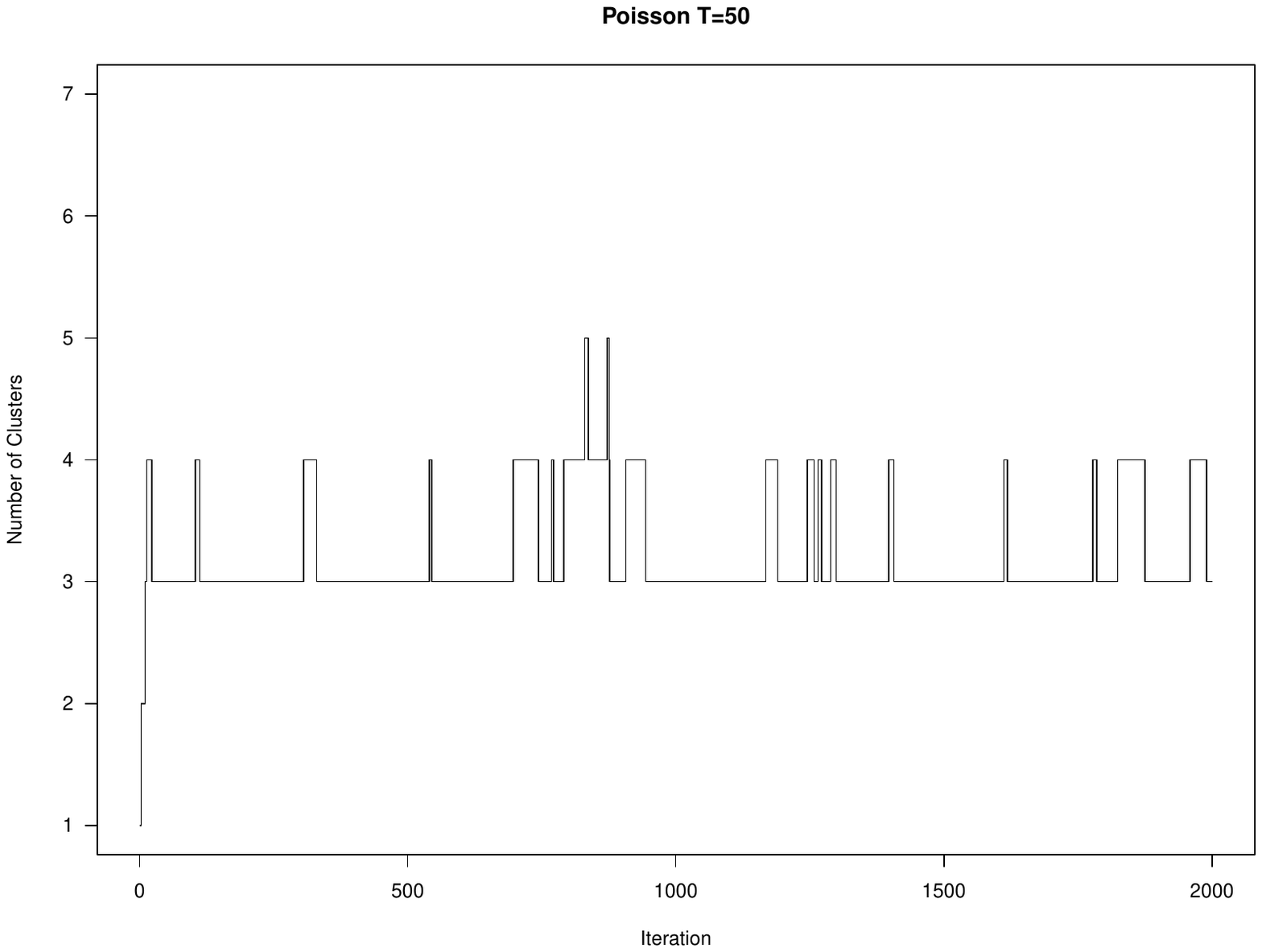}
		%\caption*{Gamma$\left(20,500\right)$}
	\end{minipage}
	\hfill
	\begin{minipage}[b]{0.425\textwidth}
		\includegraphics[width=\textwidth]{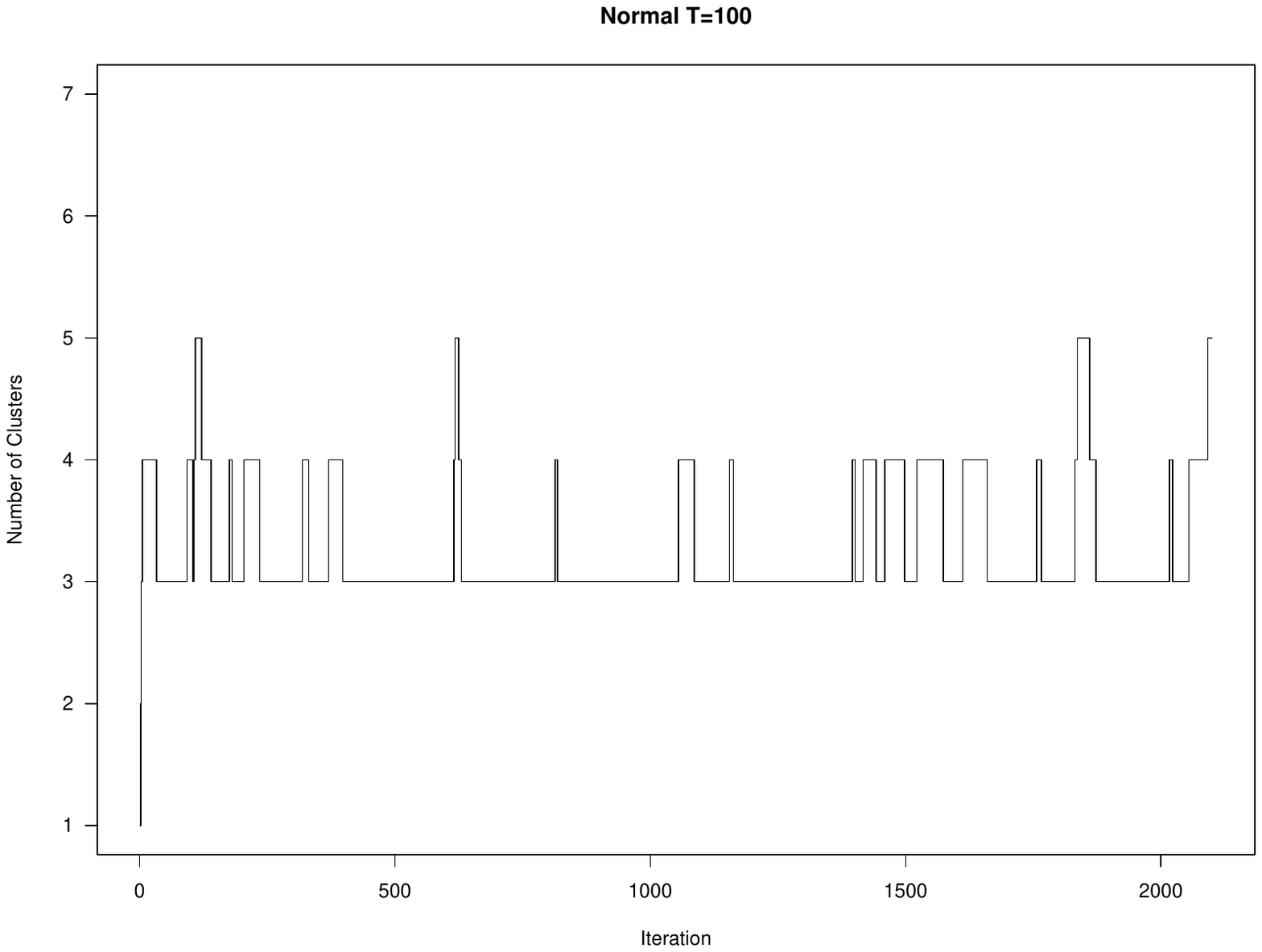}
		%\caption*{Gamma$\left(20,500\right)$}
	\end{minipage}
	\hfill
	\begin{minipage}[b]{0.425\textwidth}
		\includegraphics[width=\textwidth]{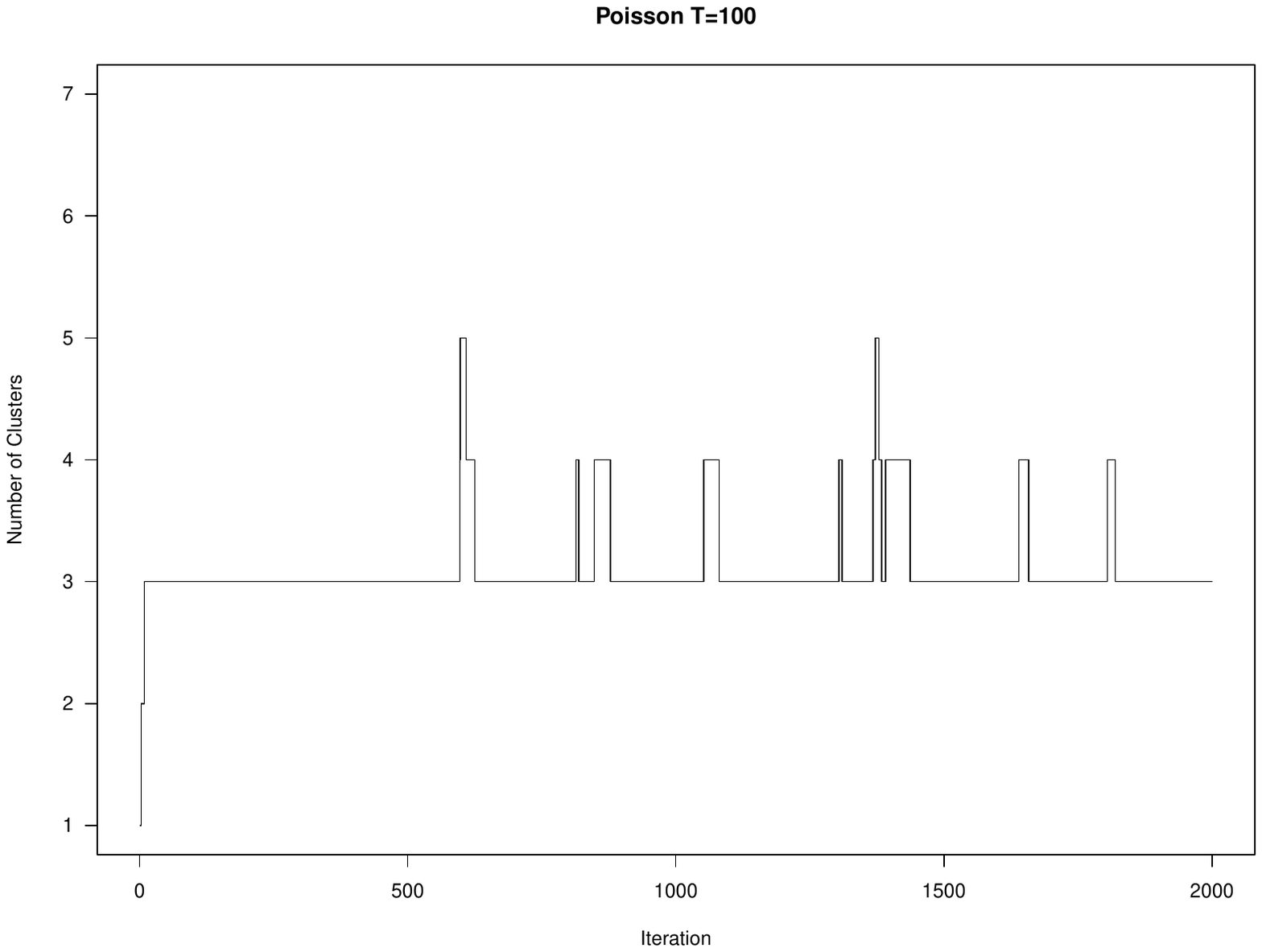}
		%\caption*{Gamma$\left(20,500\right)$}
	\end{minipage}
	\label{tracesimRJ}
\end{figure}

\begin{figure}[ht]
	\centering
	\caption{Example 1: Trace plots of the number of clusters via a mixture of finite mixture models using the split-merge algorithm models for Normal (left) and Poisson (right) models and for $T=30, 50$ and $100$ observations per subject (top, middle, bottom rows respectively).  The plot reveals that as the amount of data increases, the posterior distribution on the number of clusters becomes more concentrated on the three cluster model.}
	\begin{minipage}[b]{0.425\textwidth}
		\includegraphics[width=\textwidth]{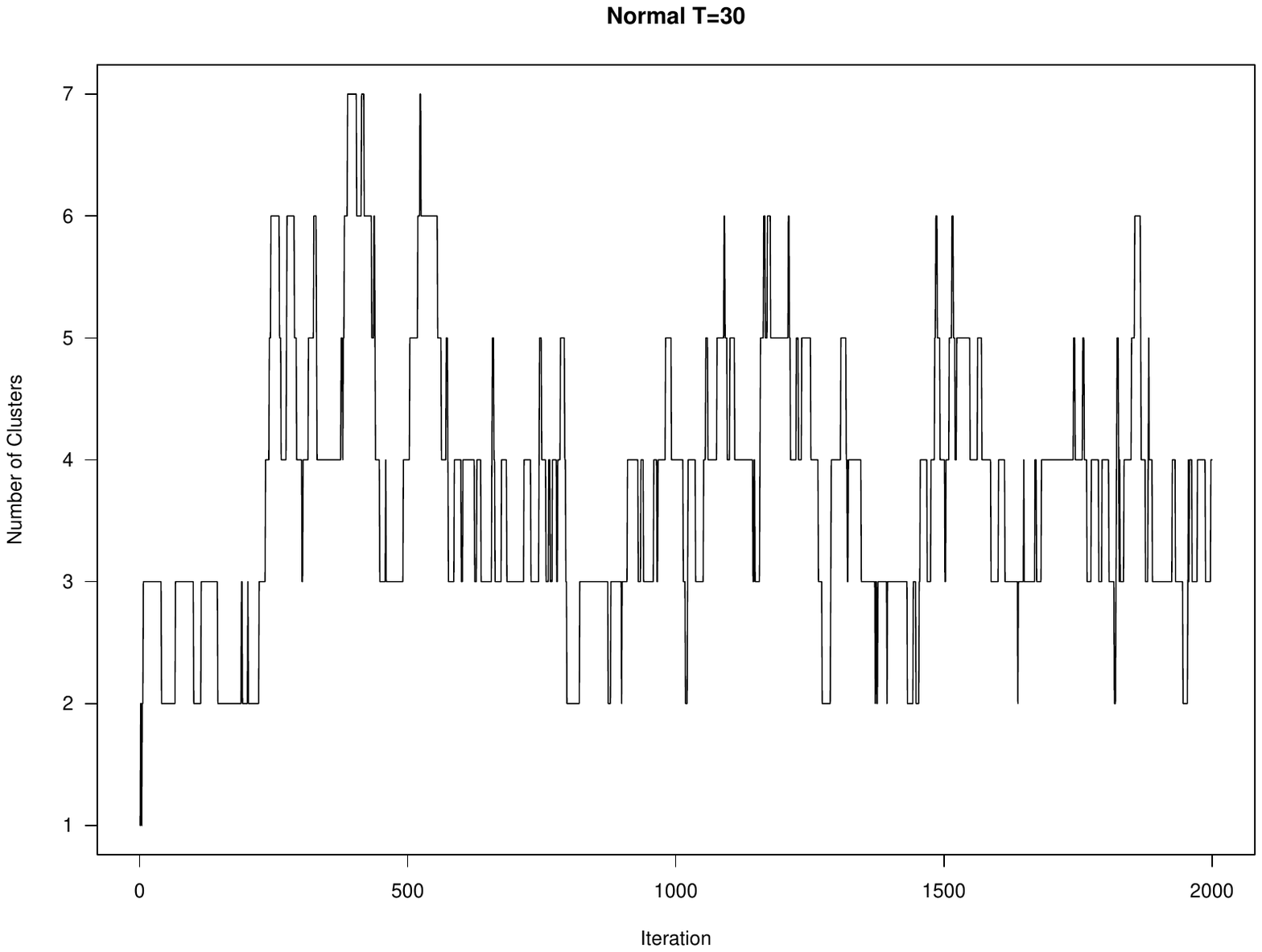}
		%\caption*{Gamma$\left(20,500\right)$}
	\end{minipage}
	\hfill
	\begin{minipage}[b]{0.425\textwidth}
		\includegraphics[width=\textwidth]{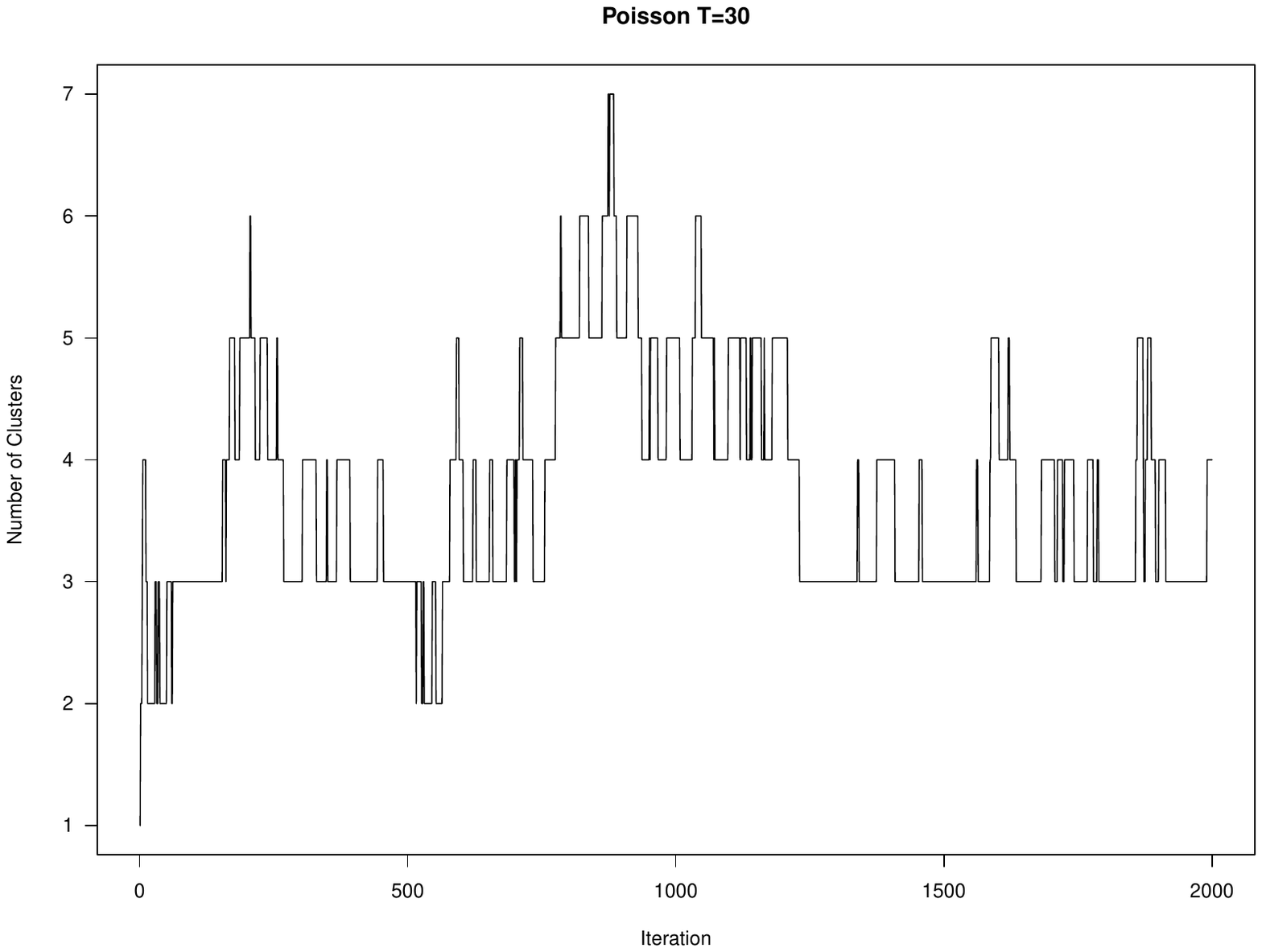}
		%\caption*{Gamma$\left(20,500\right)$}
	\end{minipage}
	\begin{minipage}[b]{0.425\textwidth}
		\includegraphics[width=\textwidth]{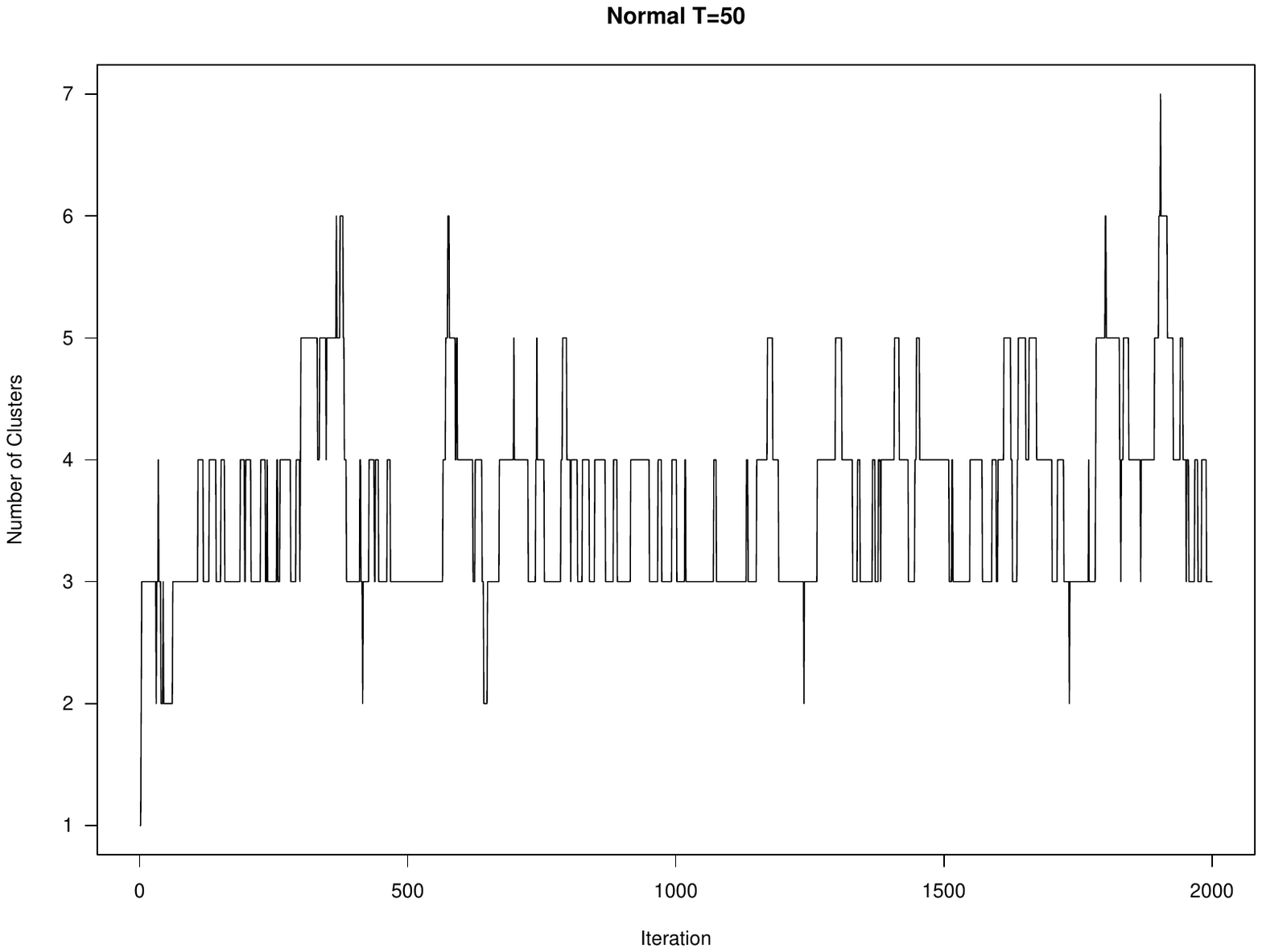}
		%\caption*{Gamma$\left(2,50\right)$}
	\end{minipage}
	\hfill
	\begin{minipage}[b]{0.425\textwidth}
		\includegraphics[width=\textwidth]{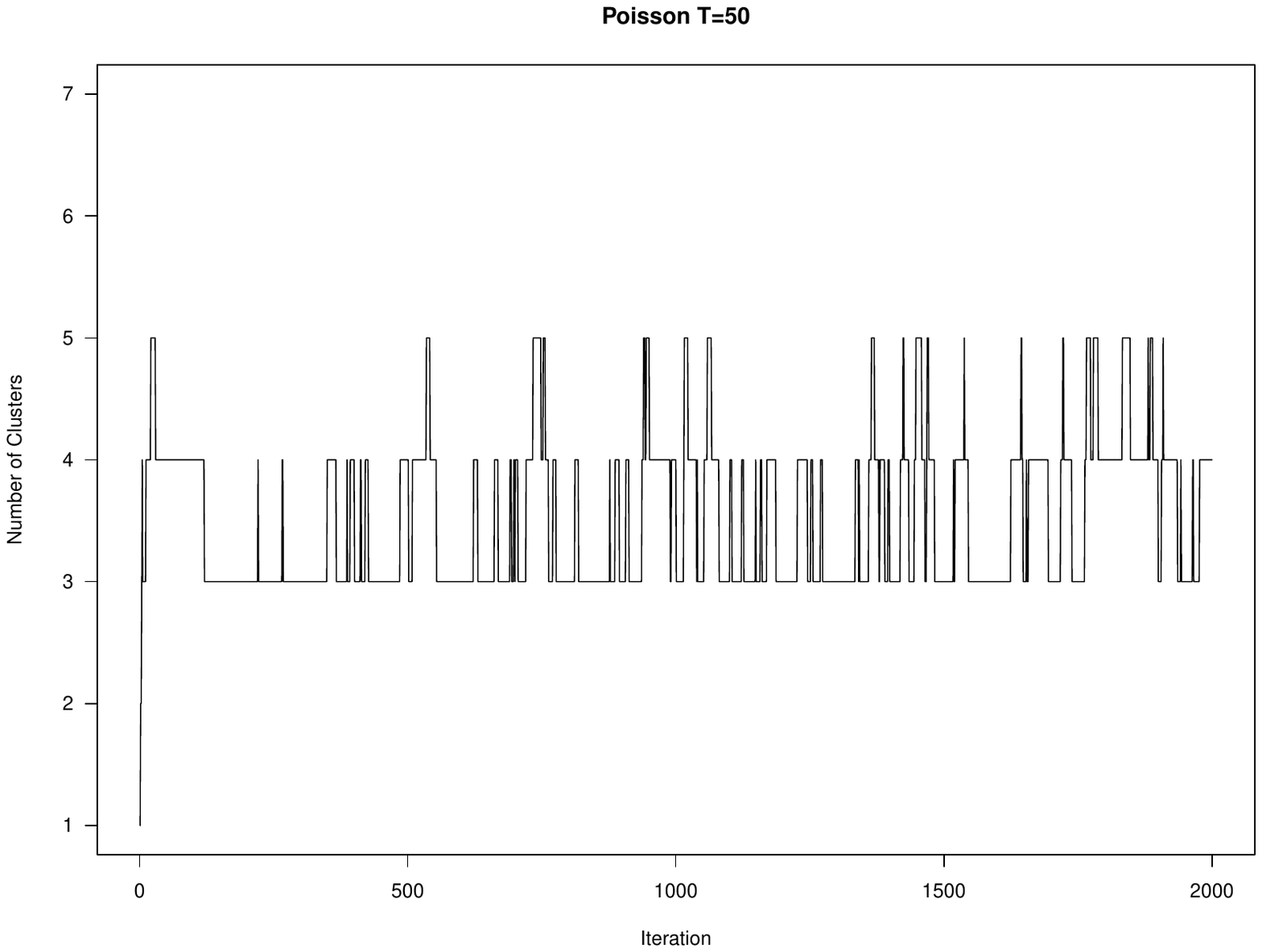}
		%\caption*{Gamma$\left(20,500\right)$}
	\end{minipage}
	\hfill
	\begin{minipage}[b]{0.425\textwidth}
		\includegraphics[width=\textwidth]{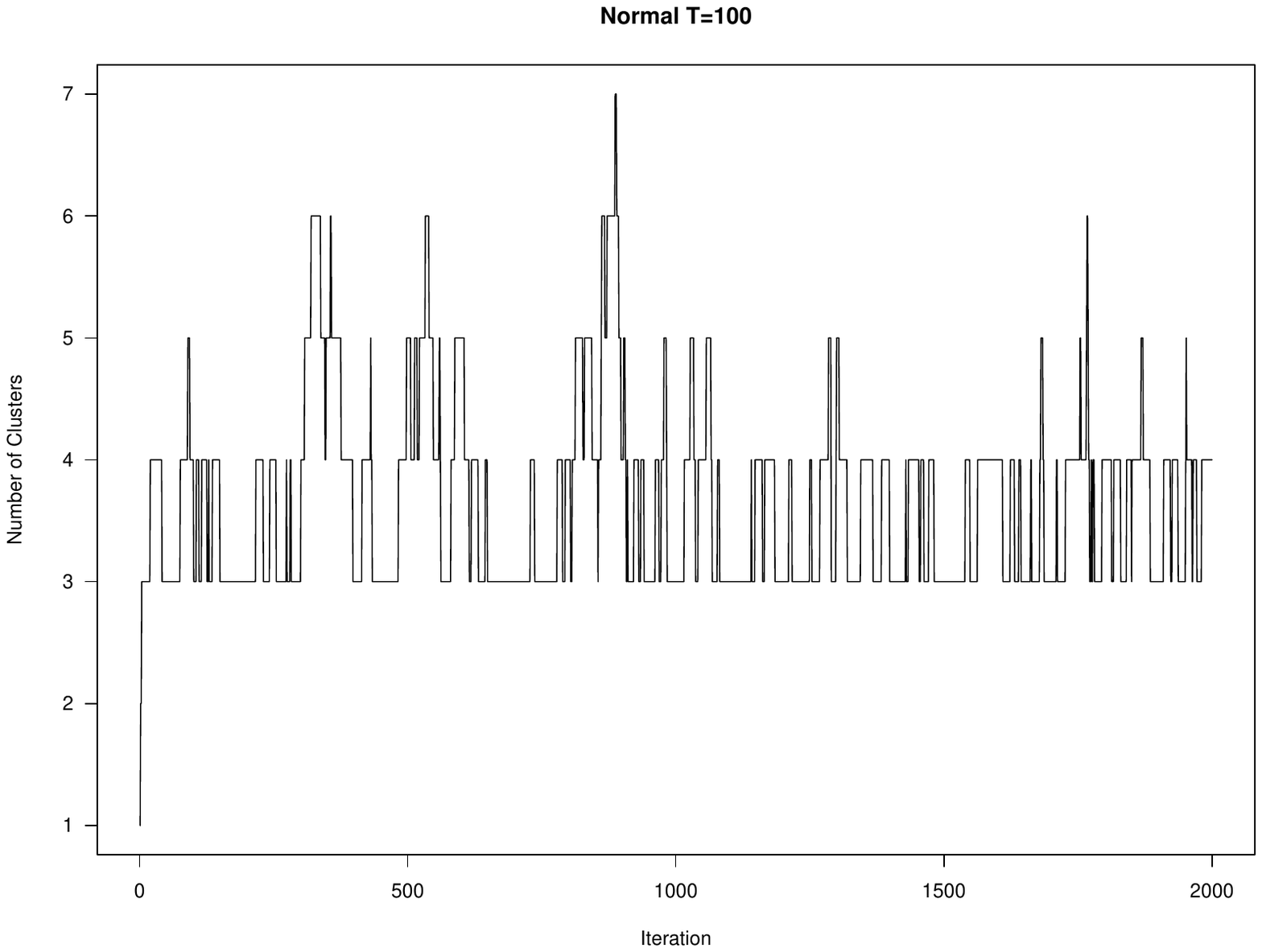}
		%\caption*{Gamma$\left(20,500\right)$}
	\end{minipage}
	\hfill
	\begin{minipage}[b]{0.425\textwidth}
		\includegraphics[width=\textwidth]{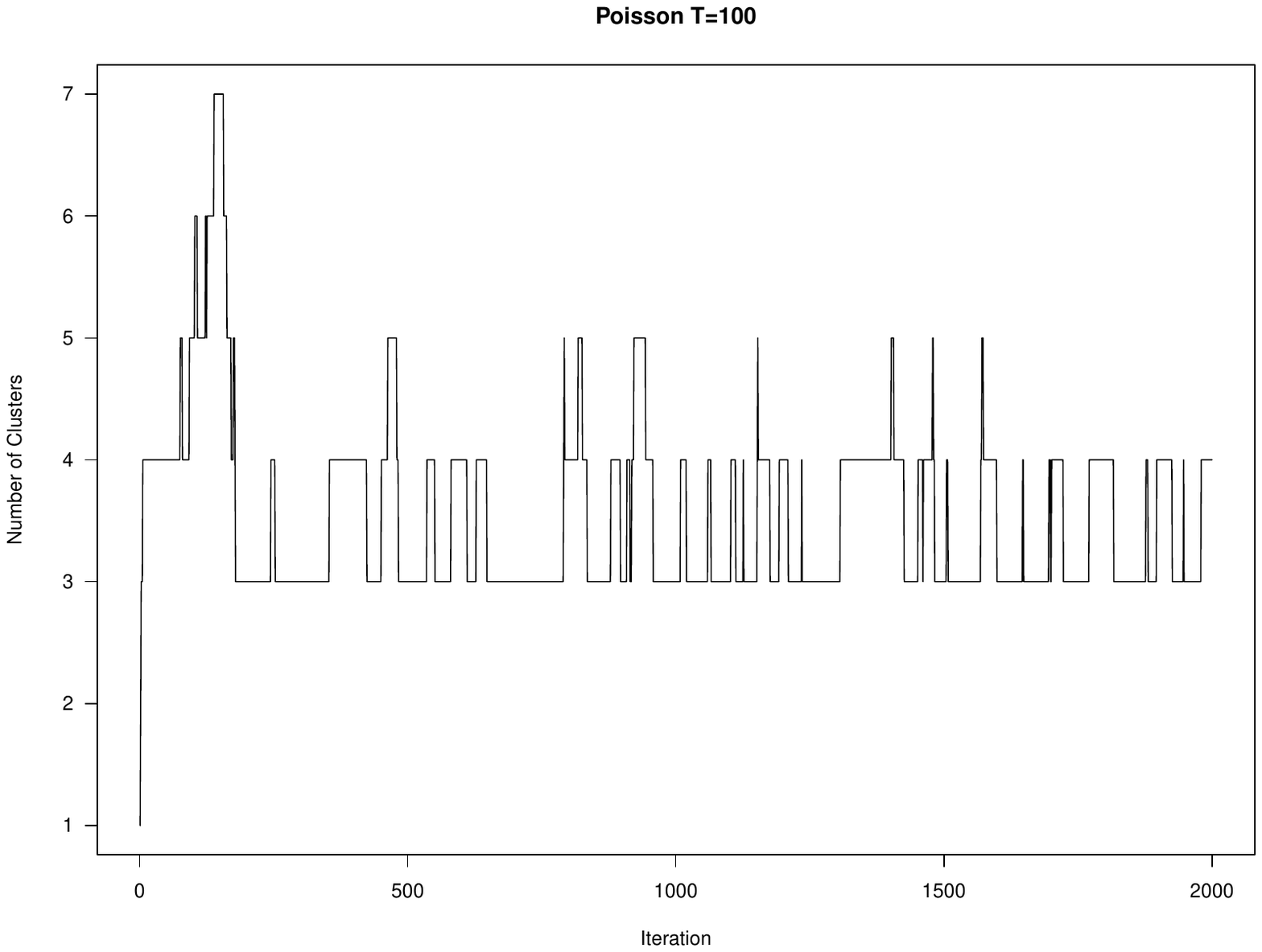}
		%\caption*{Gamma$\left(20,500\right)$}
	\end{minipage}
	\label{tracesimMFM}
\end{figure}

\pagebreak

\subsection{Example 2}

\begin{table}[ht]
	\caption{\label{sim91}Example 2: Simulation study with three clusters.  Each cluster has three latent states, the same $B$ and $\pi$ parameters, but different $Q$ matrices.}
	
	{
		\begin{tabular*}{40pc}{@{\hskip5pt}@{\extracolsep{\fill}}c@{}c@{}c@{}c@{}c@{}c@{}c@{}c@{}c@{}c@{\hskip3pt}}
			
			\hline 
			%\multicolumn{6}{c}{EM}\\\cline{1-6}
			& \multicolumn{3}{c}{MFM-RJ} &\multicolumn{3}{c}{DMM}&\multicolumn{3}{c}{MFM-SM}\\ \cline{2-10}
			%\midrule[2pt]
			%\multicolumn{8}{c}{EM algorithm with tolerance 0.05 }\\\cline{1-8}
			&	$\sigma=0.5$ &$\sigma=1$&$\sigma=2$ & $\sigma=0.5$ &$\sigma=1$&$\sigma=2$& $\sigma=0.5$ &$\sigma=1$&$\sigma=2$\\
			\hline 
			$\left\| {\pi-\hat \pi} \right\|$ & 0.03&0.03 & 0.06& 0.01&  0.01&0.07& 0.06 & 0.06& 0.05\\
			$\left\| {B-\hat B} \right\|$ & 0.02& 0.05 &0.01& 0.02  &0.02 &0.01& 0.01 & 0.01& 0.08\\
			$\left\| {Q_1-\hat Q_1} \right\|$ &0.30 &0.26 &0.78 &0.32 &0.51 &0.77  &0.79& 0.51 & 1.08\\
			$\left\| {Q_2-\hat Q_2} \right\|$ &0.08 & 0.10&0.49&0.38 & 0.34 & 0.36 & 0.21 & 0.42 & 0.73 \\
			$\left\| {Q_3-\hat Q_3} \right\|$ &0.05  &0.07 &0.87&0.14&0.16& 0.64& 0.68 & 0.85& 0.92  \\
			\% of 3-cluster iterations& 58.05\%&61.05\%& 44.10\%&80.90\%  &95.75\%  & 25.90\% & 54.10\% & 46.30\%& 37.20\%  \\
			\% of Misclassification&4.90\% &9.50\%  &24.20\% &19.20\%  & 22.13\% & 24.70\%& 15.40\% & 19.30\% &  25.30\% \\
			\hline 
	\end{tabular*}}
\end{table}

\begin{figure}[ht]
	\centering
	\caption{Example 2: Trace plots of the number of clusters via DP mixture model for Normal models where $T= 50$ observations per subject and $\sigma=0.5,1$ and $2$ with component parameter $Q$ only.}
	\begin{minipage}[b]{0.425\textwidth}
		\includegraphics[width=\textwidth]{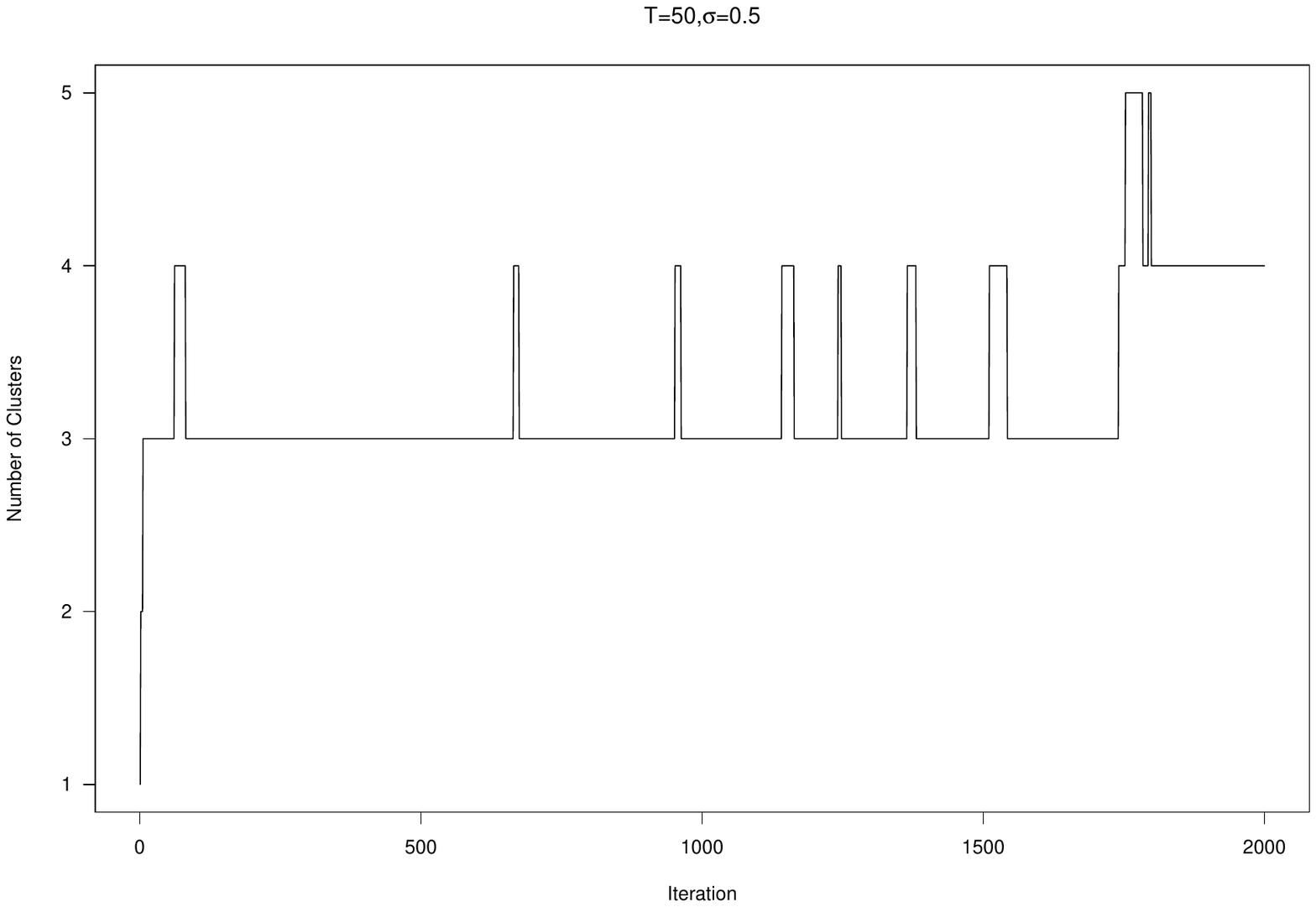}
		%\caption*{Gamma$\left(20,500\right)$}
	\end{minipage}
	\hfill
	\begin{minipage}[b]{0.425\textwidth}
		\includegraphics[width=\textwidth]{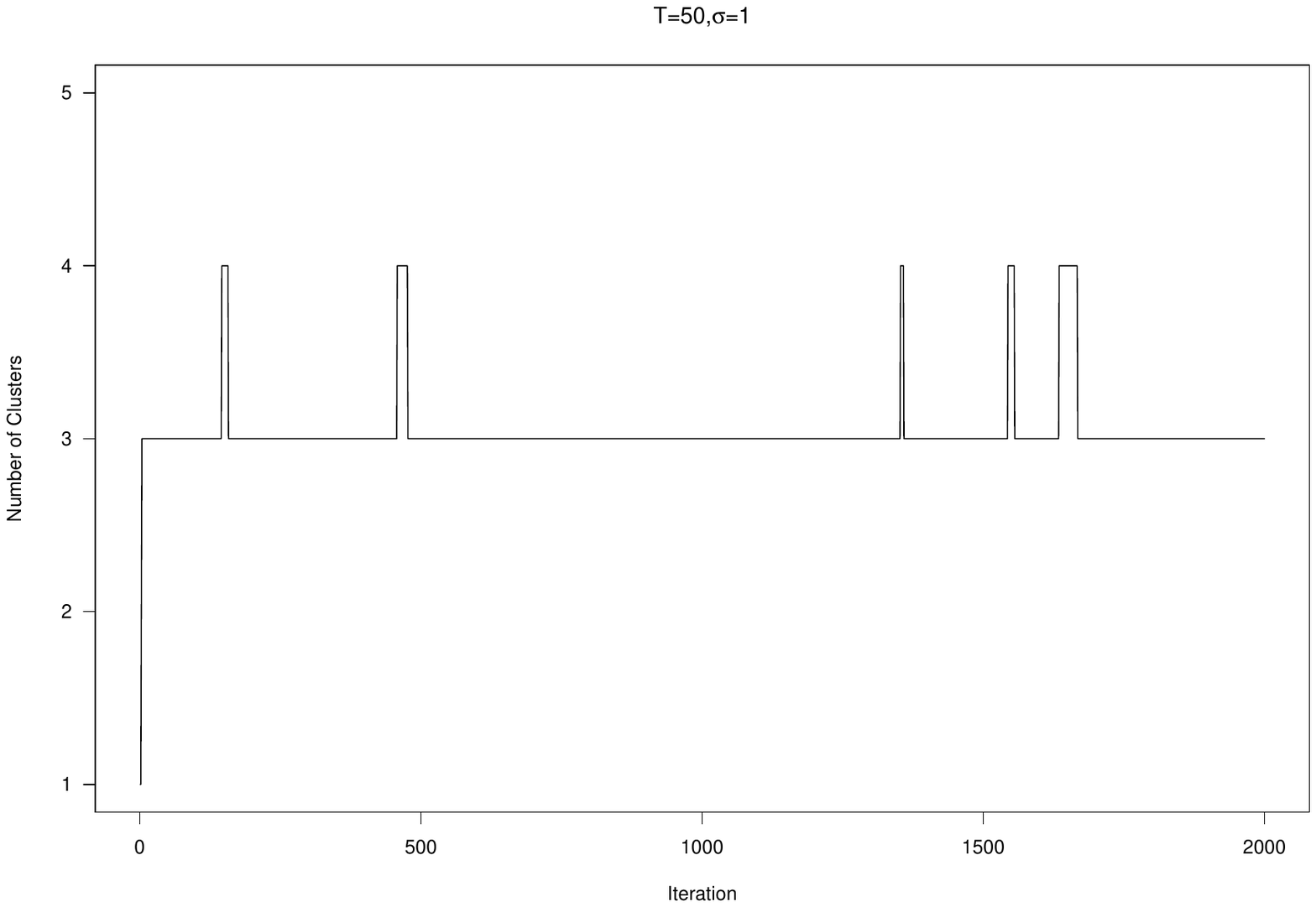}
		%\caption*{Gamma$\left(20,500\right)$}
	\end{minipage}
	\hfill
	\begin{minipage}[b]{0.425\textwidth}
		\includegraphics[width=\textwidth]{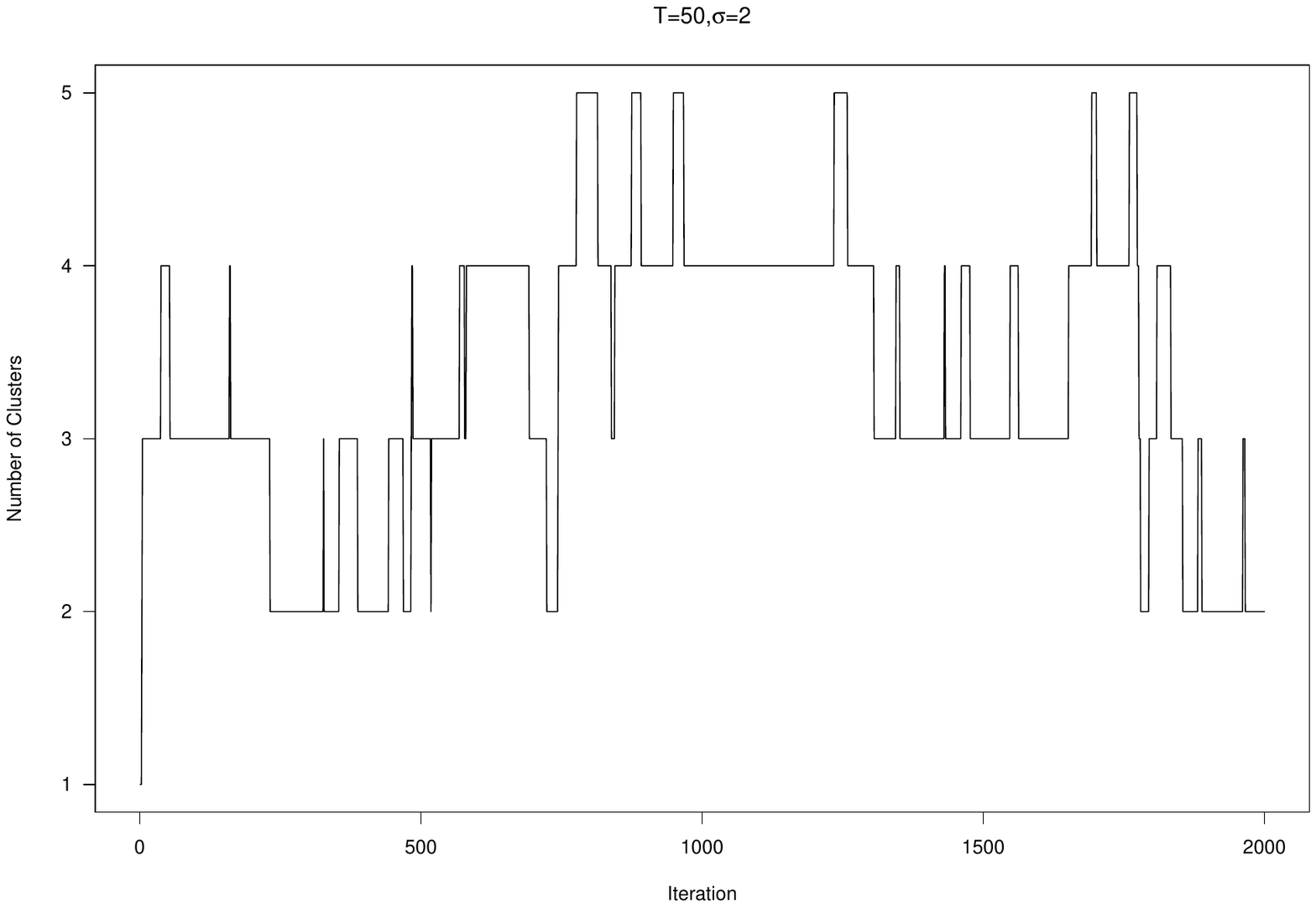}
		%\caption*{Gamma$\left(2,50\right)$}
	\end{minipage}
	\label{tracedirex2}
\end{figure}

\begin{figure}[ht]
	\centering
	\caption{Example 2: Trace plots of the number of clusters via a mixture of finite mixture models  using reversible-jump MCMC for Normal models where $T= 50$ observations per subject and $\sigma=0.5,1$ and $2$ with component parameter $Q$ only.}
	\begin{minipage}[b]{0.425\textwidth}
		\includegraphics[width=\textwidth]{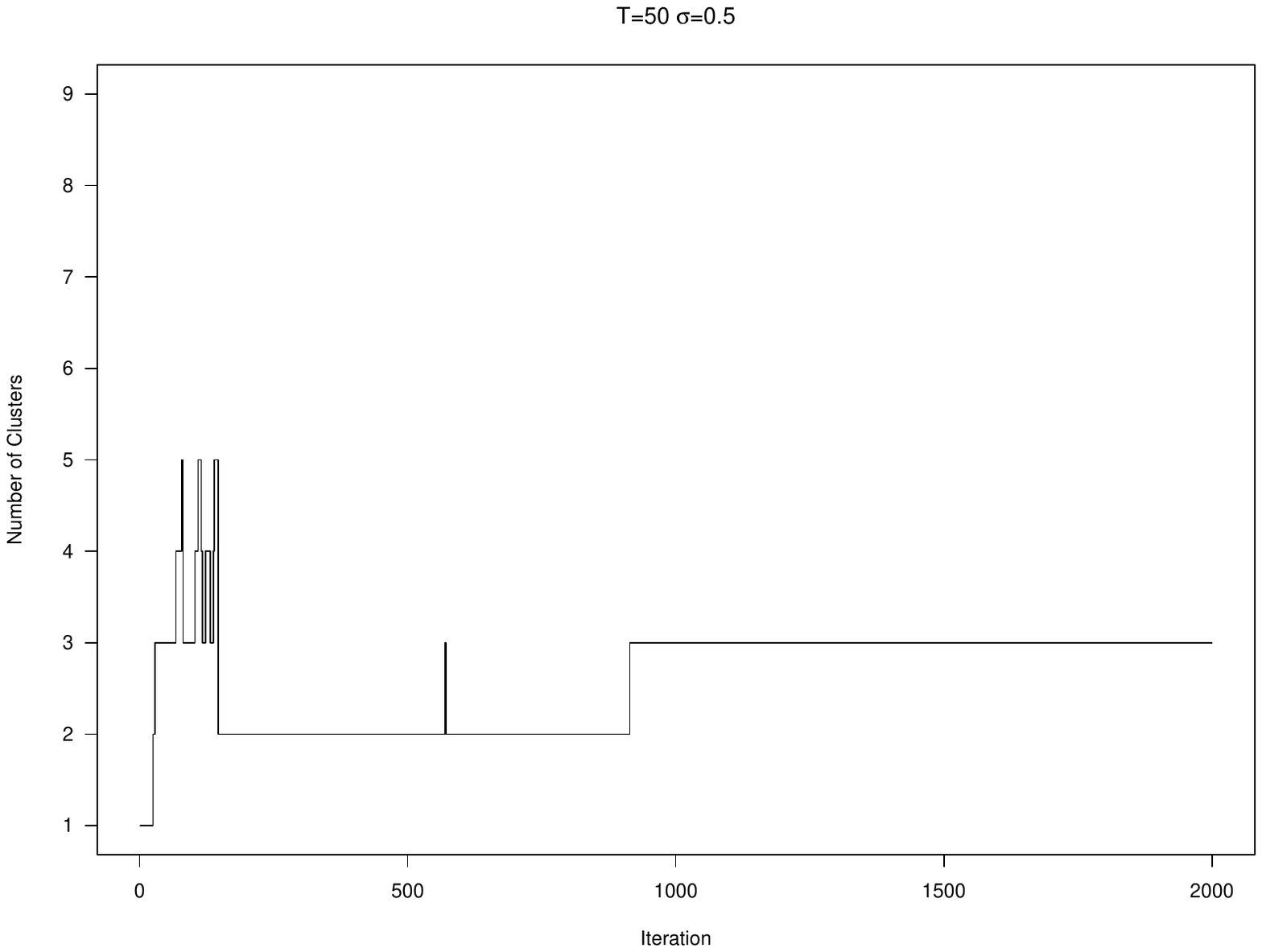}
		%\caption*{Gamma$\left(20,500\right)$}
	\end{minipage}
	\hfill
	\begin{minipage}[b]{0.425\textwidth}
		\includegraphics[width=\textwidth]{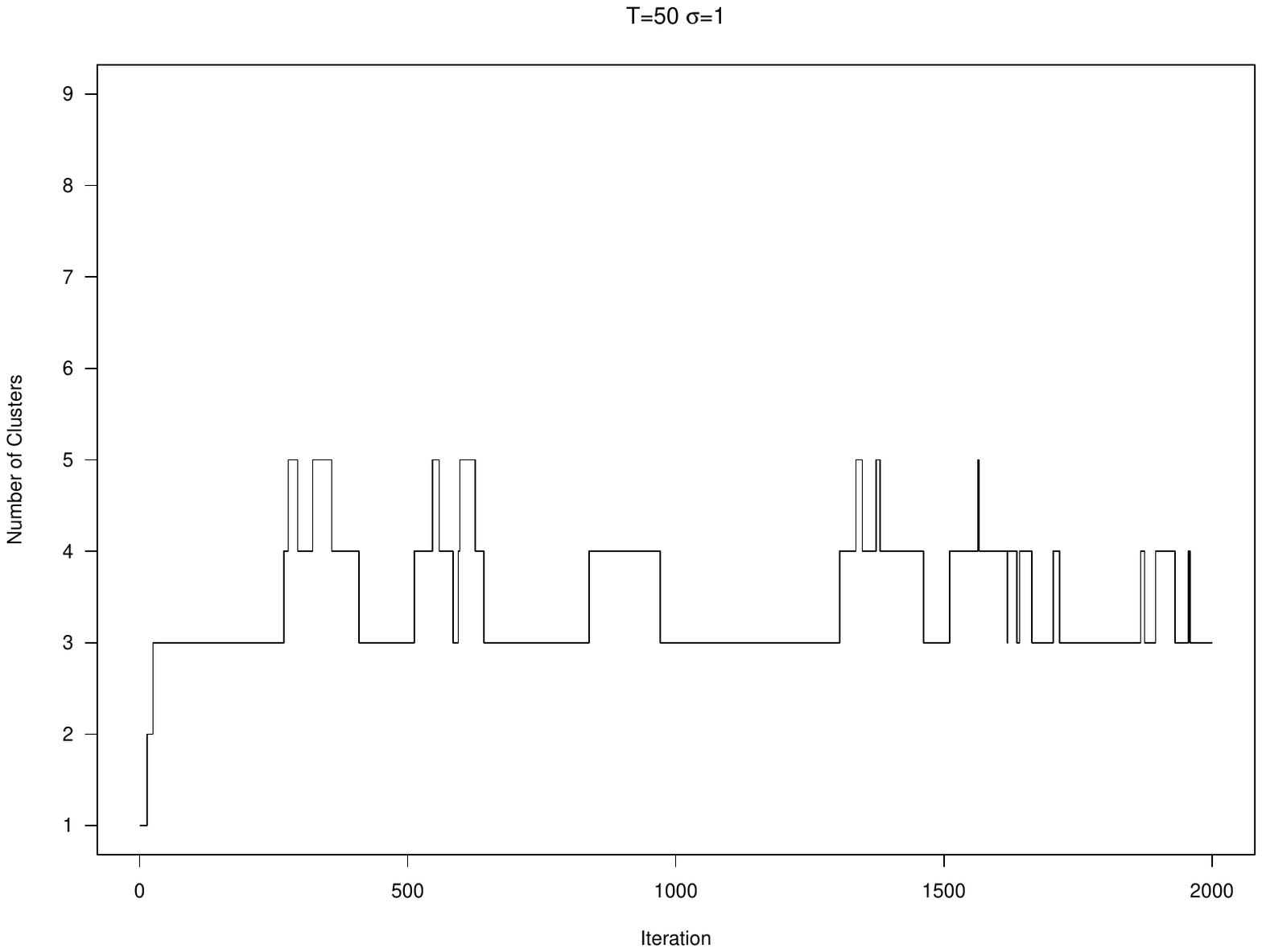}
		%\caption*{Gamma$\left(20,500\right)$}
	\end{minipage}
	\hfill
	\begin{minipage}[b]{0.425\textwidth}
		\includegraphics[width=\textwidth]{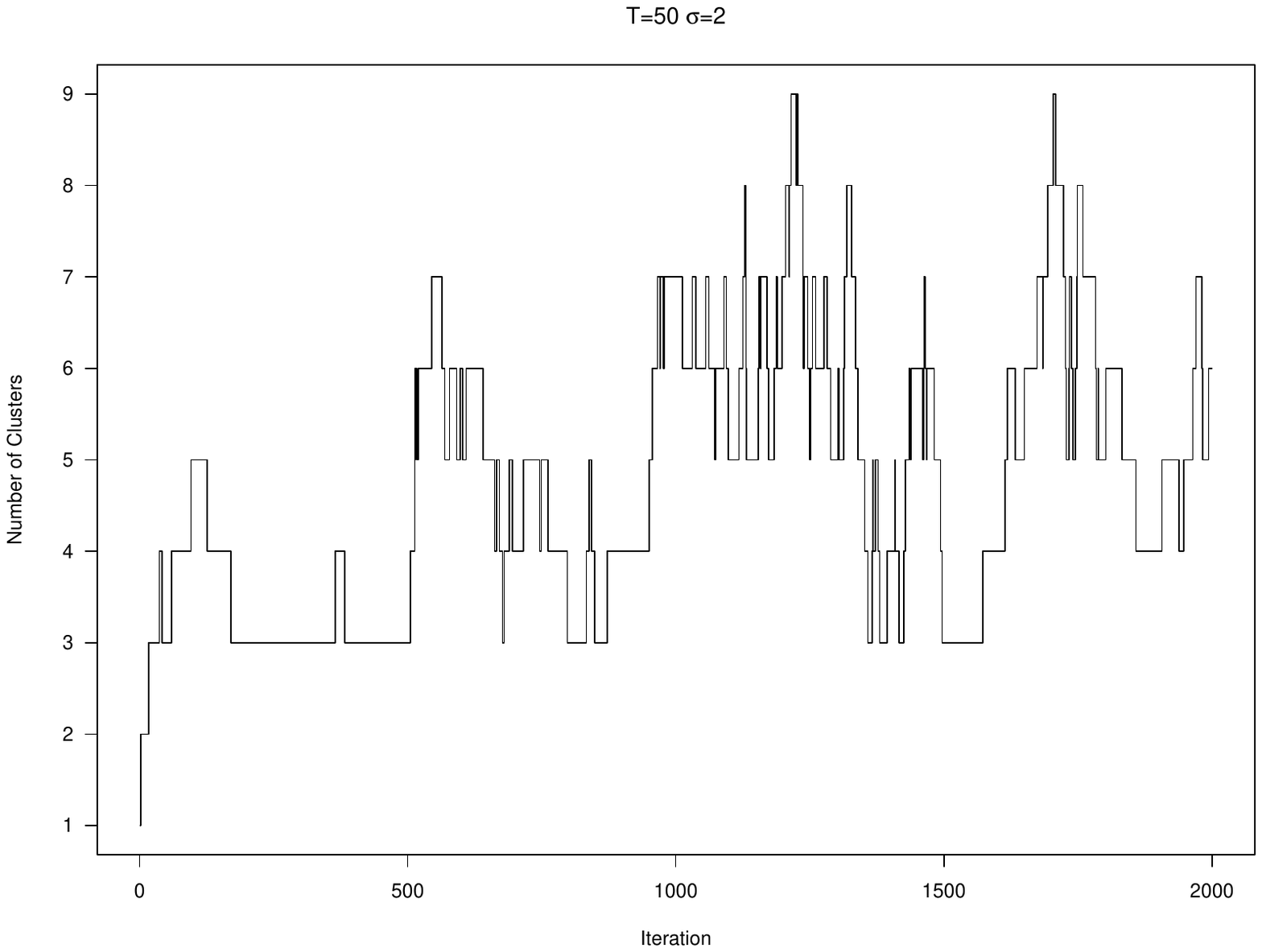}
		%\caption*{Gamma$\left(2,50\right)$}
	\end{minipage}
	\label{tracedirexRJ}
\end{figure}
\begin{figure}[ht]
	\centering
	\caption{Example 2: Trace plots of the number of clusters via a mixture of finite mixture models using the split-merge algorithm for Normal models where $T= 50$ observations per subject and $\sigma=0.5,1$ and $2$ with component parameter $Q$ only.}
	\begin{minipage}[b]{0.425\textwidth}
		\includegraphics[width=\textwidth]{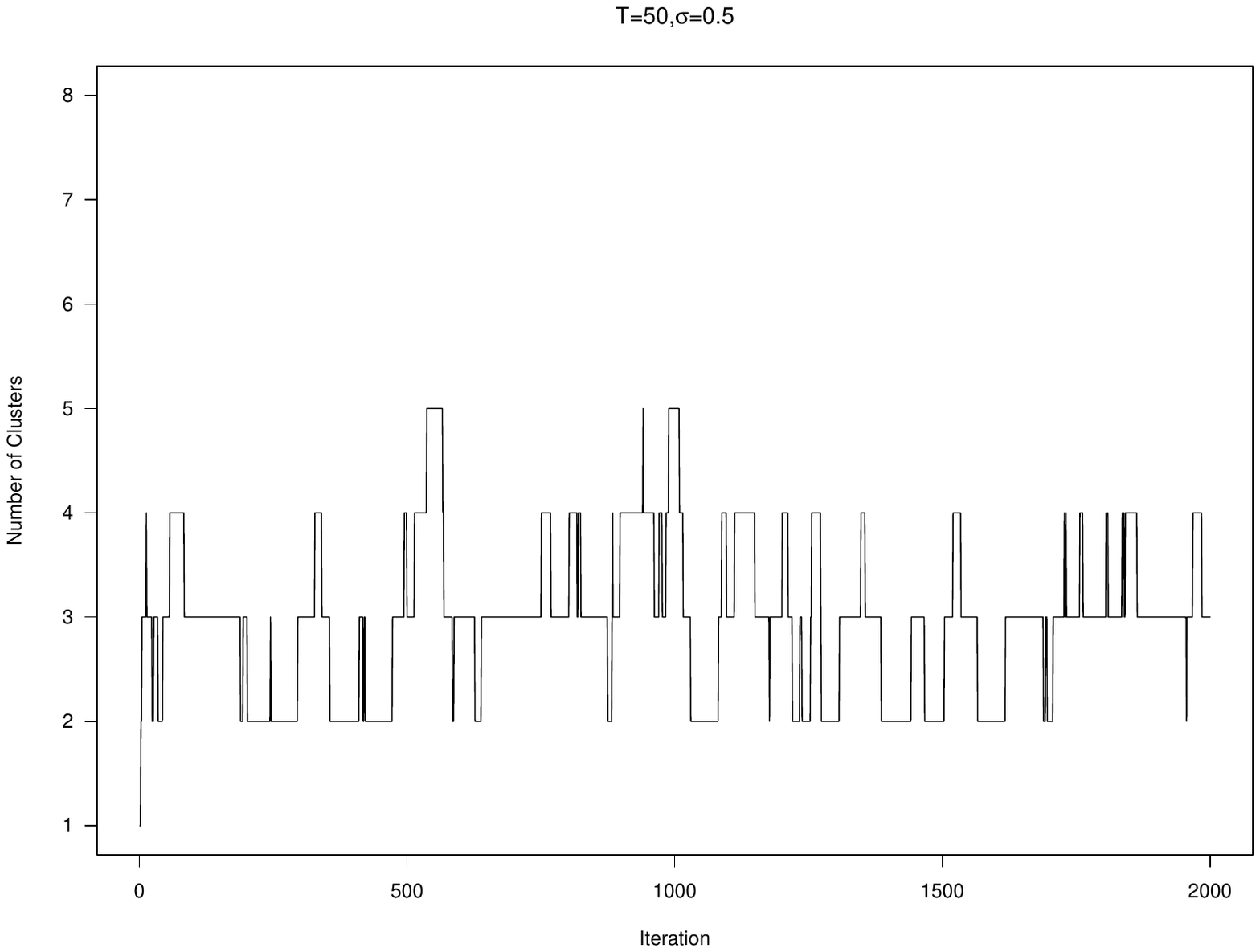}
		%\caption*{Gamma$\left(20,500\right)$}
	\end{minipage}
	\hfill
	\begin{minipage}[b]{0.425\textwidth}
		\includegraphics[width=\textwidth]{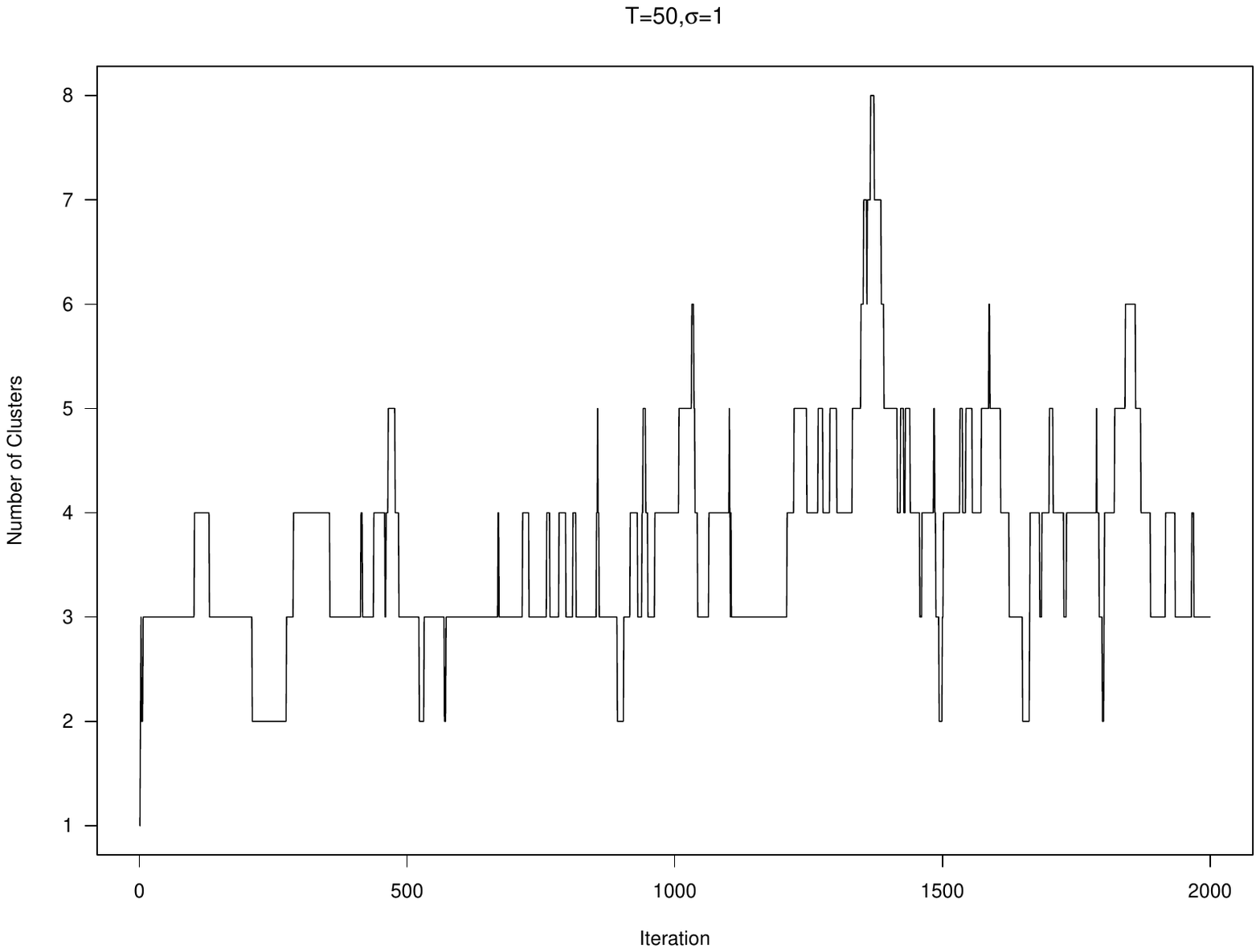}
		%\caption*{Gamma$\left(20,500\right)$}
	\end{minipage}
	\hfill
	\begin{minipage}[b]{0.425\textwidth}
		\includegraphics[width=\textwidth]{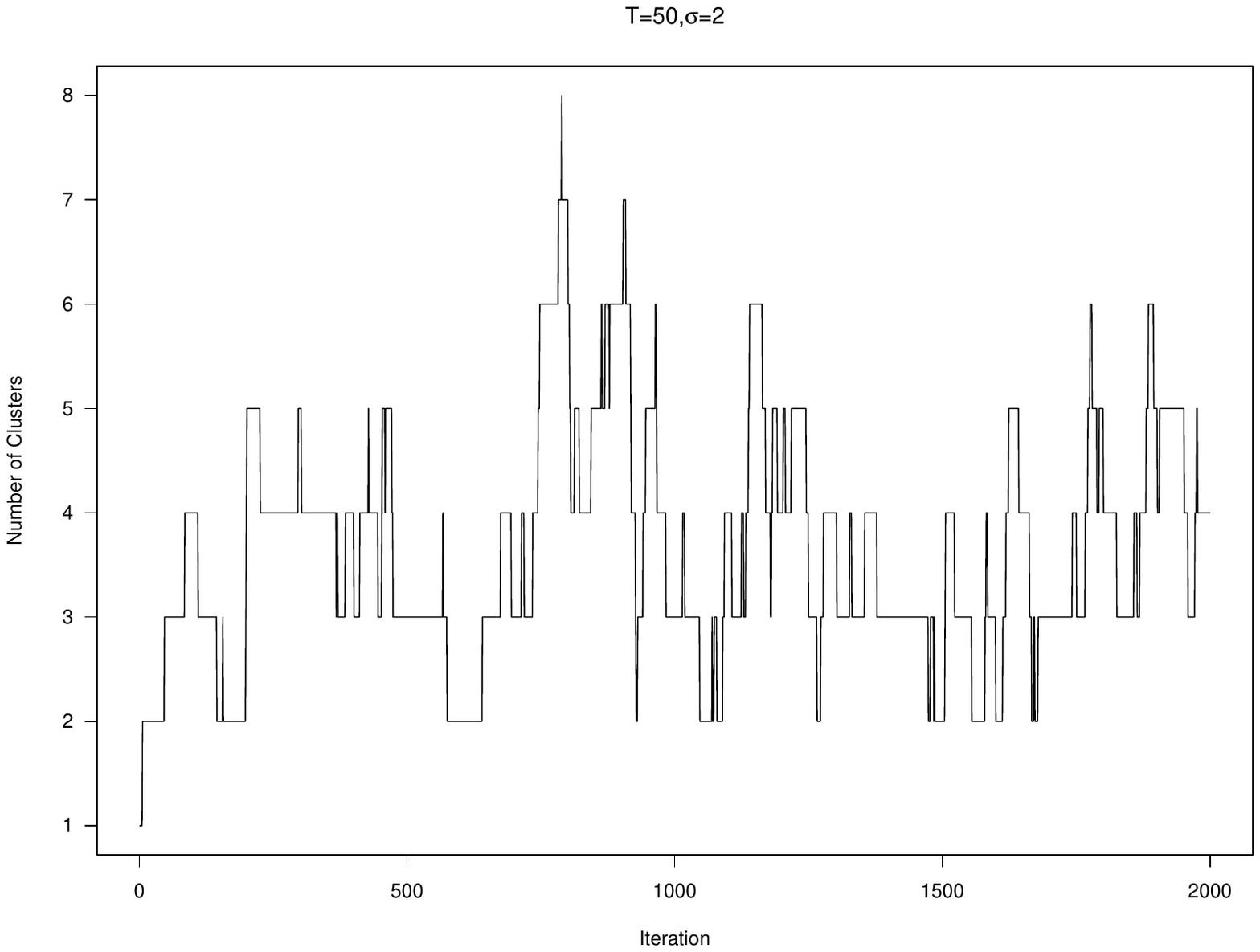}
		%\caption*{Gamma$\left(2,50\right)$}
	\end{minipage}
	\label{tracemfmQ}
\end{figure}

\pagebreak

\subsection{Example 3}

\begin{table}[ht]
	\caption{\label{simex31}Example 3: Simulation study with three clusters with one factor time-varying covariate. Each cluster has three latent states and the observation process is generated from Poisson distribution.}
	
	{\centering
		\begin{tabular*}{41pc}{@{\hskip5pt}@{\extracolsep{\fill}}c@{}c@{}c@{}c@{}@{}c@{}c@{}c@{}c@{}c@{}c@{\hskip5pt}}
			
			\hline 
			%\multicolumn{6}{c}{EM}\\\cline{1-6}
			%\midrule[2pt]
			%\multicolumn{8}{c}{EM algorithm with tolerance 0.05 }\\\cline{1-8}
			& \multicolumn{3}{c}{MFM-RJ} &\multicolumn{3}{c}{DMM}&\multicolumn{3}{c}{MFM-SM}\\ \cline{2-10}
			%\midrule[2pt]
			%\multicolumn{8}{c}{EM algorithm with tolerance 0.05 }\\\cline{1-8}
			&$T=30$ &$T=50$  & $T=100$ 	&$T=30$ &$T=50$  & $T=100$ &$T=30$ &$T=50$  & $T=100$\\
			\hline
			$\left\| {\pi_1-\hat \pi_1} \right\|$  & 0.01 &0.02 & 0.02& 0.08& 0.06 & 0.07&0.07 & 0.07 & 0.08 \\
			$\left\| {B_1-\hat B_1} \right\|$ &  0.75& 0.69 & 0.14& 0.99 & 0.70 &0.08 & 1.15  & 1.06 & 0.07 \\
			$\left\| {Q_1-\hat Q_1} \right\|$ & 0.34 & 0.10& 0.15 &0.34 & 0.94&0.37 & 0.73 & 0.61& 0.13\\
			$\left\| {\pi_2-\hat \pi_2} \right\|$& 0.06 & 0.03& 0.02 & 0.10 & 0.05& 0.04 & 0.09& 0.08 & 0.03\\
			$\left\| {B_2-\hat B_2} \right\|$ & 0.06& 0.22 &0.03& 0.21& 0.11 &0.04& 0.20& 0.17 & 0.11\\
			$\left\| {Q_2-\hat Q_2} \right\|$ & 0.11 & 0.14 & 0.16&0.47 & 0.47&0.39 & 0.48 & 0.31& 0.26\\
			$\left\| {\pi_3-\hat \pi_3} \right\|$ & 0.02 & 0.02&0.06 & 0.12 & 0.10 & 0.12 & 0.16  & 0.08 & 0.13\\
			$\left\| {B_3-\hat B_3} \right\|$& 0.41& 0.30& 0.12 & 0.42 & 0.36 &0.19 & 0.51 & 0.68 & 0.09\\
			$\left\| {Q_3-\hat Q_3} \right\|$& 0.12& 0.12& 0.08&0.39 & 0.39&0.31 & 0.43 & 0.64 & 0.22\\
			\% of 3-cluster iterations&97.55\%  & 98.65\% & 98.55\% & 37.20\%&  85.05\% & 95.60\%& 65.90\% & 65.90\% &87.50\%\\
			Misclassification rate&3.50\% & 1.90\%  & 0.60\% &11.50\% &6.20\% &2.90\% & 12.90\% &7.00\%  &3.60\%\\
			\hline 
		\end{tabular*}
	}
\end{table}

\begin{figure}[ht]
	\centering
	\caption{Example 3: Trace plots of the number of clusters via a DP mixture model for Poisson models where $T= 30, 50$ and $100$ observations per subject (top, middle, bottom rows respectively) with a factor time-varying covariate.}
	\includegraphics[width=\textwidth]{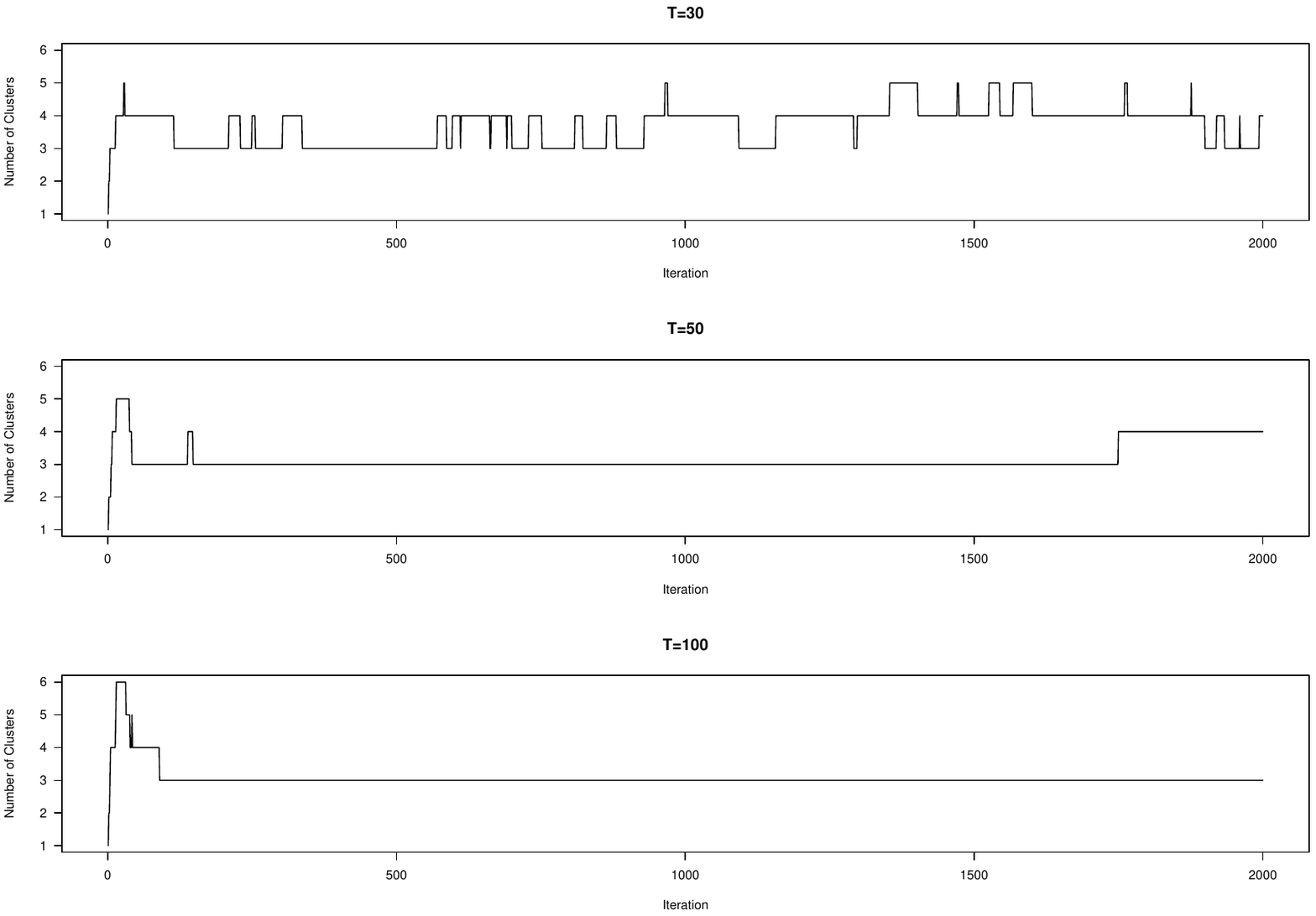}
	%\caption*{Gamma$\left(2,50\right)$}
	\label{tracedirex3}
\end{figure}

\begin{figure}[ht]
	\centering
	\caption{Example 3: Trace plots of the number of clusters via a mixture of finite mixture models using reversible-jump MCMC for Poisson models where $T= 30, 50$ and $100$ observations per subject (top, middle, bottom rows respectively) with a factor time-varying covariate.}
	\includegraphics[width=\textwidth]{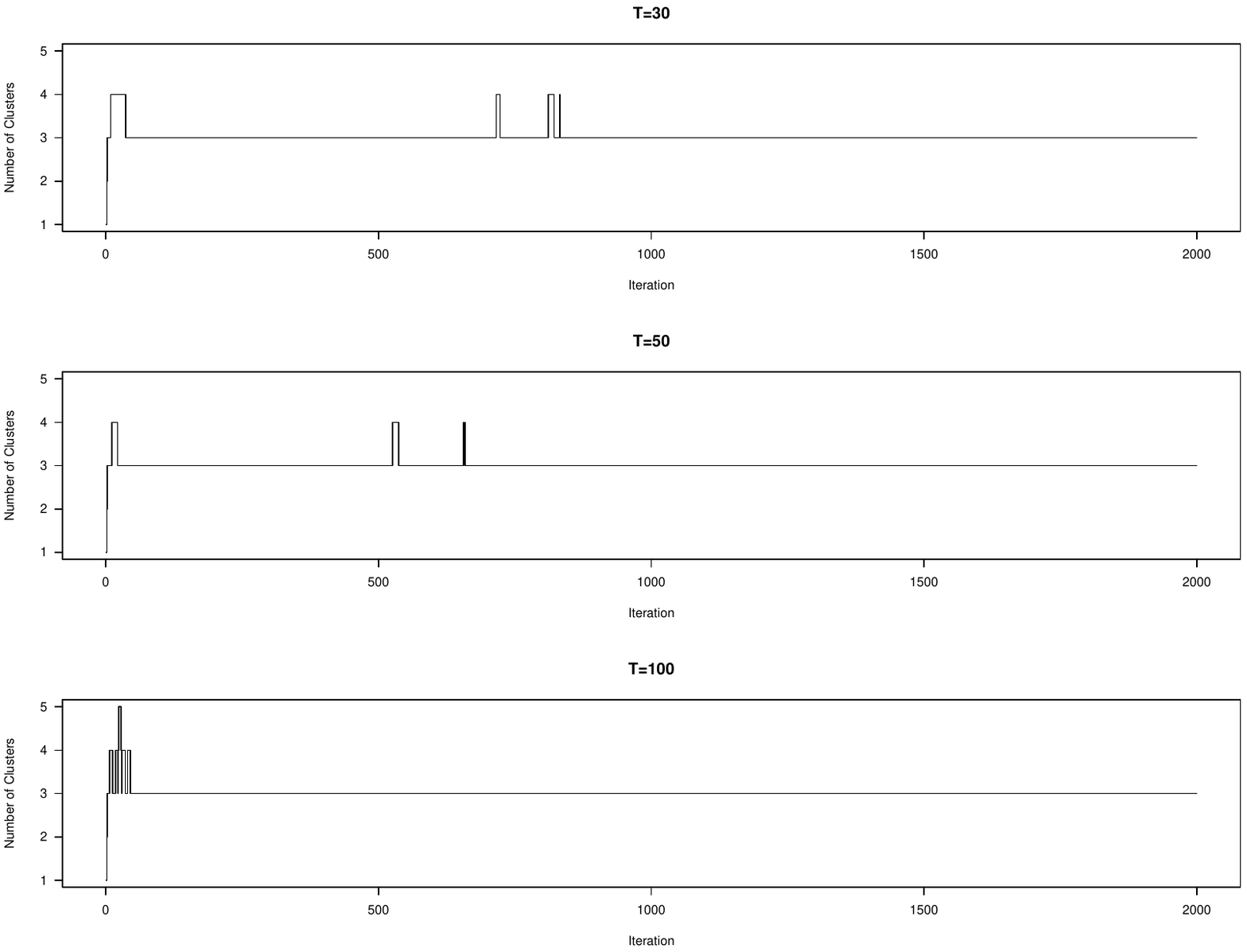}
	%\caption*{Gamma$\left(2,50\right)$}
	\label{tracedirRJ}
\end{figure}

\begin{figure}[ht]
	\centering
	\caption{Example 3: Trace plots of the number of clusters via a mixture of finite mixture models using the split-merge for Poisson models where $T= 30, 50$ and $100$ observations per subject (top, middle, bottom rows respectively) with a factor time-varying covariate.}
	\includegraphics[width=\textwidth]{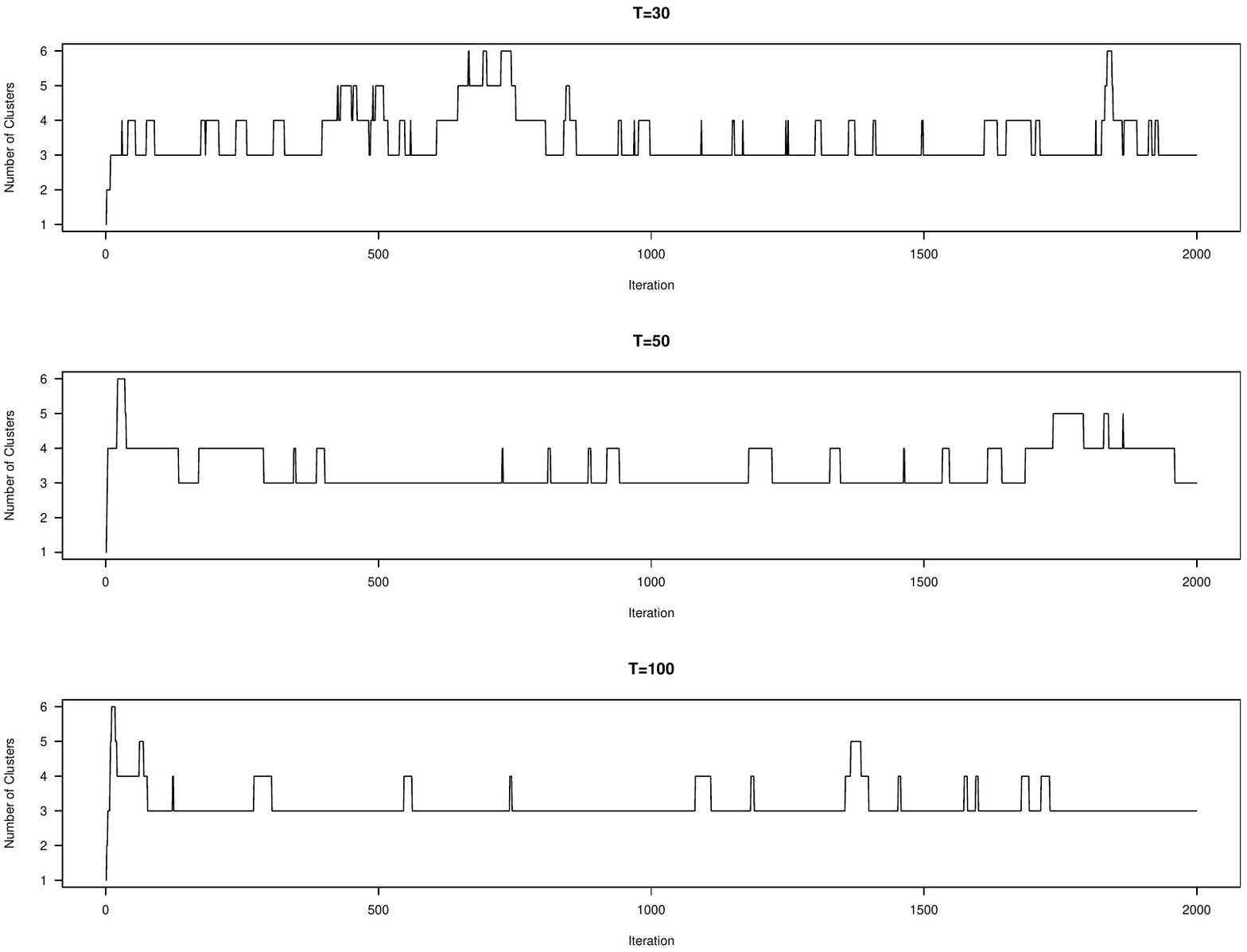}
	%\caption*{Gamma$\left(2,50\right)$}
	\label{tracedirmfm}
\end{figure}

\begin{figure}[ht]
	\centering
	\caption{Example 3: Trace plots of the number of clusters via a finite mixture model (left) and a DP mixture model (right) for Poisson models where $T= 100$ observations per subject with 10 times more precise prior for $Q$ than the example in Figures \ref{tracedirex3} and \ref{tracedirRJ}.  This prior encourages more clusters, but this plot indicates that the posterior distribution still becomes concentrated at three clusters.}
	\begin{minipage}[b]{0.425\textwidth}
		\includegraphics[width=\textwidth]{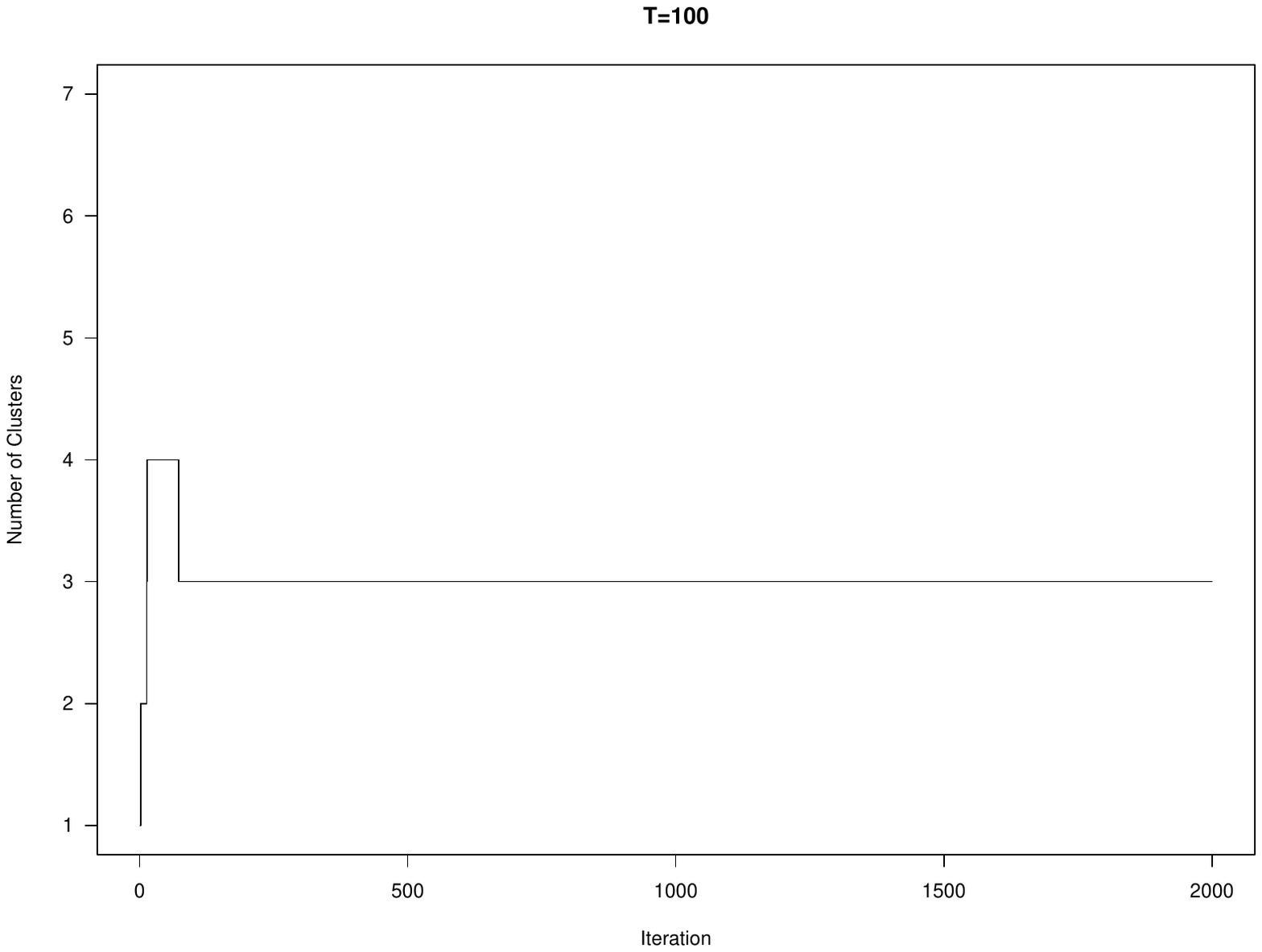}
		%\caption*{Gamma$\left(20,500\right)$}
	\end{minipage}
	\hfill
	\begin{minipage}[b]{0.465\textwidth}
		\includegraphics[width=\textwidth]{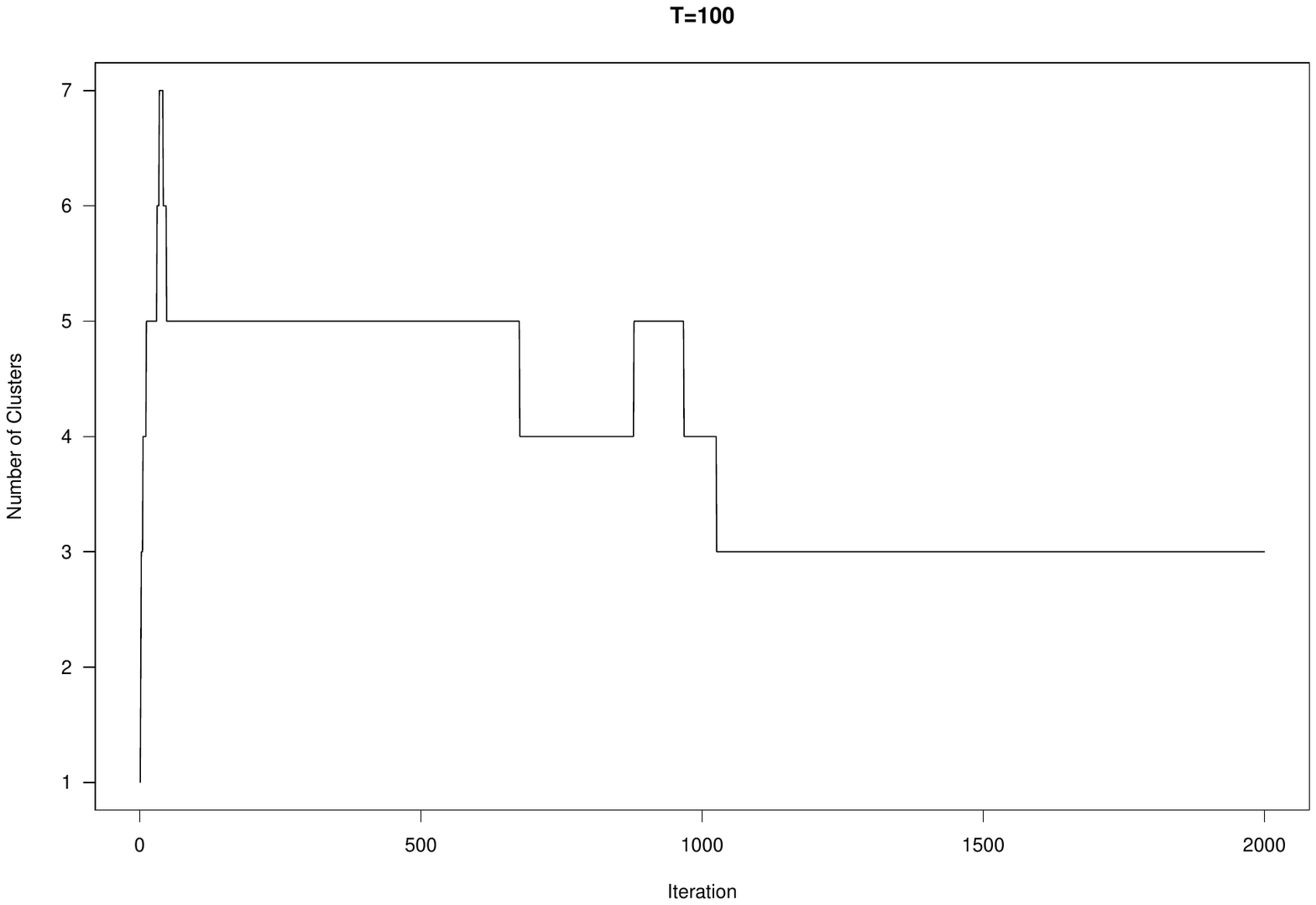}
		%\caption*{Gamma$\left(20,500\right)$}
	\end{minipage}
	\label{tracedirex310t}
\end{figure}

\section{Real example: Health surveillance of COPD patients}
The following figures and tables demonstrate adequacy of the MCMC approaches for the Bayesian posterior calculations for the real data.
\begin{figure}[ht]
	\centering
	\caption{Real data analysis: finite mixture model (left) and  DP mixture model (right), trace plot of number of clusters}
	\begin{minipage}[b]{0.425\textwidth}
		\includegraphics[width=\textwidth]{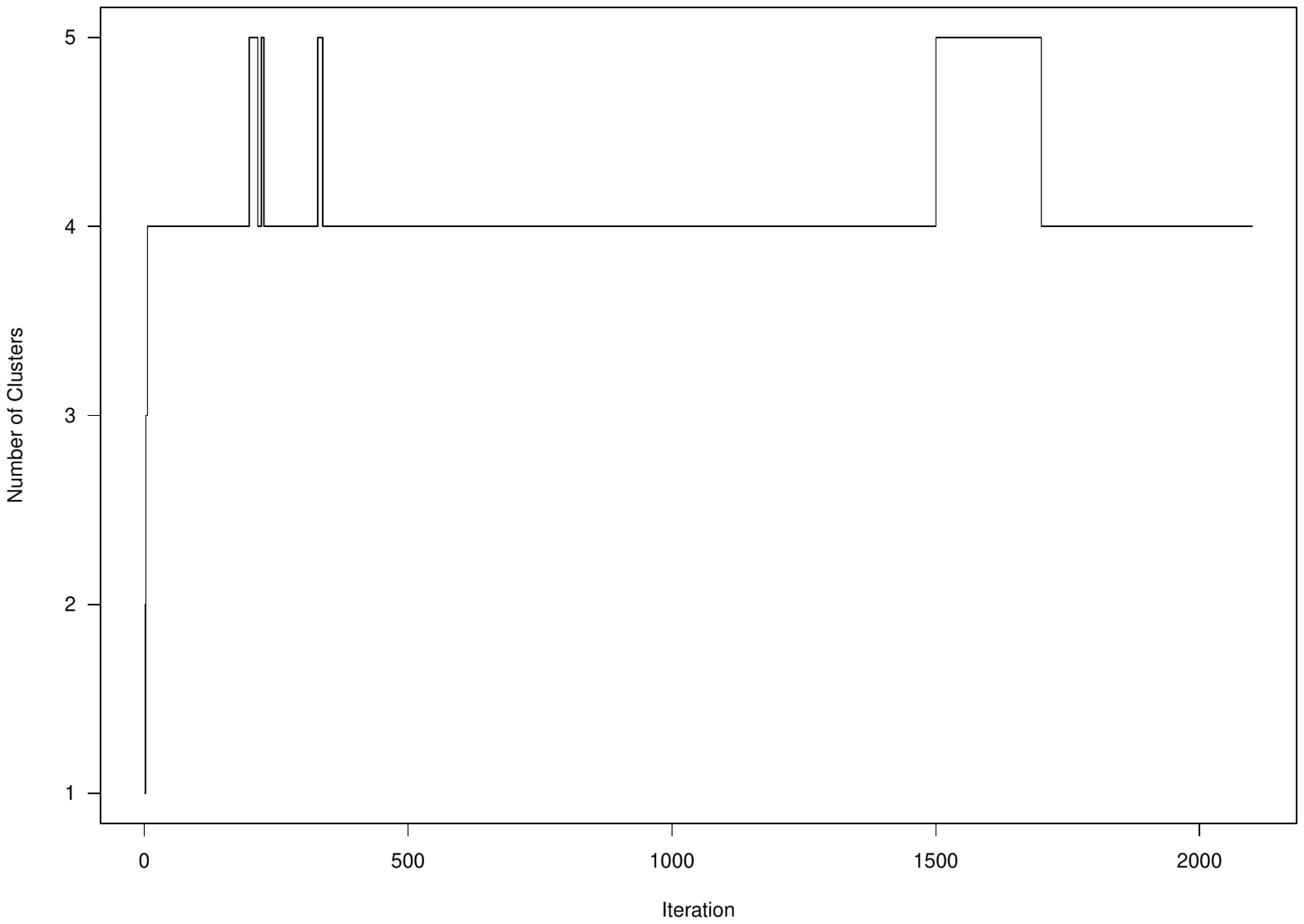}
		%\caption*{Gamma$\left(20,500\right)$}
	\end{minipage}
	\hfill
	\begin{minipage}[b]{0.425\textwidth}
		\includegraphics[width=\textwidth]{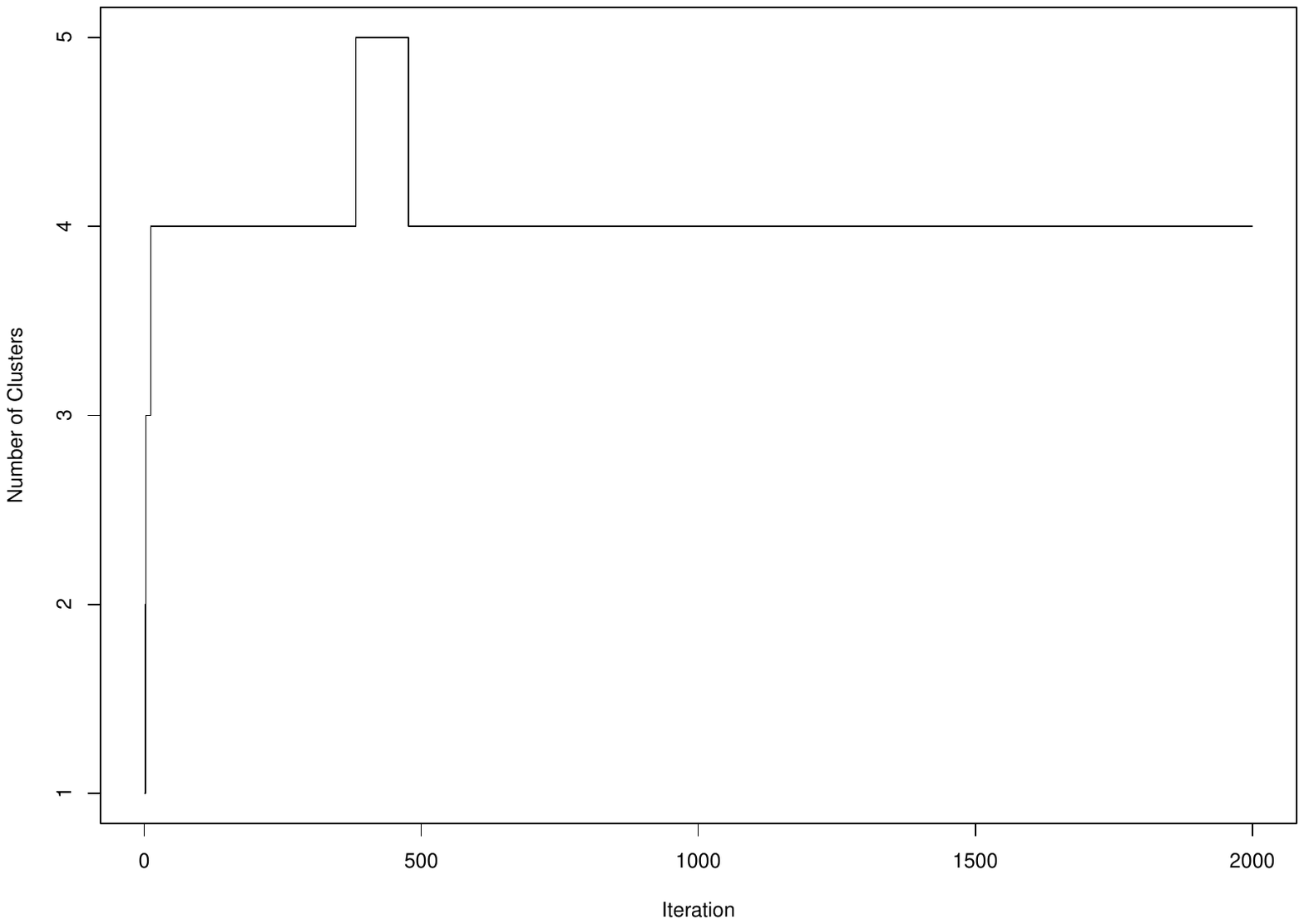}
		%\caption*{Gamma$\left(20,500\right)$}
	\end{minipage}
	\hfill	\label{trace}
\end{figure}

\begin{figure}[ht]
	\centering
	\caption{24,712 COPD Patients: Finite mixture model, trace plots of diagonal parameters in $Q$ conditional on the four cluster model}
	\includegraphics[scale=0.5]{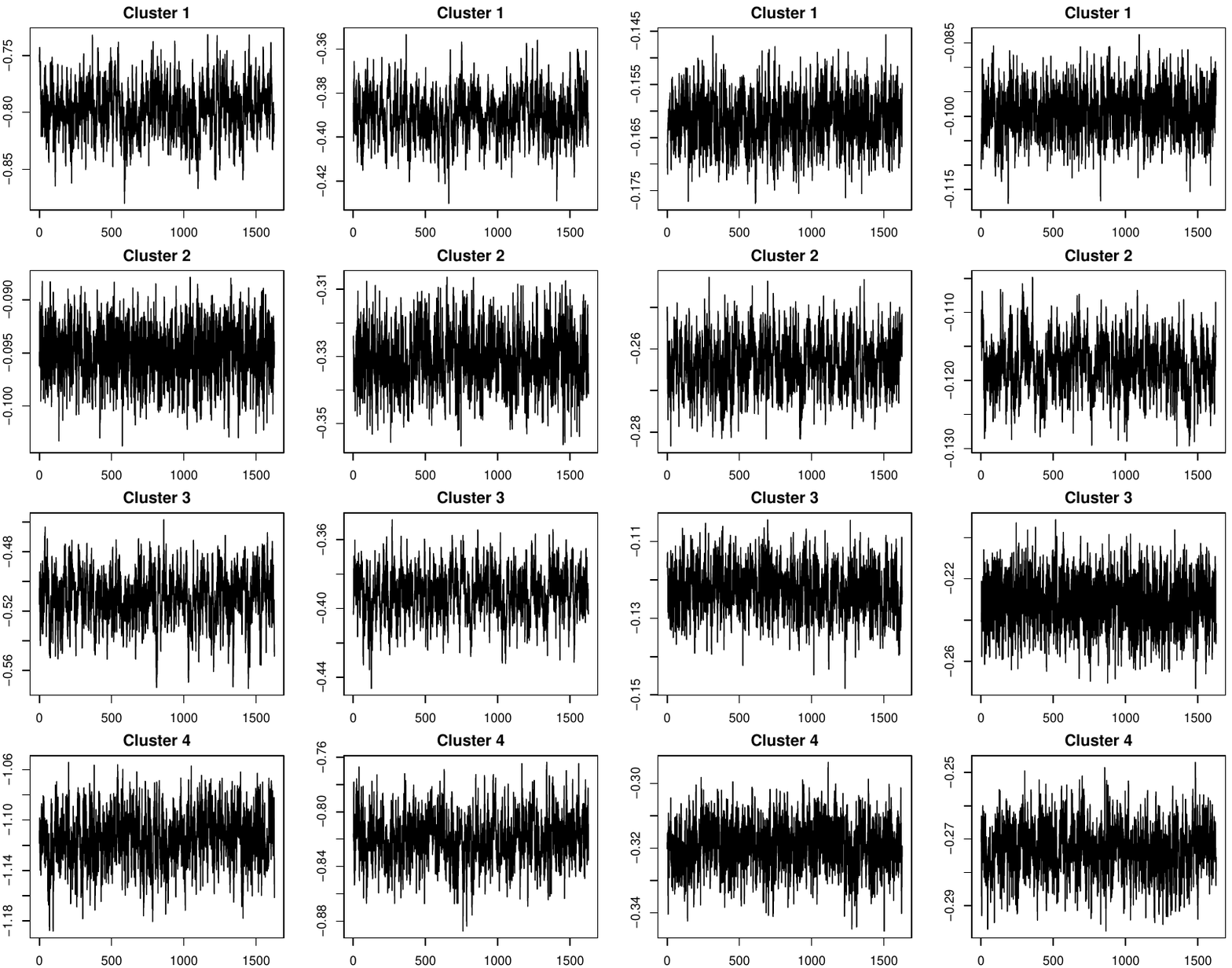}
	\label{tracefin1}
\end{figure}

\begin{figure}[ht]
	\centering
	\caption{24,712 COPD Patients: DP mixture model, trace plots of diagonal parameters in $Q$ conditional on the four cluster model}
	\includegraphics[scale=0.48]{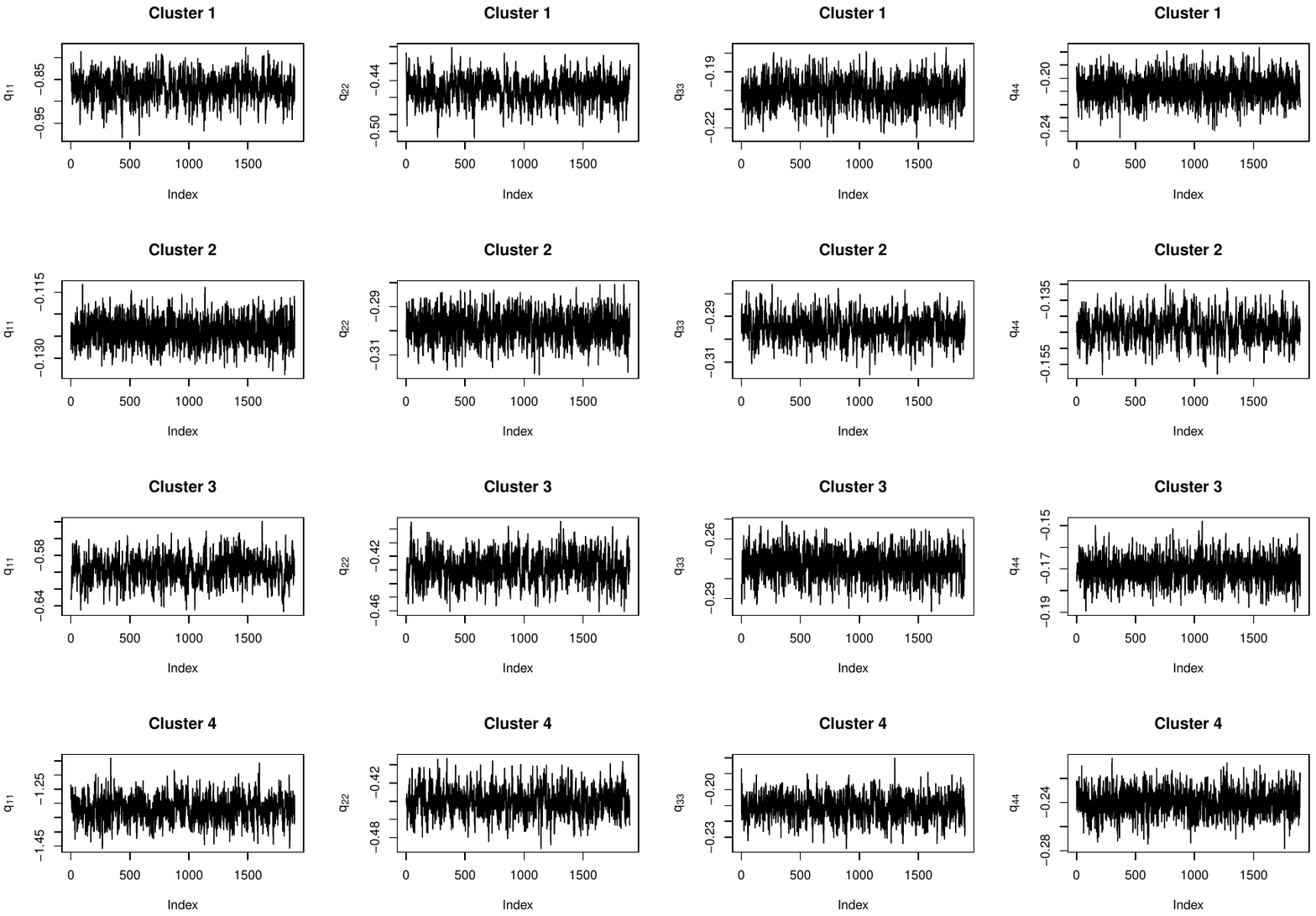}
	\label{traceQ1}
\end{figure}

\begin{table}[ht]
	\centering
	\caption{\label{app3332}24,712 COPD Patients: Finite mixture model analysis: exponential of coefficients for the three cluster Poisson model with 95\% credible intervals in relative-risk parameterization with GP as the baseline group.	}
	{
		\begin{tabular*}{40pc}{@{\hskip5pt}@{\extracolsep{\fill}}c@{}c@{}c@{}c@{}c@{}c@{\hskip5pt}}			
			\hline
			& Predictor & State 1 &  State 2 &  State 3 & State 4  \\
			\hline
			\multirow{3}{*}{Cluster 1}&\multirow{2}{*}{Intercept} & 0.29 & 1.35& 3.18& 10.71\\
			&  &(0.28,0.31) &  (1.33,1.37) & (3.16,3.21) & (10.62,10.79)\\
			&\multirow{2}{*}{ER}  & 1.08& 1.12& 1.13&  1.02  \\
			& &(0.96,1.21) & (1.08,1.17)&(1.11,1.16)& (1.00,1.03)\\
			\multirow{4}{*}{$\left(N=2108\right)$ }	&\multirow{2}{*}{HOSP} & 1.71& 1.19& 1.07&  0.99\\
			&  &(1.53,1.92) & (1.14,1.25)& (1.04,1.10)& (0.97,1.00)\\
			&\multirow{2}{*}{SEPC} &0.72 & 0.93& 0.99&  0.99\\
			&  &(0.62,0.83) & (0.90,0.97)& (0.97,1.02)& (0.97,1.01)\\
			\hline	
			\multirow{3}{*}{Cluster 2}&\multirow{2}{*}{Intercept} &0.00& 2.96& 4.89& 7.08\\
			&  &(0.00,0.00) & (2.94,2.99)&(4.86,4.91)& (7.05,7.10)\\
			&\multirow{2}{*}{ER}   & 1.49& 0.94& 1.00& 1.02 \\
			&  &(0.76,2.76) &(0.93,0.96)& (0.99,1.00) & (1.01,1.03)\\
			\multirow{4}{*}{$\left(N=12244\right)$}	&\multirow{2}{*}{HOSP} & 4.10& 0.91& 0.96& 0.99\\
			& &(2.33,7.02) & (0.90,0.93)&(0.95,0.97) & (0.98,1.00)\\
			&\multirow{2}{*}{SEPC} &4.04& 0.96& 0.95& 0.98\\
			&  &(2.40,6.79) & (0.95,0.97)&(0.94,0.96) & (0.97,0.98)\\
			\hline
			\multirow{3}{*}{Cluster 3}&\multirow{2}{*}{Intercept} & 1.02& 2.25& 4.21& 8.33\\
			&  &(1.00,1.05) & (2.22,2.27)& (4.19,4.22) & (8.29,8.37)\\
			&\multirow{2}{*}{ER} &  0.37& 1.06& 1.02& 1.03 \\
			&  &(0.35,0.40) & (1.04,1.08)& (1.01,1.03)& (1.02,1.04)\\
			\multirow{4}{*}{$\left(N=6521\right)$}	&\multirow{2}{*}{HOSP} & 0.37& 1.06& 1.00 &1.00 \\
			& &(0.33,0.40) & (1.03,1.09)& (0.98,1.01)& (0.98,1.01)\\
			&\multirow{2}{*}{SEPC} & 0.82& 0.97& 0.99& 1.00\\
			&  &(0.77,0.87) & (0.95,0.99)& (0.98,1.00)& (0.99,1.01)\\
			\hline
			\multirow{3}{*}{Cluster 4}&\multirow{2}{*}{Intercept} & 0.10& 1.81& 5.86& 9.16\\
			&  &(0.09,0.11) & (1.79,1.93)& (5.83,5.89) & (9.12,9.20)\\
			&\multirow{2}{*}{ER} & 0.09& 0.95& 0.98& 1.03\\
			&  &(0.06,0.12) & (0.93,0.98)& (0.97,0.99)& (1.02,1.04)\\
			\multirow{4}{*}{$\left(N=3839\right)$}	&\multirow{2}{*}{HOSP} & 0.17& 0.90& 0.94 & 1.00 \\
			& &(0.12,0.24) & (0.87,0.93)& (0.92,0.95)& (0.99.1.01)\\
			&\multirow{2}{*}{SEPC} & 0.44& 0.99& 0.98 & 0.99\\
			&  &(0.34,0.56) & (0.96,1.02)& (0.96,0.99)& (0.98,1.01)\\
			\hline
	\end{tabular*}}
\end{table}

\begin{table}[ht]
	\centering
	\caption{\label{app3331}24,712 COPD Patients: Dirichlet process mixture model analysis: exponential of coefficients for the three cluster Poisson model with 95\% credible intervals in relative-risk parameterization with GP as the baseline group.	}
	{
		\begin{tabular*}{40pc}{@{\hskip5pt}@{\extracolsep{\fill}}c@{}c@{}c@{}c@{}c@{}c@{\hskip5pt}}			
			\hline
			& Predictor & State 1 &  State 2 &  State 3 & State 4  \\
			\hline
			\multirow{3}{*}{Cluster 1}&\multirow{2}{*}{Intercept} & 0.24& 4.63&3.84&10.06 \\
			&  &(0.21,0.26) & (4.61,4.66)& (3.81,3.87)& (10.00,10.12)\\
			&\multirow{2}{*}{ER}  & 0.05& 1.10& 1.03&1.02   \\
			& &(0.02,0.08) & (1.06,1.14)&(1.02,1.05)& (1.01,1.03)\\
			\multirow{4}{*}{$\left(N=2846\right)$ }	&\multirow{2}{*}{HOSP} & 0.34& 1.12 &1.00& 0.99 \\
			&  &(0.25,0.44) & (1.08,1.18)& (0.98,1.02)& (0.98,1.01)\\
			&\multirow{2}{*}{SEPC} &0.41& 1.01& 0.99& 0.98\\
			&  &(0.32,0.53) & (0.97,1.04)& (0.98,1.01)& (0.97,1.00)\\
			\hline	
			\multirow{3}{*}{Cluster 2}&\multirow{2}{*}{Intercept} &0.00 & 2.61& 5.08& 7.28\\
			&  &(0.00,0.00) & (2.59,2.62)&(5.06,5.10)& (7.26,7.30)\\
			&\multirow{2}{*}{ER}   &  0.86 & 0.93&1.00&1.03 \\
			&  &(0.36,1.80) &(0.92,0.94)& (0.99,1.01) & (1.02,1.04)\\
			\multirow{4}{*}{$\left(N=13698\right)$}	&\multirow{2}{*}{HOSP} & 1.40& 0.88& 0.96&1.00\\
			& &(0.53,3.13) & (0.87,0.90)&(0.96,0.97) & (0.99,1.00)\\
			&\multirow{2}{*}{SEPC} &0.38 &0.94& 0.97& 0.98\\
			&  &(0.07,1.29) & (0.93,0.95)&(0.96,0.98) & (0.98,0.99)\\
			\hline
			\multirow{3}{*}{Cluster 3}&\multirow{2}{*}{Intercept} & 0.14& 1.44&4.52& 8.69 \\
			&  &(0.13,0.15) & (1.43,1.46)& (4.50,4.55) & (8.65,8.72)\\
			&\multirow{2}{*}{ER} &  0.17& 1.06& 1.01& 1.03 \\
			&  &(0.13,0.22) & (1.03,1.08)& (1.00,1.02)& (1.01,1.04)\\
			\multirow{4}{*}{$\left(N=5362\right)$}	&\multirow{2}{*}{HOSP} &  0.22&1.07& 0.97& 1.00 \\
			& &(0.16,0.30) & (1.04,1.11)& (0.95,0.98)& (0.99,1.01)\\
			&\multirow{2}{*}{SEPC} & 0.33& 0.96& 0.96&  0.99\\
			&  &(0.25,0.42) & (0.93,0.98)& (0.95,0.97)& (0.98,1.00)\\
			\hline
			\multirow{3}{*}{Cluster 4}&\multirow{2}{*}{Intercept} & 0.23& 1.98&4.01& 8.54 \\
			&  &(0.21,0.25) & (1.95,2.01)& (3.98,4.03) & (8.49,8.60)\\
			&\multirow{2}{*}{ER} &  0.08& 0.99& 1.06&1.03 \\
			&  &(0.05,0.13) & (0.96,1.02)& (1.04,1.08)& (1.01,1.04)\\
			\multirow{4}{*}{$\left(N=2806\right)$}	&\multirow{2}{*}{HOSP} & 0.07&0.95& 1.02& 0.99 \\
			& &(0.02,0.16) & (0.91,0.99)& (1.00,1.04)& (0.98.1.01)\\
			&\multirow{2}{*}{SEPC} & 0.33& 0.96&  0.98 &1.01\\
			&  &(0.23,0.46) & (0.92,0.99)& (0.97,1.00)& (0.99,1.02)\\
			\hline
	\end{tabular*}}
\end{table}

\begin{table}[h]
	\caption{\label{finess}24,712 COPD Patients: Finite mixture model, effective sample sizes (ESSs) for parameters in $Q$  conditional on 1636 four-cluster iterations}
	{
		\begin{tabular*}{40pc}{@{\hskip5pt}@{\extracolsep{\fill}}c@{}c@{}c@{}c@{}c@{}c@{}c@{}c@{\hskip5pt}}
			\hline
			Parameter & ESS& Parameter & ESS & Parameter & ESS &  Parameter & ESS   \\
			\hline
			\multicolumn{8}{c}{Cluster 1}\\
			\hline
			$q_{12}$ &  300.61 &$q_{21}$& 217.26 &$q_{31}$ & 265.46 &$q_{41}$&  1042.76\\
			$q_{13}$&  288.48&$q_{23}$& 515.78& $q_{32}$ & 217.66 &$q_{42}$& 591.93\\
			$q_{14}$ &851.18&$q_{24}$& 377.74&$q_{34}$ &973.16 &$q_{43}$& 483.49 \\
			\hline
			\multicolumn{8}{c}{Cluster 2}\\
			\hline
			$q_{12}$ & 309.20&$q_{21}$&562.45&$q_{31}$ &431.60&$q_{41}$&  683.10 \\
			$q_{13}$& 404.99&$q_{23}$& 516.30 & $q_{32}$ & 238.04&$q_{42}$& 287.36\\
			$q_{14}$ &461.82 &$q_{24}$&156.82&$q_{34}$ & 370.57 &$q_{43}$& 185.15\\
			
			\hline
			\multicolumn{8}{c}{Cluster 3}\\
			\hline
			$q_{12}$ &384.46&$q_{21}$& 217.28&$q_{31}$ &345.52 &$q_{41}$&  1165.85\\
			$q_{13}$& 467.94&$q_{23}$& 511.19 & $q_{32}$ & 289.33 &$q_{42}$&  900.17\\
			$q_{14}$ &1038.50 &$q_{24}$& 714.76&$q_{34}$ &  684.81 &$q_{43}$& 854.58\\
			\hline
			\multicolumn{8}{c}{Cluster 4}\\
			\hline
			$q_{12}$ &390.26 &$q_{21}$&376.44&$q_{31}$ &410.11 &$q_{41}$& 664.70\\
			$q_{13}$& 360.58&$q_{23}$& 569.69 & $q_{32}$ & 295.72&$q_{42}$& 366.11\\
			$q_{14}$ &635.86 &$q_{24}$& 298.25&$q_{34}$ &  720.29 &$q_{43}$& 282.69\\
			\hline
	\end{tabular*}}
\end{table}

\begin{table}[h]
	\caption{\label{xie11}24,712 COPD Patients: DP mixture model, effective sample sizes (ESSs) for parameters in $Q$  conditional on 1894 four-cluster iterations}
	{
		\begin{tabular*}{40pc}{@{\hskip5pt}@{\extracolsep{\fill}}c@{}c@{}c@{}c@{}c@{}c@{}c@{}c@{\hskip5pt}}
			\hline
			Parameter & ESS& Parameter & ESS & Parameter & ESS &  Parameter & ESS   \\
			\hline
			\multicolumn{8}{c}{Cluster 1}\\
			\hline
			$q_{12}$ &  413.62 &$q_{21}$& 286.26 &$q_{31}$ & 468.32 &$q_{41}$&  1161.28\\
			$q_{13}$&  508.90&$q_{23}$& 741.59& $q_{32}$ & 374.72 &$q_{42}$& 685.85\\
			$q_{14}$ &964.17 &$q_{24}$& 570.95&$q_{34}$ &1271.31 &$q_{43}$& 1059.18 \\
			\hline
			\multicolumn{8}{c}{Cluster 2}\\
			\hline
			$q_{12}$ & 876.39&$q_{21}$&715.00&$q_{31}$ &398.21 &$q_{41}$&  819.26 \\
			$q_{13}$& 452.95&$q_{23}$& 750.99 & $q_{32}$ & 411.18&$q_{42}$& 366.51\\
			$q_{14}$ &586.39 &$q_{24}$&298.37&$q_{34}$ & 1356.51 &$q_{43}$& 240.72\\
			
			\hline
			\multicolumn{8}{c}{Cluster 3}\\
			\hline
			$q_{12}$ &306.68 &$q_{21}$& 292.76&$q_{31}$ &407.86 &$q_{41}$&  1031.25\\
			$q_{13}$& 516.50&$q_{23}$& 794.15 & $q_{32}$ & 338.10 &$q_{42}$& 449.77\\
			$q_{14}$ &762.50 &$q_{24}$& 446.57&$q_{34}$ &  1052.53 &$q_{43}$& 712.01\\
			\hline
			\multicolumn{8}{c}{Cluster 4}\\
			\hline
			$q_{12}$ &372.38 &$q_{21}$&390.12&$q_{31}$ &407.82 &$q_{41}$& 820.60\\
			$q_{13}$& 291.55&$q_{23}$& 400.30 & $q_{32}$ & 303.40 &$q_{42}$& 550.24\\
			$q_{14}$ &743.86 &$q_{24}$& 576.38&$q_{34}$ &  1052.53 &$q_{43}$& 719.27\\
			\hline
	\end{tabular*}}
\end{table}

24,712 COPD Patients: Finite mixture model analysis, effective sample sizes (out of 1636) from the MCMC analysis.
\begin{equation*}
\text{}B_1= \left(\begin{array}{cccc}
345.13 & 363.50& 1043.26 & 186.97\\
659.54& 842.31&  951.77 &1500.14\\
702.02& 965.89 &1232.28 &1238.58\\
724.61 &878.36 &1580.02 & 904.11\\
\end{array}\right)
\end{equation*}
\begin{equation*}
\text{}
B_2= \left(\begin{array}{cccc}
990.85 & 494.09& 237.19 & 265.83\\
896.22 &1061.16 &827.78 &1196.76\\
1015.40 & 915.37 &962.76 &1303.17\\
711.15 &1031.49 &941.85 &1340.41\\
\end{array}\right)
\end{equation*}
\begin{equation*}
\text{}
B_3= \left(\begin{array}{cccc}
317.83 & 455.61 & 975.34 &1633.40\\
573.53 &917.85 &1256.76 &1626.00\\
709.59 &923.34 &1232.51 &1626.00\\
653.84& 908.40 &1223.54 &1897.06\\
\end{array}\right)
\end{equation*}
\begin{equation*}
\text{}
B_4= \left(\begin{array}{cccc}
359.48  & 725.70 & 933.28& 1073.54\\
532.56  &948.04 &1170.37 &1339.97\\
729.75 & 980.18 &1137.84 &1283.59\\
653.06 &1062.22 & 937.84& 1316.31\\
\end{array}\right)
\end{equation*}

24,712 COPD Patients: DP mixture model analysis, effective sample sizes (out of 1894) from the MCMC analysis.
\begin{equation*}
\text{}B_1= \left(\begin{array}{cccc}
299.61 & 376.56& 1098.02 & 1487.14\\
458.36&  748.37& 1389.51& 1612.77\\
731.24& 1066.26& 1416.24& 1668.70\\
625.73&  884.41& 1435.79& 1710.12\\
\end{array}\right)
\end{equation*}

\[
B_2= \left(\begin{array}{cccc}
935.95&  918.36&  273.36&  394.80\\
888.59&1174.44& 942.72& 1259.58\\
808.25& 1327.58 & 1080.74 & 1334.23 \\
254.06& 1098.53&1005.48& 1230.68\\

\end{array}\right)
\]
\[
B_3= \left(\begin{array}{cccc}
391.79& 483.80&  864.93& 1229.37\\
509.37 & 1167.20& 1248.01& 1573.17\\
603.78& 1005.89& 1285.82 & 1598.57\\
603.20& 1075.36& 1331.98& 1675.99\\
\end{array}\right)
\]

\[
B_4= \left(\begin{array}{cccc}
331.92&  450.80&  740.91& 1259.97\\
700.80 & 1029.96 & 1268.15& 1526.97\\
148.08 & 1100.08 & 1384.23& 1460.41\\
614.14& 989.11 & 1304.94& 1428.55\\
\end{array}\right)
\]

\begin{figure}[ht]
	\centering
	\includegraphics[scale=0.5]{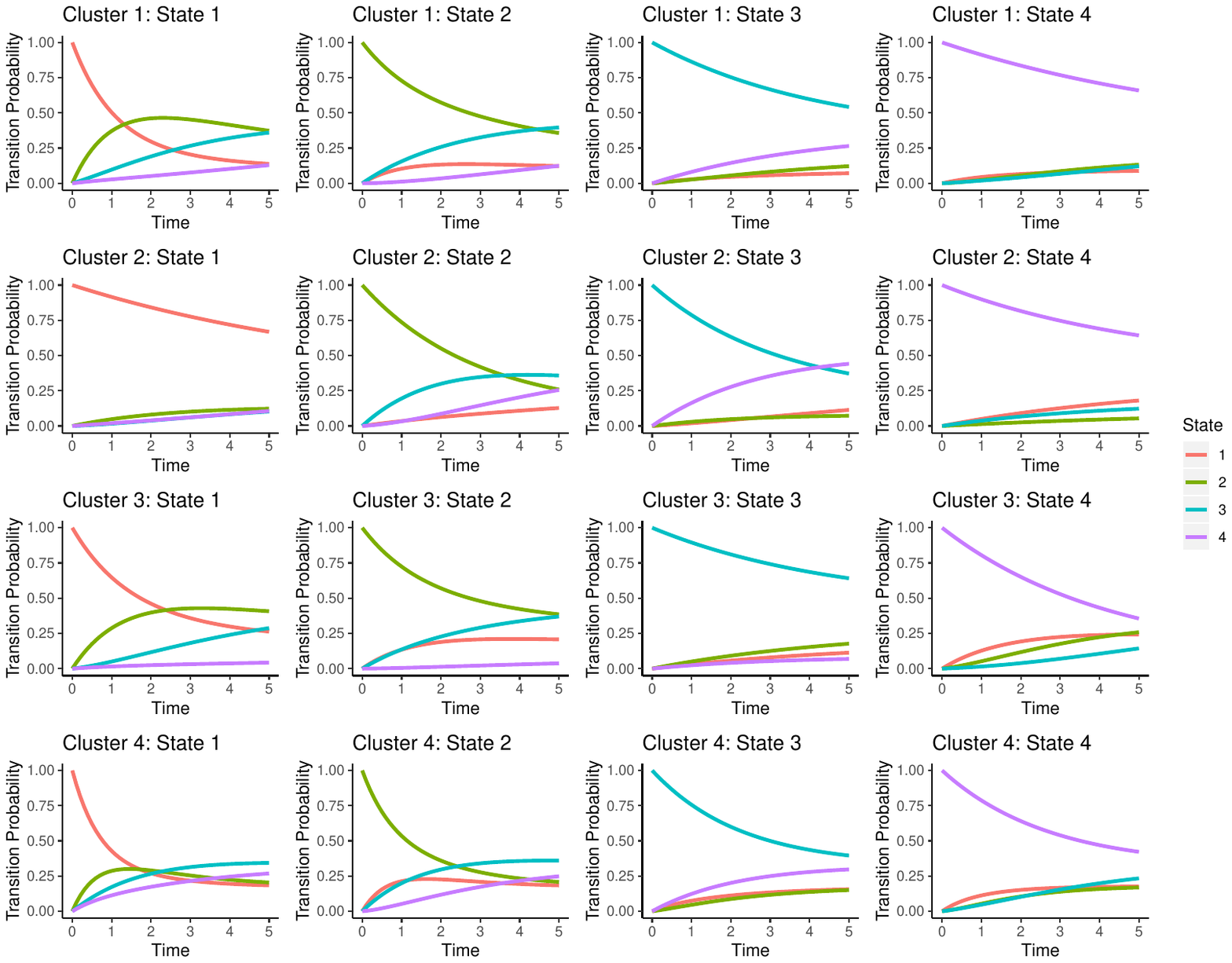}
	\caption{24,712 COPD Patients: Finite mixture model, transition probability over time for four clusters}
	\label{clufin}
\end{figure}

\begin{figure}[ht]
	\centering
	\includegraphics[scale=0.5]{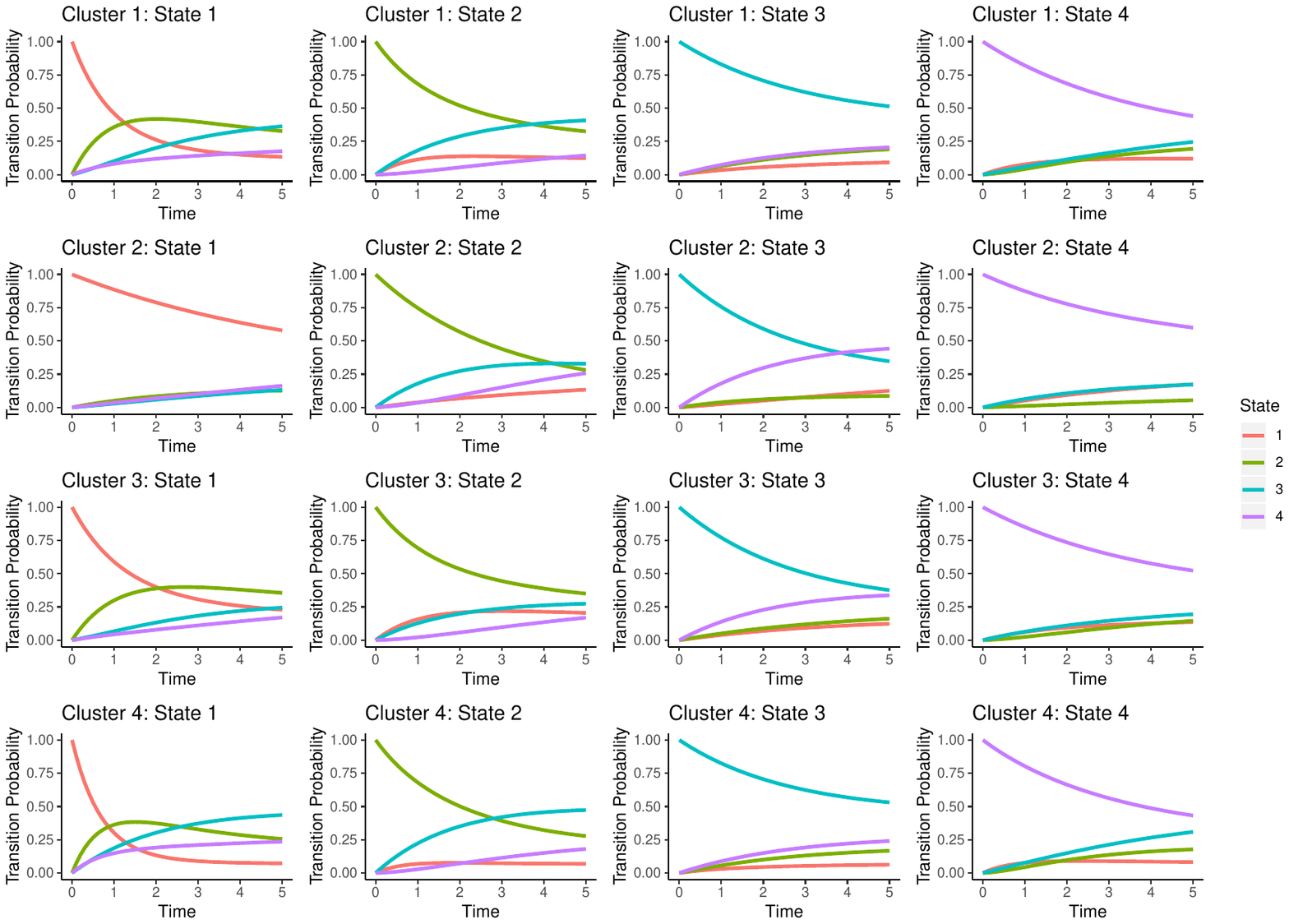}
	\caption{24,712 COPD Patients: Dirichlet process mixture model, transition probability over time for four clusters}
	\label{cluDir1}
\end{figure}

\begin{figure}[ht]
	\centering
	\caption{24,712 COPD Patients: Finite mixture model analysis, box plots of posterior samples of elements of $Q$.  Within each panel, boxes 1, 2, 3 and 4 correspond to the four clusters.}
	\includegraphics[scale=0.5]{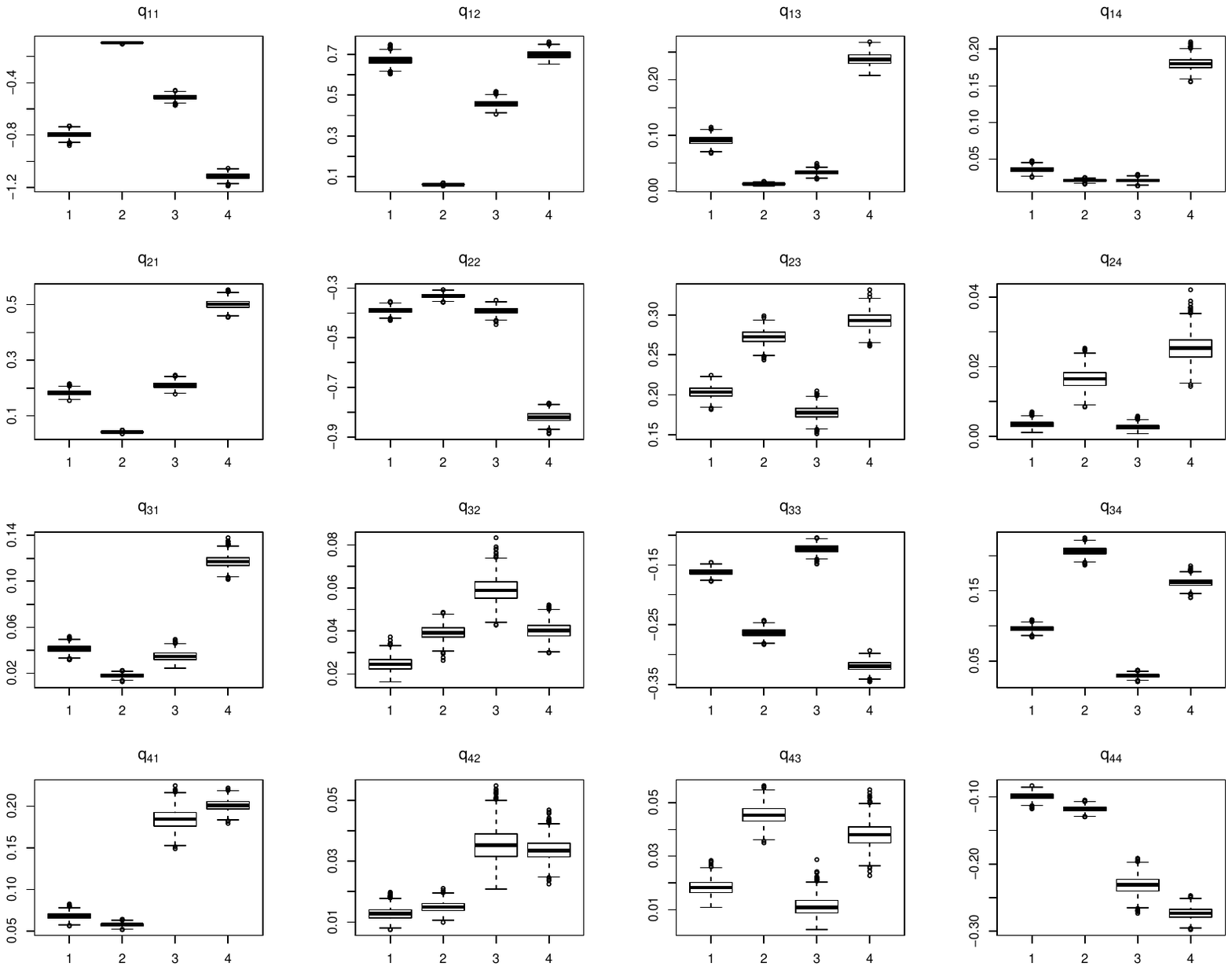}
	\label{boxfinQ1}
\end{figure}

\begin{figure}[ht]
	\centering
	\caption{24,712 COPD Patients: DP mixture model analysis, box plots of posterior samples of elements of $Q$.  Within each panel, boxes 1, 2, 3 and 4 correspond to the four clusters.}
	\includegraphics[scale=0.5]{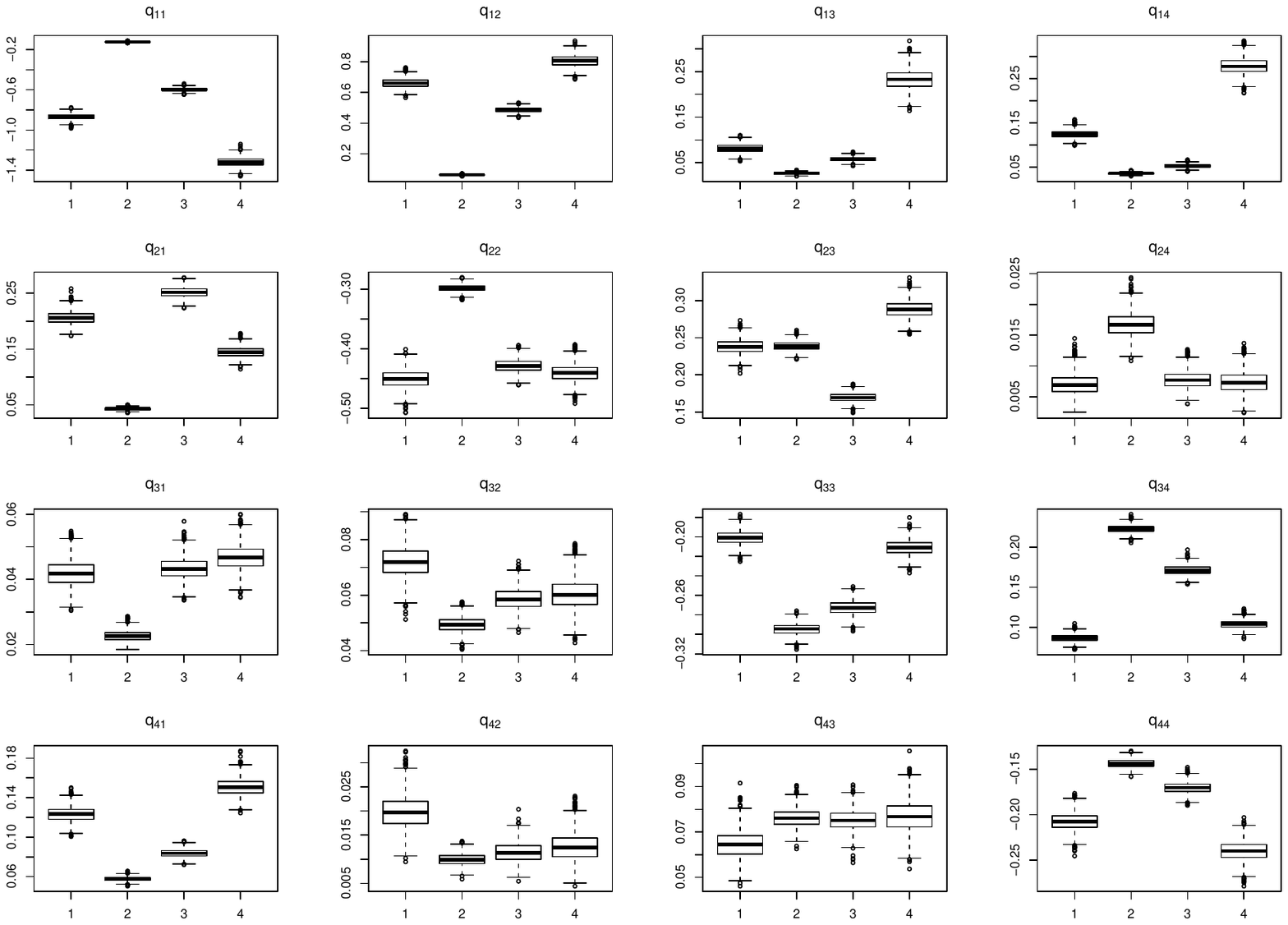}
	\label{boxQ1}
\end{figure}

\begin{figure}[ht]
	\centering
	\caption{24,712 COPD Patients: Finite mixture model analysis, eigenvalues of posterior samples for $Q$.}
	\includegraphics[scale=0.48]{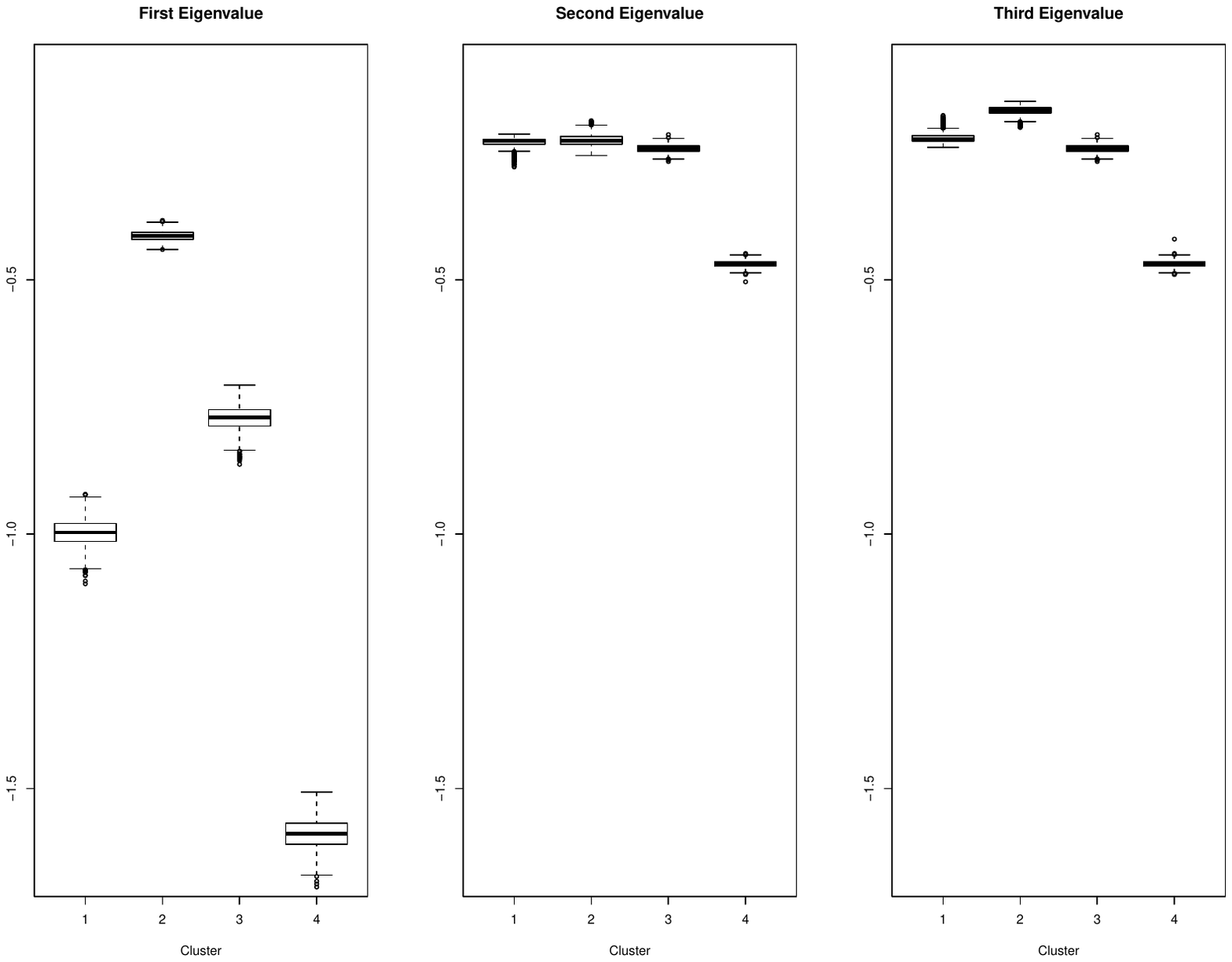}
	\label{eigenfinQ1}
\end{figure}

\begin{figure}[ht]
	\centering
	\caption{24,712 COPD Patients: DP mixture model analysis, eigenvalues of posterior samples for $Q$.}
	\includegraphics[scale=0.48]{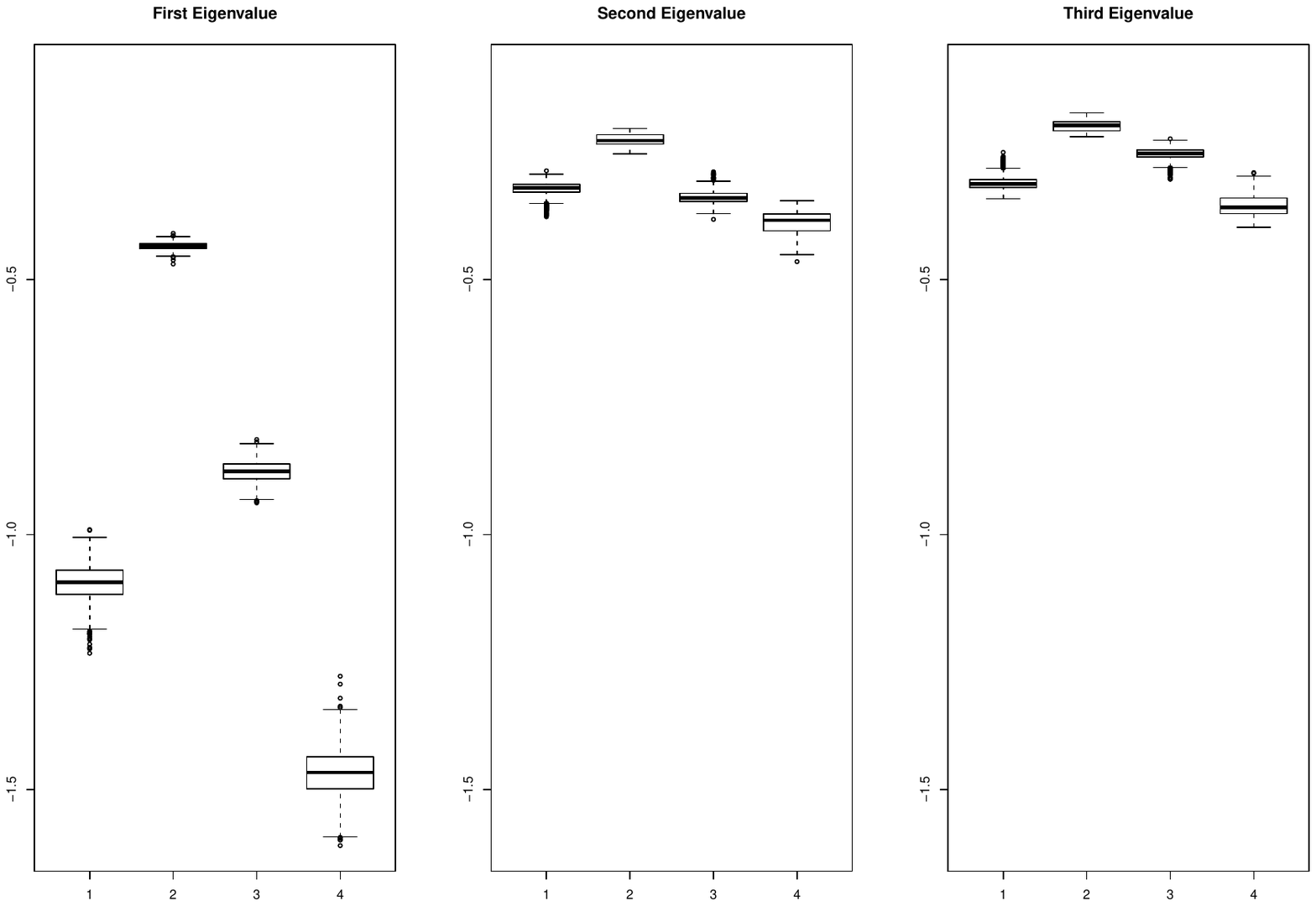}
	\label{eigenQ1}
\end{figure}

\end{document}